\documentclass[a4paper,12pt]{article}

\usepackage[left=24mm, right=24mm, top=30mm, bottom=30mm]{geometry}
\usepackage[utf8]{inputenc}
\usepackage[english]{babel}
\usepackage{amsmath}
\usepackage{amsfonts}
\usepackage{amssymb}
\usepackage{mathrsfs}
\usepackage{dsfont}
\usepackage{graphicx}
\usepackage{cite}
\usepackage{xcolor}
\usepackage[colorlinks, allcolors=blue, linktocpage=true]{hyperref}

\newcommand{\ket}[1]{\left| #1 \right\rangle}
\newcommand{\bra}[1]{\left\langle #1 \right|}
\newcommand{\Lagr}{\mathscr{L}}
\renewcommand{\O}{\mathcal{O}}
\DeclareMathOperator{\im}{\text{im}}
\newcommand{\SO}{\text{SO}}

\numberwithin{equation}{section}

\newcounter{exercise}[section]
\newenvironment{exercise}[1][]%
	{\refstepcounter{exercise}\bigskip
	\noindent\textbf{Exercise~\thesection.\theexercise #1} \rmfamily
	\itshape}
  	{\bigskip}

\title{Conformal field theory for particle physicists}

\author{Marc Gillioz}

\date{SISSA, via Bonomea 265, 34136 Trieste, Italy%
\footnote{Former affiliation. Current correspondence address: \emph{firstname.lastname@gmail.com}.}}

\begin{document} 

\maketitle

\vspace{20mm}

\begin{abstract}
	This is a set of introductory lecture notes on conformal field theory. Unlike most existing reviews on the subject, CFT is presented here from the perspective of a unitary quantum field theory in Minkowski space-time. It begins with a non-perturbative formulation of quantum field theory (Wightman axioms), and then gradually focuses on the implications of scale and special conformal symmetry, all the way to the modern conformal bootstrap. This approach includes topics often left out, such as subtleties of conformal transformations in Minkowski space-time, the construction of Wightman functions and time-ordered correlators both in position- and momentum-space, unitarity bounds derived from the spectral representation, and the appearance of UV and IR divergences.

	\vspace{15mm}

	\emph{These notes were created for a graduate class
	on conformal field theory at
	the Institute for Theoretical Physics of the University of Bern,
	taught in the spring semester of 2022. 
	They are aimed at physicists with a good basic knowledge
	of quantum field theory but without prior experience
	in conformal field theory.}
	
	\vspace{15mm}
	
	\centering
	\noindent\fbox{\parbox{0.94\linewidth}{%
	This is a preprint of the following work: Marc Gillioz,
	``Conformal Field Theory for Particle Physicists:
	From QFT Axioms to the Modern Conformal Bootstrap'', 
	2023, SpringerBriefs in Physics,
	reproduced with permission of Springer Nature Switzerland AG.
	The final authenticated version is available online at:
	\url{https://doi.org/10.1007/978-3-031-27086-4}.}}
\end{abstract}

%%%%%%%%%%%%%%%%%%%%%%%%%%%%%%%%%%%%%%%%%%%%%%%%%%%%%%%%%%%%%%%%%%%%%%%

\newpage

\tableofcontents

%%%%%%%%%%%%%%%%%%%%%%%%%%%%%%%%%%%%%%%%%%%%%%%%%%%%%%%%%%%%%%%%%%%%%%%

\newpage
\section{Introduction}

Conformal field theory is a ubiquitous subject in modern theoretical physics. Every local quantum field theory approaches a CFT in the large- and small-distance limits,%
\footnote{Massless free theories are conformal. The ``empty'', low-energy limit of massive theories can also be viewed as a special kind of CFT.}
and it even plays a key role in the study of quantum gravity through the AdS/CFT correspondence. CFT is also one of the rare frameworks in which quantum field theory can be studied outside the realm of perturbation theory.
There a several excellent modern reviews on the subject~\cite{Rychkov:2016iqz, Simmons-Duffin:2016gjk, Poland:2018epd, Chester:2019wfx, Qualls:2015qjb},
and large parts of these lecture notes are directly inspired by them.

Most introductory courses on CFT treat the conformal group as a whole. This approach usually requires working in Euclidean space (we will see why),% 
\footnote{An exception is Slava Rychkov's unpublished lecture notes on \emph{Lorentzian methods in conformal field theory}, available at \url{https://courses.ipht.fr/node/226}.
}
and the connection with ``traditional'' quantum field theory appears very late in the course, if at all.
This can be frustrating for a particle physicist.
We propose to start here instead with the following definition:
\begin{equation}
\begin{aligned}
	\text{CFT}
	&= \text{relativistic QFT} 
	\\
	& \quad
	+ \text{scale symmetry}
	\\
	& \quad
	+ \text{special conformal symmetry},
\end{aligned}
\end{equation}
and go through each part one after the other.
The first part is quantum field theory in flat Minkowski space-time. We expect the reader to be familiar with it, but our attention will not be restricted to theories that have a nice classical limit and a perturbative definition.
We will therefore go through some of the basics of QFT in section~\ref{sec:quantum}, namely the Wightman axioms, but without unnecessary mathematical rigor.

Scale symmetry will come next. Particle physicists typically also have a good physical intuition of it from the renormalization group. The main novelty in CFT is that scale symmetry forbids the existence of particles with a definite mass; instead, in a scale-invariant theory, there are states of \emph{any} energy (unless we are in the very special case of a free theory).

Finally, special conformal symmetry will be discussed last. It nearly always comes along with scale symmetry, but it is vastly more powerful. This will lead us to the study of conformal correlation functions in section~\ref{sec:correlators}, to the discussion of the operator product expansion (OPE) and its very nice features in section~\ref{sec:OPE}, and finally to the conformal bootstrap in section~\ref{sec:bootstrap}, including an example of a strongly-coupled theory that has been solved by symmetry principles only.

Before diving into the \emph{quantum} theory, however, we begin the course with a look at \emph{classical} conformal transformations in section~\ref{sec:classical}. It is convenient to work in $d$-dimensional space-time and think of $d = 4$ as a particular realization of the more general framework.
Lorentz indices $\mu$, $\nu$, $\ldots$ therefore run between $0$ and $d - 1$, with $x^0$ being the time component of the vector $x^\mu$, and $p^0$ the energy component of the momentum $p^\mu$.
Even though we ultimately care about unitary quantum field theory in Minkowski space-time, we will need to establish a connection with the same theory defined in flat Euclidean space.
For this reason, we (unfortunately) work with the mostly-plus metric convention $(-, +, +, \dots, +)$,
so that going from Minkowski space-time to Euclidean space is simply achieved by a rotation of the time coordinate in the complex plane.

%%%%%%%%%%%%%%%%%%%%%%%%%%%%%%%%%%%%%%%%%%%%%%%%%%%%%%%%%%%%%%%%%%%%%%

\section{Classical conformal transformations}
\label{sec:classical}

One of the most fundamental principles of physics is the independence of the reference frame: observers living at different points might have different perspectives, but the underlying physical laws are the same.
This is true in space (invariance under translations and rotations), but also in space-time (invariance under Lorentz boosts).

\subsection{Infinitesimal transformations}

In mathematical language, this means that if we have a coordinate system $x^\mu$, the laws of physics do not change under a transformations 
\begin{equation}
	x^\mu \to x'^\mu.
	\label{eq:diffeo}
\end{equation}
This principle applies to all maps that are invertible (isomorphisms) and differentiable (smooth transformations), hence it is usually called \emph{diffeomorphism} invariance.
Being differentiable, the transformation \eqref{eq:diffeo} can be Taylor-expanded to write
\begin{equation}
	x^\mu \to x'^\mu = x^\mu + \varepsilon^\mu(x),
	\label{eq:diffeo:infinitesimal}
\end{equation}
in terms of an infinitesimal vector $\varepsilon^\mu$ (meaning that we will always ignore terms of order $\varepsilon^2$).

In addition to the coordinate system, the description of a physical system requires a way of measuring distances that is provided by a metric $g_{\mu\nu}(x)$. Distances are measured integrating an infinitesimal line element whose square equals
\begin{equation}
	ds^2 = g_{\mu\nu}(x) dx^\mu dx^\nu.
\end{equation}
Since all observers should agree on the measure of distances, we must have
\begin{equation}
	g'_{\mu\nu}(x') dx'^\mu dx'^\nu = g_{\mu\nu}(x) dx^\mu dx^\nu,
\end{equation}
Here $g_{\mu\nu}$ could be the Euclidean metric $\delta_{\mu\nu}$ or the Minkowski metric $\eta_{\mu\nu}$; for simplicity we only consider the case in which $g_{\mu\nu}$ is flat, i.e.~$\partial_\alpha g_{\mu\nu} = 0$.
In this case, we can write
\begin{equation}
\begin{aligned}
	g'_{\mu\nu} &= g_{\alpha\beta}
	\frac{\partial x^\alpha}{\partial x'^\mu}
	\frac{\partial x^\beta}{\partial x'^\nu}
	\\
	&= g_{\alpha\beta}
	\left( \delta^\alpha_\mu - \partial_\mu \varepsilon^\alpha \right)
	\left( \delta^\beta_\nu - \partial_\nu \varepsilon^\beta \right)
	\\
	&= g_{\mu\nu}
	- \left( \partial_\mu \varepsilon_\nu
	+ \partial_\nu \varepsilon_\mu \right).
\end{aligned}
\label{eq:metric:transformation}
\end{equation}
If we require the different observers to also agree on the metric, then  we must have $g'_{\mu\nu} = g_{\mu\nu}$, which gives a constraint on what kind of coordinate transformations are possible: they must satisfy
\begin{equation}
	\partial_\mu \varepsilon_\nu
	+ \partial_\nu \varepsilon_\mu = 0.
\end{equation}
This condition admits as most general solution
\begin{equation}
	\varepsilon^\mu = a^\mu + \omega^\mu_{~\nu} x^\nu,
\end{equation}
where $a^\mu$ is a constant vector and $\omega_{\mu\nu} = g_{\mu\rho} \omega^\rho_{~\nu}$ an antisymmetric tensor, i.e.~$\omega_{\nu\mu} = - \omega_{\mu\nu}$.
The transformation
\begin{equation}
	x^\mu \xrightarrow{P} x^\mu + a^\mu
\end{equation}
is obviously a translation and
\begin{equation}
	x^\mu \xrightarrow{M}
	\left( \delta^\mu_\nu + \omega^\mu_{~\nu} \right) x^\nu
\end{equation}
a rotation/Lorentz transformation around the origin $x = 0$: the matrix $R^\mu_{~\nu} = \delta^\mu_\nu + \omega^\mu_{~\nu}$ satisfies $R^\mu_{~\alpha} g^{\alpha\beta} R^{T\nu}_\beta = g^{\mu\nu}$. The composition of these two operations generates the Poincaré group.
This is the fundamental symmetry of space-time underlying all relativistic quantum field theory. It is a symmetry of nature to a very good approximation, at least up to energy scales at which quantum gravity becomes important.

However, one can also consider the situation in which the two observers use different systems of units, i.e.~they disagree on the overall definition of scale but agree otherwise on the metric being flat.
In this case we must have $g'_{\mu\nu} \propto g_{\mu\nu}$, and therefore the constraint \eqref{eq:metric:transformation} becomes
\begin{equation}
	\partial_\mu \varepsilon_\nu
	+ \partial_\nu \varepsilon_\mu = 2 \lambda g_{\mu\nu},
\end{equation}
for some positive real number $\lambda$,
with the most general solution
\begin{equation}
	\varepsilon^\mu = a^\mu + \omega^\mu_{~\nu} x^\nu
	+ \lambda x^\mu.
\end{equation}
The new infinitesimal transformation is
\begin{equation}
	x^\mu \xrightarrow{D} (1 + \lambda) x^\mu.
\end{equation}
It is a scale transformation, also known as dilatation. Note that scale symmetry is not a good symmetry of nature: there is a fundamental energy scale on which all observers must agree (this can be for instance chosen to be the mass of the electron).
Nevertheless, there are systems in which this is a very good approximate symmetry, making it worth studying.

If one pushes this logic further, in a scale-invariant world in which observers have no physical means of agreeing on a fundamental scale, they might even decide to change their definition of scale as they walk around, or as time passes. This would correspond to a situation in which the metric $g'_{\mu\nu}$ of one observer can differ from the original metric $g_{\mu\nu}$ by a function of space-time:
\begin{equation}
	g'_{\mu\nu}(x) = \Omega(x) g_{\mu\nu}.
\end{equation}
Note that we are not saying that $g'_{\mu\nu}$ is completely arbitrary: at every point in space-time, it is related to the flat metric by a scale transformation. But the scale factor is different at every point.
The condition on $\varepsilon^\mu$ becomes in this case
\begin{equation}
	\partial_\mu \varepsilon_\nu
	+ \partial_\nu \varepsilon_\mu = 2 \sigma g_{\mu\nu},
	\label{eq:conformalkillingeq}
\end{equation}
where $\sigma$ is the infinitesimal version of $\Omega$, with the conventional relation $\Omega(x) = e^{-2 \sigma(x)} \approx 1 - 2\sigma(x)$. To find the most general solution to this equation, note that contracting the indices with $g^{\mu\nu}$ gives
\begin{equation}
	\partial_\mu \varepsilon^\mu  = d \sigma,
\end{equation}
where $d$ is the space(-time) dimension,
while acting with $\partial^\nu$ gives
\begin{equation}
	\partial_\mu \partial_\nu \varepsilon^\nu
	+ \partial^2 \varepsilon_\mu
	= 2 \partial_\mu\sigma,
\end{equation}
so that we get
\begin{equation}
	\partial^2 \varepsilon_\mu
	= (2 - d) \partial_\mu\sigma.
\end{equation}
Acting once again with $\partial^\mu$, we arrive at 
\begin{equation}
	(d - 1) \partial^2 \sigma = 0,
\end{equation}
while acting with $\partial^\nu$ and symmetrizing the indices yields
\begin{equation}
	(2 - d) \partial_\mu\partial_\nu \sigma
	= g_{\mu\nu} \partial^2 \sigma.
\end{equation}
The condition $\partial^2 \sigma = 0$ must therefore be satisfied in all dimensions ($d > 1$), and in $d > 2$ the additional condition $\partial_\mu \partial_\nu \sigma = 0$, which is solved by 
\begin{equation}
	\sigma(x) = \lambda + 2 b \cdot x.
\end{equation}
The corresponding value of $\varepsilon^\mu$ is
\begin{equation}
	\varepsilon^\mu = a^\mu + \omega^\mu_{~\nu} x^\nu
	+ \lambda x^\mu
	+ 2 (b \cdot x) x^\mu - x^2 b^\mu.
	\label{eq:conformalkillingvec}
\end{equation}
Therefore, in addition to the transformations found before, we also have
\begin{equation}
	x^\mu \xrightarrow{K} x^\mu +  2 (b \cdot x) x^\mu - x^2 b^\mu,
\end{equation}
which is called \emph{special conformal transformation}.
If we examine the Jacobian for this transformation, we find
\begin{equation}
	\frac{\partial x'^\mu}{\partial x^\nu}
	= \left(1 + 2 b \cdot x \right) \delta^\mu_\nu
	+ 2 \left( b_\nu x^\mu - x_\nu b^\mu \right)
	\approx \left(1 + 2 b \cdot x \right)
	R^\mu_{~\nu}(x).
\end{equation}
We have written this as a position-dependent scale factor $(1 + 2 b \cdot x)$, multiplying an orthogonal matrix
\begin{equation}
	R^\mu_{~\nu}(x) = \delta^\mu_\nu
	+ 2 \left( b_\nu x^\mu - x_\nu b^\mu \right).
\end{equation}
This shows that special conformal transformations act locally as the composition of a scale transformation and a rotation (or Lorentz transformation). This also shows that conformal transformations preserve angles, which is the origin of their name.
Eq.~\eqref{eq:conformalkillingeq} is called the conformal Killing equation and its solutions \eqref{eq:conformalkillingvec} the Killing vectors.

Note that in our derivation the original metric $g_{\mu\nu}$ was flat, but the new metric $g'_{\mu\nu}$ is not. It is however \emph{conformally flat}: it is always possible to make a change of coordinate after which it is flat again. In general, transformations
\begin{equation}
	g_{\mu\nu}(x) \to \Omega(x) g_{\mu\nu}(x)
\end{equation}
are called Weyl transformations. They change the geometry of space-time. We found that Weyl transformations that are at most quadratic in $x$ can be compensated by a change of coordinates to go back to flat space. The corresponding flat-space transformation is called conformal transformation.%
\footnote{This implies that the group of conformal transformation is a subgroup of diffeomorphisms. It is in fact the largest finite-dimensional subgroup.}

In $d = 2$ the situation is a bit different: the conditions $\partial^2 \sigma = 0$ is sufficient to ensure that the Killing equation has a solution. This is most easily seen in light-cone coordinates,
\begin{equation}
	x^+ = \frac{x^0 + x^1}{2},
	\qquad
	x^- = \frac{x^0 - x^1}{2},
\end{equation}
in terms of which
\begin{equation}
	\partial^2 \sigma = \partial_+ \partial_- \sigma.
\end{equation}
This is satisfied by taking for $\sigma$ the sum of an arbitrary function of the coordinate $x^+$ and of another arbitrary function of $x^-$. In fact, if we write $\varepsilon^\pm = \varepsilon^0 \pm \varepsilon^1$, we can take arbitrary functions $\varepsilon^+(x^+)$ and $\varepsilon^-(x^-)$, and verify that eq.~\eqref{eq:conformalkillingeq} is satisfied with $\sigma = \frac{1}{2} \left( \partial_+ \varepsilon_+ + \partial_- \varepsilon_- \right)$.
In Euclidean space, we define
\begin{equation}
	z = \frac{x^1 + i x^2}{2},
	\qquad
	\bar{z} = \frac{x^1 - i x^2}{2},
\end{equation}
complex-conjugate to each other, and the same logic follows: we can apply arbitrary holomorphic and anti-holomorphic transformations on $z$ and $\bar{z}$, and the conformal Killing equation is always satisfied. This shows that there are (infinitely) many more conformal transformations in $d = 2$ than in $d > 2$, and also that there is no significant difference between Euclidean and Minkowski conformal transformations in $d = 2$, as the transformation acts essentially on the two light-cone/holomorphic coordinates independently.

\subsection{The conformal algebra}

The conformal Killing equation \eqref{eq:conformalkillingvec} determines the most general form of \emph{infinitesimal} conformal transformations. \emph{Finite} conformal transformations follow from a sequence of infinitesimal transformations.
However, one has to bear in mind that infinitesimal conformal transformations do not commute: for instance, a translation followed by a rotation is not the same as the opposite.
The conformal transformations form a \emph{group}: the composition of conformal transformations is again a conformal transformation.

As we know from quantum field theory, a group is characterized by its \emph{generators} and their commutation relations (the \emph{algebra}). A generator $G$ describes an infinitesimal transformation in some direction, and finite transformations are obtained by exponentiation, $e^{i \theta G}$, with parameter $\theta$ (the factor of $i$ is a physicist's convention that makes the generators Hermitian).
A representation of the conformal group can be obtained from smooth functions of the coordinates, $f(x)$. For instance, under an infinitesimal translation, we have
\begin{equation}
	f(x) \xrightarrow{P} f(x') = f(x + a) \approx f(x)
	+ a^\mu \partial_\mu f(x)
\end{equation}
and we require this to be equal to $e^{-i a_\mu P^\mu} f(x)$, which means
\begin{equation}
	P_\mu = i \partial_\mu.
	\label{eq:P:fcts}
\end{equation}
Performing the same analysis for the other infinitesimal transformations given in eq.~\eqref{eq:conformalkillingvec}, we obtain for the other generators%
\footnote{The sign of these generators is an arbitrary convention. It defines once and for all the commutations relations that we will derive next. After that, we will always refer to the commutation relations as defining the generators for other representations of the conformal group.}
\begin{align}
	& \text{rotations/Lorentz transformations:} \quad &
	M^{\mu\nu} &= i \left( x^\mu \partial^\nu
	- x^\nu \partial^\mu \right)
	\label{eq:M:fcts}
	\\
	& \text{scale transformations:} & 
	D &= i x^\mu \partial_\mu,
	\label{eq:D:fcts}
	\\
	& \text{special conformal transformations:} &
	K^\mu &= i \left( 2 x^\mu x^\nu \partial_\nu 
	- x^2 \partial^\mu \right).
	\label{eq:K:fcts}
\end{align}
The number of generators matches that of the Killing vectors: there are $d$ translations, $d$ special conformal transformations, $d (d-1)/2$ rotations/Lorentz transformations ($M^{\mu\nu}$ is a $d \times d$ antisymmetric matrix), and one scale transformation.
Therefore the total number of generators, i.e.~the dimension of this group, is $(d + 1)(d + 2)/2$. In $d = 4$ space-time dimensions, the conformal group has 15 generators.

Using the above definition, one can verify that the following commutation relations are satisfied,
\begin{equation}
\begin{aligned}
	\left[ M^{\mu\nu}, M^{\rho\sigma} \right]
	&= -i \left( g^{\mu\rho} M^{\nu\sigma} - g^{\mu\sigma} M^{\nu\rho}
	- g^{\nu\rho} M^{\mu\sigma} + g^{\nu\sigma} M^{\mu\rho} \right)
	\\
	\left[ M^{\mu\nu}, P^\rho \right]
	&= -i \left( g^{\mu\rho} P^\nu - g^{\nu\rho} P^\mu \right)
	\\
	\left[ M^{\mu\nu}, K^\rho \right]
	&= -i \left( g^{\mu\rho} K^\nu - g^{\nu\rho} K^\mu \right)
	\\
	\left[ D, P^\mu \right] &= -i P^\mu 
	\\
	\left[ D, K^\mu \right] &= i K^\mu 
	\\
	\left[ P^\mu, K^\nu \right]
	&= 2i \left( g^{\mu\nu} D - M^{\mu\nu} \right)
\end{aligned}
\label{eq:conformalalgebra}
\end{equation}
while all other commutators vanish:
\begin{equation}
	\left[ M^{\mu\nu}, D \right]
	= \left[ P^\mu, P^\nu \right] 
	= \left[ K^\mu, K^\nu \right] 	
	= 0.
\end{equation}
The first two relations in eq.~\eqref{eq:conformalalgebra} are the familiar Poincaré algebra. 
The next one states that $K^\mu$ transforms like a vector (as $P^\mu$ does), whereas $D$ is obviously a scalar. The next two relations remind us that $K^\mu$ and $P^\mu$ have respectively the dimension of length and inverse length. 

Even though it is not immediately obvious, this algebra is isomorphic to that of the group $\SO(d+1, 1)$ if $g^{\mu\nu}$ is the Euclidean metric, or $\SO(d, 2)$ if it is the Minkowski metric.
To see that it is the case, let us introduce a $(d + 2)$-dimensional space with coordinates
\begin{equation}
	X^\mu, \quad X^{d+1}, \quad X^{d+2},
\end{equation}
and a metric defined by the line element
\begin{equation}
	ds^2 = g_{\mu\nu} dX^\mu dX^\nu + dX^{d+1} dX^{d+1}
	- dX^{d+2} dX^{d+2}
	\equiv \eta_{MN} dX^M dX^N.
\end{equation}
$X^{d+1}$ is a new spatial coordinate, and $X^{d+2}$ a new time.
Then we can write all conformal commutation relations as being defined by the Lorentzian algebra
\begin{equation}
	\left[ J^{MN}, J^{RS} \right]
	= -i \left( \eta^{MR} J^{NS} - \eta^{MS} J^{NR}
	- \eta^{NR} J^{MS} + \eta^{NS} J^{MR} \right),
\end{equation}
provided that we identify the antisymmetric generators $J^{MN}$ with the conformal generators as follows:
\begin{equation}
\begin{aligned}
	M^{\mu\nu} &= J^{\mu\nu},
	\\
	P^\mu &= J^{\mu, d+1} + J^{\mu, d+2},
	\\
	K^\mu &= J^{\mu, d+1} - J^{\mu, d+2},
	\\
	D &= J^{d+1, d+2}.
\end{aligned}
\label{eq:embeddingspacealgebra}
\end{equation}

\subsection{Finite transformations}

We just saw that the infinitesimal conformal transformations generate a group. But how can we describe finite conformal transformations? Let us see how each generator exponentiates into an element of the group; the most general conformal transformation can then be obtained as a composition of such finite transformations.

In some cases the exponentiation is trivial. For instance, with translations we obtain immediately
\begin{equation}
	x^\mu \xrightarrow{P} x^\mu + a^\mu,
\end{equation}
where $a$ is now any $d$-dimensional vector, not necessarily small.
The same is true of scale transformations,
\begin{equation}
	x^\mu \xrightarrow{D} \lambda x^\mu
\end{equation}
with finite $\lambda$.
Rotations or Lorentz transformations exponentiate as
\begin{equation}
	x^\mu \xrightarrow{M} \Lambda^\mu_{~\nu} x^\nu
\end{equation}
where $\Lambda^\mu_{~\nu}$ is a $\SO(d)$ or $\SO(1, d-1)$ matrix, depending whether the metric is Euclidean or Minkowski.
All of this is standard in quantum field theory.

On the contrary, special conformal transformations do not exponentiate trivially. The easiest way to derive their finite form is to make the following observation: recall that in infinitesimal form we have
\begin{equation}
	x'^\mu = x^\mu + 2 (b \cdot x) x^\mu - x^2 b^\mu,
\end{equation}
which implies $x'^2 = \left( 1 + 2 b \cdot x \right) x^2$, and therefore (as always neglecting terms of order $b^2$)
\begin{equation}
	\frac{x'^\mu}{x'^2}
	= \frac{x^\mu}{x^2} - b^\mu.
	\label{eq:K:inversion}
\end{equation}
The ratio $x^\mu/x^2$ appearing on both side of the equation is the \emph{inverse} of the coordinate $x^\mu$, respectively $x'^\mu$: the inversion is defined by
\begin{equation}
	x^\mu \xrightarrow{I} \frac{x^\mu}{x^2}.
\end{equation}
This transformation does not have an infinitesimal form, but otherwise it shares the essential properties of a conformal transformation: its Jacobian is
\begin{equation}
	\frac{\partial x'^\mu}{\partial x^\nu}
	= \frac{1}{x^2}
	\left[ \delta^\mu_\nu - 2 \frac{x^\mu x_\nu}{x^2} \right],
\end{equation}
which is the product of a position-dependent scale factor ($x^{-2}$) with an orthogonal matrix. To understand what this transformation does globally, let us consider a Euclidean point $\vec{x} = (a, 0, \ldots 0) \in \mathbb{R}^d$. Then the matrix in square brackets is diagonal, and equates \linebreak $\text{diag}(-1, 1, \ldots, 1)$. This is an orthogonal matrix with determinant $-1$, which is part of $\text{O}(d)$ but not $\SO(d)$. This shows that the inversion is a discrete transformation not connected to the identity.
A conformally-invariant theory might be invariant under inversions, but it needs not be.

Eq.~\eqref{eq:K:inversion} shows that infinitesimal special conformal transformations are obtained taking an inversion followed by a translation, followed by an inversion again. Since this process involves the inversion twice, and since inversion is its own inverse, it does not matter whether inversion is a true symmetry of the system or not.
The advantage of this representation is that it can easily be exponentiated: the composition of (infinitely) many infinitesimal special conformal transformation can be written as an inversion followed by a finite translation, followed by an inversion again. In other words, eq.~\eqref{eq:K:inversion} holds for finite $b^\mu$.
This can be used to show that
\begin{equation}
	x^\mu \xrightarrow{K}
	x'^\mu = \frac{x^\mu - x^2 b^\mu}
	{1 - 2 b \cdot x + b^2 x^2}.
	\label{eq:K:finite}
\end{equation}

\begin{exercise}
	Use eq.~\eqref{eq:K:inversion} to show \eqref{eq:K:finite}.
\end{exercise}

What do special transformation do globally?
Let us look specifically at Euclidean space. There are some special points:
\begin{itemize}

\item
The origin of the coordinate system $x = 0$ is mapped onto itself.

\item
The point $b^\mu / b^2$ is mapped to $\infty$. 

\item
Conversely, the ``point'' $x \to \infty$ is mapped to the finite value $-b^\mu/b^2$.

\end{itemize}
These properties can be understood from the fact that special conformal transformations and translations are related by inversion:
special conformal transformations keep the origin fixed but move every other point, including $\infty$; translations move every point except $\infty$. The other two transformations, rotations and scale transformations, keep both $0$ and $\infty$ fixed.

An essential property of conformal transformations is that they let us map any 3 points $(x_1, x_2, x_3)$ onto another triplet $( x'_1, x'_2, x'_3)$. This can be seen as follows: first, apply a translation to place $x_1$ at the origin, followed by a special conformal transformation that takes $x_3$ to $\infty$, after which the image of the original triplet is $(0, x_2'', \infty)$; then use rotations and scale transformations to move $x''_2$ to another point $x'''_2$, while keeping $0$ and $\infty$ fixed; finally apply again a special conformal transformation that takes $\infty$ to $x'_3 - x'_1$, and a translation by $x'_1$ to reach the configuration $( x'_1, x'_2, x'_3)$.
This property has an immediate physical consequence: in correlation functions involving 2 or 3 local operators (see next sections for a definition), all kinematics is fixed by conformal symmetry. The only freedom is about the operators themselves, not about their position in space.

Another interesting property of conformal transformations is that they map spheres to spheres: this is an obvious property of translations, rotations, and scale transformations, but it is also true of special conformal transformations. 

\begin{exercise}
	Show that under the special conformal transformation \eqref{eq:K:finite}, a sphere centered at the point $a^\mu$ and with radius $R$ gets mapped to a sphere centered at the point 
	$$
	a'^\mu = \frac{a^\mu - (a^2 - R^2) b^\mu}
	{1 - 2 a \cdot b + (a^2 - R^2) b^2}
	$$
	and with radius
	$$
	R' = \frac{R}{\left| 1 - 2 a \cdot b + (a^2 - R^2) b^2 \right|}.
	$$
	In the special case in which $b^\mu / b^2$ lies on the surface of the original sphere, show that the sphere gets mapped to a plane orthogonal to the vector $a^\mu + (R^2 - a^2) b^\mu$.
	Note that a plane is a sphere of infinite radius.
\end{exercise}

In $d = 2$, the additional conformal transformations (infinitely many of them!) mean that (nearly) any shape can be mapped onto another. This is known as the Riemann mapping theorem.

\subsection{Compactifications}

We mentioned earlier that conformal symmetry is a symmetry of flat space(-time).
It is true, as we have just seen, provided that we treat the point $\infty$ as being part of the space.
This is quite straightforward in Euclidean space, but much more subtle in Minkowski space-time, as there are different, inequivalent ways of reaching $\infty$ there.

\begin{figure}
	\centering
	\vspace{-12mm}
	\includegraphics[width=0.75\linewidth]{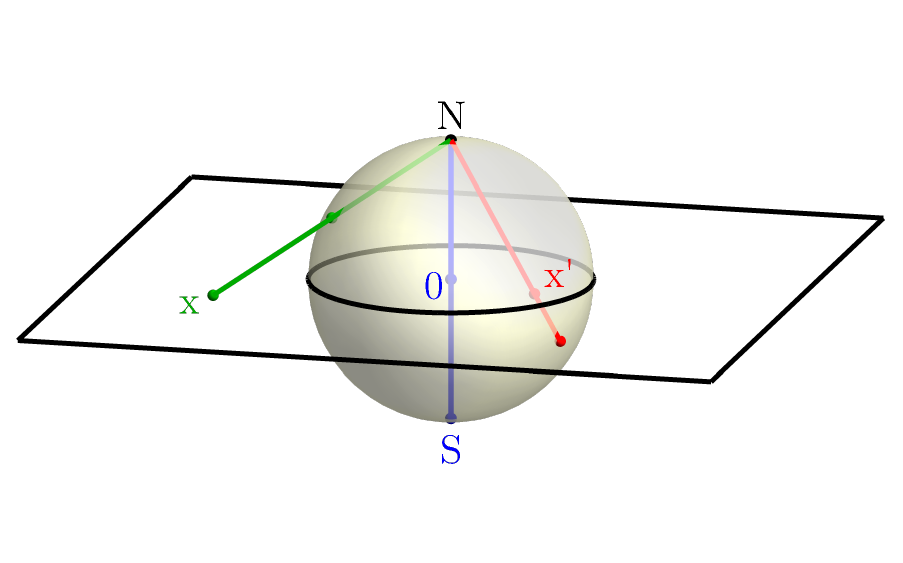}
	\vspace{-12mm}
	\caption{Inverse stereographic projection of the Euclidean space
	$\mathds{R}^d$ (represented here as the horizontal plane)
	onto the sphere $S^d$, both embedded in $\mathds{R}^{d+1}$.
	A point $x$ with $\left| x \right| > 1$ gets mapped to the
	northern hemisphere; a point $x'$
	with $\left| x' \right| < 1$ to the southern hemisphere;
	the origin is mapped to the south pole (S),
	and $\infty$ to the north pole (N).}
	\label{fig:stereographic}
\end{figure}

To gain a better understanding of this, it is useful to map the flat Euclidean space $\mathbb{R}^d$ or the Minkowski space-time $\mathbb{R}^{1,d-1}$ onto a curved manifold.
In Euclidean space, this is for instance achieved by the (inverse) stereographic projection that maps $\mathbb{R}^d \cup \{ \infty \}$ to the unit sphere $S^d$.
Geometrically, the stereographic projection is constructed as follows (see figure~\ref{fig:stereographic}): embed $\mathbb{R}^d$ as a plane in $\mathbb{R}^{d+1}$, together with a sphere of unit radius centered at the origin. Every point on the plane has an image on the sphere obtained by drawing a segment between the original point and the north pole of the sphere, and noting where it intersects the sphere. The origin is mapped to the south pole, $\infty$ to the north pole, and the sphere $S^{d-1}$ of unit radius to the equator.
Algebraically, this is achieved as follows: first write the Euclidean metric in spherical coordinates,
\begin{equation}
	ds^2 = dr^2 + r^2 d\Omega_{d-1}^2,
\end{equation}
where the solid angle is given in $d = 2$ by $d\Omega_1^2 = d\phi^2$, in $d = 3$ by $d\Omega_2^2 = d\theta^2 + \sin\theta^2 d\phi^2$, and more generically by the recursion relation $d\Omega_n^2 = d\theta^2 + \sin\theta^2 d\Omega_{n-1}^2$.
Let us perform the change of variable
\begin{equation}
	r = \frac{\sin\varphi}{1 - \cos\varphi},
\end{equation}
and interpret $\varphi \in [0, \pi]$ as the zenith angle on the sphere:
$\varphi = 0$ is the north pole, corresponding to $r \to \infty$, and $\varphi = \pi$ the south pole, corresponding to $r = 0$. In these coordinates we have
\begin{equation}
	ds^2 = \frac{1}{(1 - \cos\varphi)^2}
	\left( d\varphi^2 + \sin\varphi^2 d\Omega_{d-1}^2 \right)
	= \frac{1}{(1 - \cos\varphi)^2} d\Omega_d^2.
\end{equation}
The new metric is flat up to an overall Weyl factor that depends on $\varphi$.
In these coordinates, conformal transformations are always non-singular. 
For this reason, it is often convenient to study \emph{classical} conformal transformations on the sphere $S^d$ instead of Euclidean space.

However, this compactification is not as nice in a \emph{quantum} theory in which one wants to foliate the space along some preferred direction: if one chooses $\varphi$ as the ``Euclidean time'', then the ``space'' direction is a sphere $S^{d-1}$ whose volume depends on $\varphi$. In other words, the generator of ``time'' translations is not a symmetry of the system. This is also true of any other choice of time direction on the sphere.

Instead, another compactification is often preferred to the sphere: from the Euclidean metric in spherical coordinates, one can make the change of variable
\begin{equation}
	\tau = \log(r)
	\qquad\Leftrightarrow\qquad
	r = e^\tau,
\end{equation}
after which
\begin{equation}
	ds^2 = e^{2\tau} \left( d\tau^2 + d\Omega_{d-1}^2 \right).
	\label{eq:cylindercoordinates}
\end{equation}
This is again a flat metric, up to a Weyl factor $r^2 = e^{2\tau}$. In this case, however, the rescaled metric is independent of $\tau$.
This space has the geometry of a \emph{cylinder}: $\mathbb{R} \times S^{d-1}$. It is not fully compact: $\tau$ goes from $-\infty$ to $+\infty$. But it has an important advantage: translations in $\tau$ are generated by dilatations $D$, which will be taken to be a symmetry of the quantum theory. Foliating the space into surfaces of constant $\tau$ will later lead us to radial quantization in conformal field theory.

\begin{figure}
	\includegraphics[width=0.48\linewidth]{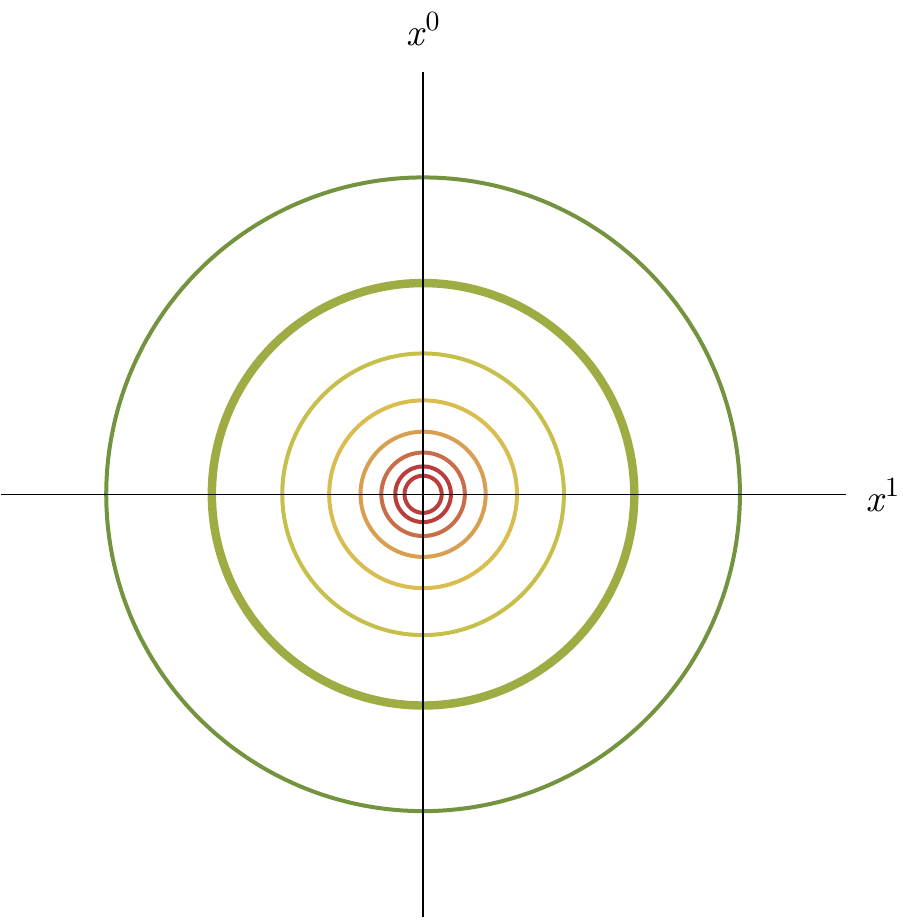}
	\hfill
	\includegraphics[width=0.48\linewidth]{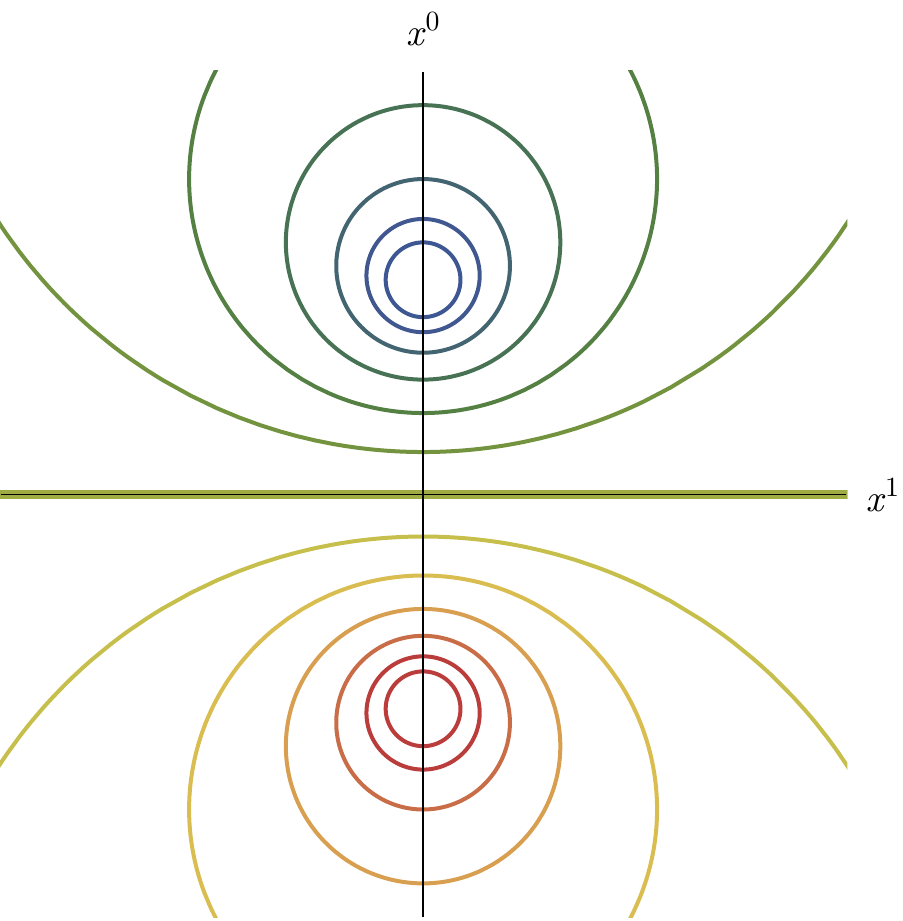}
	\caption{Foliations of Euclidean space
	(here in $d = 2$) 
	in radial (left) and N-S quantization (right).
	Circles of identical colors are mapped onto each other
	by a conformal transformation. In particular, the plane 
	$x^0 = 0$ in N-S quantization is mapped to the unit sphere 
	in radial quantization.}
	\label{fig:foliations}
\end{figure}

\begin{figure}
	\includegraphics[width=0.25\linewidth]
	{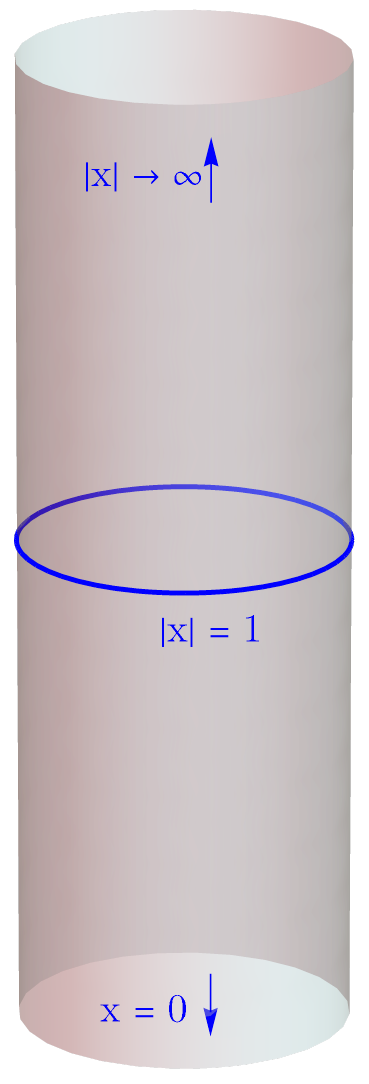}
	\hfill
	\includegraphics[width=0.25\linewidth]
	{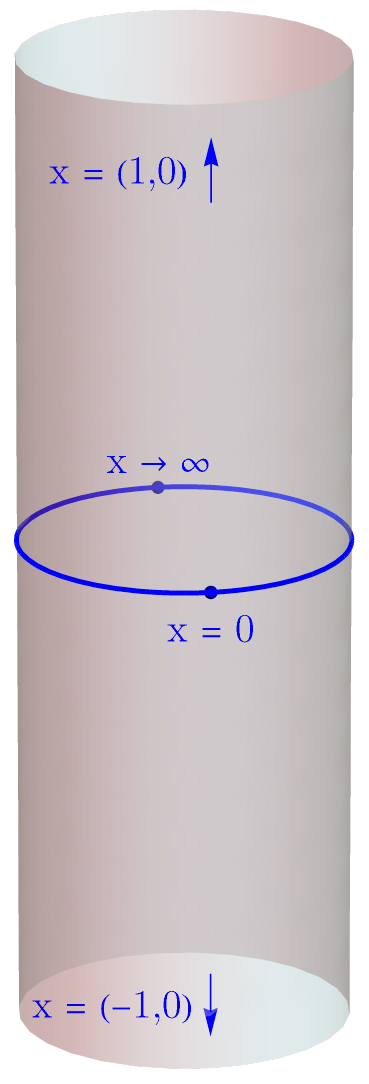}
	\hfill
	\includegraphics[width=0.25\linewidth]
	{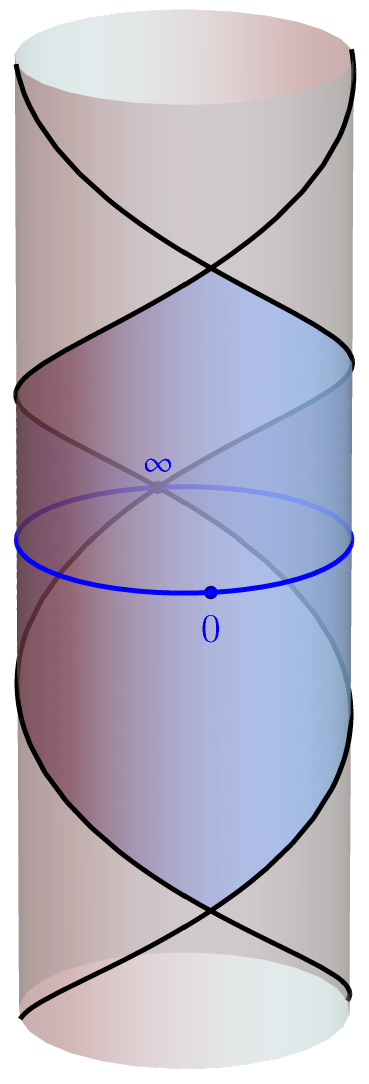}
	\caption{Left: the Euclidean cylinder corresponding the radial
	coordinates \eqref{eq:cylindercoordinates}, in which
	the evolution in the non-compact direction is given by
	the generator $D$ of scale transformations;
	the blue circle corresponds to the unit sphere.
	Middle: same Euclidean cylinder after a conformal transformation,
	so that the evolution is given by $\frac{1}{2} (P^0 - K^0)$;
	the blue circle is now the plane $x^0 = 0$, which includes
	the origin and the point at infinity.
	Right: the Lorentzian cylinder, obtained after a Wick rotation
	in $x^0$, so that the surface $x^0 = 0$ is unchanged;
	the image of this surface under time translations 
	generated by $P^0$ is the Poincaré patch, shown as 
	the blue diamond on the cylinder's surface;
	only special conformal transformations can move a point
	from inside the Poincaré patch to the outside.}
	\label{fig:cylinders}
\end{figure}

Note that in $\SO(d + 1, 1)$ language, the generator $D = J^{d+1, d+2}$ is completely equivalent to the other generators $J^{\mu, d+2} = \frac{1}{2} \left( P^\mu - K^\mu \right)$, since they are related to them by $\SO(d + 1)$ rotations. So we might as well look for a cylinder compactification in which the non-compact direction corresponds to transformation generated by
$\frac{1}{2} \left( P^0 - K^0 \right)$. This combination of generators obeys
\begin{equation}
	\frac{1}{2} \left( P^0 - K^0 \right)
	= i \left( \frac{1 - (x^0)^2 + \vec{x}^2}{2} \partial_0
	- x^0 \, \vec{x} \cdot \vec{\partial} \right),
\end{equation}
with two fixed points at $x^0 = \pm 1$ with $\vec{x} = 0$. 
The foliation of space generated by this linear combination of generators is what is used in N-S quantization,%
\footnote{The line drawn when evolving a point with this Hamiltonian looks like a magnetic field line connecting the north (N) and south (S) poles of a magnet, hence the name N-S quantization.}
discussed later in section~\ref{sec:OPE}.
Figure~\ref{fig:foliations} illustrates the two foliations of Euclidean space by $D$ and $\frac{1}{2} \left( P^0 - K^0 \right)$,
and the corresponding cylinder interpretations are shown in figure~\ref{fig:cylinders}.

\begin{exercise}
	Find the change of coordinates that makes the Euclidean metric Weyl-equivalent to a cylinder in which translations in the non-compact direction are generated by $\frac{1}{2} \left( P^0 - K^0 \right)$. \\
	Hint: Find a special conformal transformation followed by a translation that takes $(0, \infty)$ to $(-1, 1)$, and apply it to the radial coordinates.
\end{exercise}

The cylinder compactifications of Euclidean space are interesting by themselves, but they are also extremely convenient to understand the connection between Euclidean and Minkowski space-times: in this last form, performing a Wick rotation $\tau \to - i t$ defines a cylinder on which the Lorentzian conformal group $\SO(d, 2)$ acts naturally. 
But before we get there, let us go back to flat Minkowski space-time and make some general remarks.

\subsection{Minkowski space-time}

Translations and Lorentz transformations in Minkowski space-time are familiar, and even dilatations are a standard tool in the renormalization group analysis. But what do conformal transformations do?

To understand this, let us place an observer at the origin of Minkowski space-time. The presence of this observer breaks translations, but not Lorentz transformations (let us assume that the observer is point-like), nor dilatations or special conformal transformations. For the observer, space-time is split into three regions: a future light cone, a past light cone, and a   space-like region from which they know nothing.
Lorentz and scale transformation preserve this causal structure: the future and past light-cones are mapped onto themselves. In other words, if a point $x$ is space-like separated from the observer, it will remain space-like separated no matter the choice of Lorentz frame, or the definition of length. Without loss of generality, let us choose this point to be at position $x = (0, \vec{n})$, where $\vec{n}$ is a unit vector (units can be chosen so that this is the case).
Now apply a special conformal transformation with parameter $b^\mu = (-\alpha, \alpha \vec{n})$, with $\alpha$ varying between 0 and 1.
This draws a curve $y^\mu $ in space-time, parameterized by $\alpha$,
with
\begin{equation}
	y^0(\alpha) = \frac{\alpha}{1 - 2 \alpha},
	\qquad\qquad
	\vec{y}(\alpha) = \frac{1 - \alpha}{1 - 2 \alpha} \vec{n}.
\end{equation}
This curves begins at the space-like point $y = (0, \vec{n})$, and ends in the past light cone at $y = (-1, \vec{0})$.
Note that $y$ never crosses a light cone: the image of a point $x$ is never null under a special conformal transformation unless $x$ is itself null, since $x'^2 = x^2 / (1 - 2 x \cdot b + x^2 b^2)$.
Instead, we have
\begin{equation}
	y^2(\alpha) = \frac{1}{1 - 2\alpha} \neq 0.
\end{equation}
What happens is that the point travels all the way to space-like infinity at $\alpha = \frac{1}{2}$, and comes back from past infinity.
Clearly, special conformal transformation break causality!

The resolution of this puzzle is that conformal transformations do not act directly on Minkowski space, but rather on its universal cover that is isomorphic to the Lorentzian cylinder described in figure~\ref{fig:cylinders}.
Evolution on that cylinder is given by the Hamiltonian $H = \frac{1}{2} (P^0 - K^0)$, but this differs from the Minkowski time evolution generated by $P^0$. On any given slice of the Lorentzian cylinder, space is compactified in such a way that the notion of infinite distance is unequivocal: space-like infinity corresponds to a point on the sphere, antipodal to the origin. If one takes any other point of that sphere and applies finite translations using the generator $P^\mu$, then this defines a compact \emph{Poincaré patch}. The full Lorentzian cylinder is a patchwork of Poincaré patches, but every local observer only has access to one.%
\footnote{As Lüscher and Mack put it: \emph{In picturesque language, [the superworld] consists of Minkowski space, infinitely many ``spheres of heaven'' stacked above it and infinitely many ``circles of hell'' below it}~\cite{Luscher:1974ez}.}

The lesson that we must learn is that \emph{only the infinitesimal form of special conformal transformations can be used in Minkowski space-time}: any finite special conformal transformation brings part of space-time into another patch on the cylinder.
This is sometimes called \emph{weak conformal invariance}.

\subsection{Conformal symmetry in classical field theory}

So far we have only been discussing conformal transformations of the coordinates. The next step is to consider a field theory (for the moment a classical one) that has conformal symmetry built in. The simplest example is the free, massless scalar field, defined by the action
\begin{equation}
	S = \int d^dx
	\left[ - \frac{1}{2} \partial_\mu \phi \partial^\mu \phi \right].
	\label{eq:freescalaraction}
\end{equation}
We shall see in the next section that the action principle can in fact be dropped in CFT, but for now it is a convenient starting point.

In this context, a conformal transformation is a transformation \emph{of the fields}.
There are two distinct and complementary perspectives one can adopt. It is often convenient to think of the metric tensor as a field in its own right, and to define a conformal transformation as a (position-dependent) scale transformation of the field
\begin{equation}
	\phi(x) \to e^{\Delta \sigma(x)} \phi(x),
	\label{eq:freescalar:conformaltransformation:1}
\end{equation}
combined with a Weyl transformation of the metric
\begin{equation}
	g_{\mu\nu}(x) \to e^{2 \sigma(x)} g_{\mu\nu}(x).
	\label{eq:freescalar:conformaltransformation:2}
\end{equation}
$\Delta$ is the \emph{scaling dimension} of the field $\phi$. In a free theory it coincides with the dimension of the field in units of energy (inverse units of length), namely
\begin{equation}
	\Delta = \frac{d-2}{2}.
\end{equation}
$\sigma(x)$ is an infinitesimal scale factor that satisfies $\partial_\mu \partial_\nu \sigma = 0$.
The advantage of this perspective is that the conformal transformations are simple, multiplicative transformations of the fields. The disadvantage is that it requires thinking of the theory in curved space-time. This means that the metric that is implicit in the action \eqref{eq:freescalaraction} must be made explicit, but also that the action can be supplemented with a term depending on the scalar curvature tensor $R$ as
\begin{equation}
	S = \int d^dx \, \sqrt{\left| g \right|}
	\left[ - \frac{1}{2} g^{\mu\nu} \partial_\mu \phi \partial_\nu \phi
	+ \alpha R \phi^2 \right].
\end{equation}
Since $R$ vanishes in flat space, it looks like this additional term could appear with an arbitrary coefficient $\alpha$ without modifying the original flat-space action, but this is not the case.

\begin{exercise}
	Verify that there is a unique value of $\alpha$ 
	for which this action is invariant
	under the infinitesimal conformal transformations
	\eqref{eq:freescalar:conformaltransformation:1} and
	\eqref{eq:freescalar:conformaltransformation:2}.
	What is it?
\end{exercise}

For this reason, it is also convenient to consider the opposite perspective in which conformal transformations are transformations of the dynamical fields and of the coordinates, but not of the metric. In this case the conformal transformation can be defined as
\begin{equation}
	\phi(x) \to e^{\Delta \sigma(x)} \phi(x + \varepsilon),
\end{equation}
where the parameters $\sigma$ and $\varepsilon$ are related by the conformal Killing equation \eqref{eq:conformalkillingeq}, i.e.~$\sigma = \frac{1}{d} \partial_\mu \varepsilon^\mu$. 
In infinitesimal form, this transformation becomes
\begin{equation}
	\phi(x) \to \left[ 1
	+ \frac{d-2}{2d} (\partial_\nu \varepsilon^\nu)
	+ \varepsilon^\nu \partial_\nu \right] \phi(x).
	\label{eq:freescalar:conformaltransformation:3}
\end{equation}

\begin{exercise}
	Show that under the transformation
	\eqref{eq:freescalar:conformaltransformation:3},
	the Lagrangian of the free scalar field
	$$
	\Lagr = - \frac{1}{2} \partial_\mu \phi \partial^\mu \phi,
	$$
	is shifted by a total derivative term
	$$
	\delta\Lagr = \partial_\mu \left( 
	- \frac{1}{2} \varepsilon^\mu \partial_\nu \phi \partial^\nu \phi
	- \frac{d-2}{2d} \partial_\nu \varepsilon^\nu
	\phi \partial^\mu \phi \right),
	$$
	hence proving that this is a symmetry of the action.
	You will need to use the fact that $\varepsilon$
	is at most quadratic in $x$.
\end{exercise}

\noindent
Note that Poincaré symmetry is a special case of this transformation, corresponding to constant $\varepsilon$ (and thus $\sigma = 0$).

By Noether's theorem, whenever an action is invariant under some transformation of the field
\begin{equation}
	\phi \to \phi + \delta_\varepsilon \phi,
\end{equation}
i.e.~whenever the Lagrangian varies by a total derivative term,
\begin{equation}
	\Lagr \to \Lagr + \partial_\mu \Lambda_\varepsilon^\mu,
\end{equation}
then there exists a conserved current
\begin{equation}
	J_\varepsilon^\mu = \Lambda^\mu_\varepsilon
	- \frac{\partial \Lagr}{\partial (\partial_\mu \phi)}
	\delta_\varepsilon \phi.
\end{equation}
In our example, this conserved current is therefore
\begin{equation}
	J_\varepsilon^\mu = \varepsilon_\nu
	\left( \partial^\mu \phi \partial^\nu \phi
	- \frac{1}{2} g^{\mu\nu} \partial_\rho \phi \partial^\rho \phi
	\right)
	\equiv \varepsilon_\nu T_c^{\mu\nu}.
	\label{eq:Noethercurrent}
\end{equation}
The 2-index tensor on the right-hand side is called the \emph{canonical energy-momentum tensor}.
Its divergence satisfies
\begin{equation}
	\partial_\nu T_c^{\mu\nu}
	= \partial^\mu \phi \partial^2 \phi,
	\label{eq:Tcanonical:conservation}
\end{equation}
therefore vanishing by the equation of motion for the free field, $\partial^2 \phi = 0$.
This implies in turn that the Noether current \eqref{eq:Noethercurrent} is conserved \emph{for constant $\varepsilon$}.
If on the contrary $\varepsilon$ depends on space(-time), then we have
\begin{equation}
	\partial_\mu J_\varepsilon^\mu
	= (\partial_\nu \varepsilon_\mu) T_c^{\mu\nu}.
\end{equation}
In our example the canonical energy-momentum tensor is symmetric in its indices $\mu$ and $\nu$, and therefore we can write
\begin{equation}
	\partial_\mu J_\varepsilon^\mu
	= \frac{1}{2} (\partial_\mu \varepsilon_\nu
	+ \partial_\nu \varepsilon_\mu) T_c^{\mu\nu}
	= \sigma g_{\mu\nu} T_c^{\mu\nu}
	= -\frac{d-2}{2} \, \sigma \, \partial_\mu \phi \partial^\mu \phi.
\end{equation}
In dimensions $d > 2$, this current is only conserved when $\sigma = 0$. This is surprising, because we just showed that scale and special conformal transformations are also symmetries of the action, so why is the Noether current not conserved?

The reason is that the version of Noether's theorem given above does not straightforwardly apply to the case of a space-time dependent parameter $\varepsilon$. In fact, the energy-momentum tensor that we computed in this example is not unique: one can always add to it a piece proportional to
\begin{equation}
	\left( \partial^\mu \partial^\nu - g^{\mu\nu} \partial^2 \right)
	\phi^2,
\end{equation}
without affecting the conservation equation \eqref{eq:Tcanonical:conservation}, but changing the value of its trace. The combination
\begin{equation}
	T^{\mu\nu} = T^{\mu\nu}_c + \frac{d-2}{2(d-1)}
	\left( \partial^\mu \partial^\nu - g^{\mu\nu} \partial^2 \right)
	\phi^2
\end{equation}
is for instance traceless in any $d$.
It turns out that it is always possible in a field theory with conformal symmetry to construct an energy-momentum tensor that is:
\begin{itemize}
\item symmetric ($T^{\mu\nu} = T^{\nu\mu}$),
\item traceless ($g_{\mu\nu} T^{\mu\nu} = 0$), and
\item conserved once the equation of motions are imposed ($\partial_\nu T^{\mu\nu} \stackrel{e.o.m.}{=} 0$).
\end{itemize}
This is a non-trivial fact, but we will skip its proof (as already mentioned, we are interested in theories that are not necessarily defined through an action).

Strictly speaking, Noether's theorem only applies to theories that have a Lagrangian description, but we will assume the existence of a traceless energy-momentum tensor in all cases (in some sense this is going to be one of the ``axioms'' of conformal field theory).
From this assumption, we can deduce that the theory is invariant under conformal transformations: a conserved current can always be built from the energy-momentum tensor and a conformal Killing vector $\varepsilon$ as
\begin{equation}
	J^\mu = \varepsilon_\nu T^{\mu\nu}
	\qquad\Rightarrow\qquad
	\partial_\mu J^\mu
	= \varepsilon_\nu \partial_\mu T^{\mu\nu}
	\stackrel{e.o.m.}{=} 0.
\end{equation}
As always, conserved charges can be constructed as the integral of the time component of a conserved current over space. The simplest example is 
\begin{equation}
	P^\mu = -i \int d^{d-1}\vec{x}\, T^{0\mu}(x).
	\label{eq:conservedcharge:P}
\end{equation}
This is in principle a function of $x^0$, but it is in fact constant over time, since
\begin{equation}
	\partial_0 P^\mu
	= -i \int d^{d-1}\vec{x}\, \partial_0 T^{0\mu}(x)
	= i \int d^{d-1}\vec{x}\, \partial_i T^{i\mu}(x) = 0.
\end{equation}
This conserved charge is the \emph{momentum}, associated with translation symmetry. Similarly, there are conserved charges associated with Lorentz transformations,
\begin{equation}
	M^{\mu\nu} = -i \int d^{d-1}\vec{x}\, \left[
	x^\mu T^{0\nu}(x) - x^\nu T^{0\nu}(x) \right]
	\label{eq:conservedcharge:M}
\end{equation}
with scale transformations,
\begin{equation}
	D = -i \int d^{d-1}\vec{x}\, x_\mu T^{0\mu}(x),
	\label{eq:conservedcharge:D}
\end{equation}
and with special conformal transformations
\begin{equation}
	K^\mu = -i \int d^{d-1}\vec{x} \,
	\left[ 2 x^\mu x_\nu T^{0\nu}(x) - x^2 T^{0\mu}(x) \right].
	\label{eq:conservedcharge:K}
\end{equation}

\begin{exercise}
	Show that the charges \eqref{eq:conservedcharge:M},
	\eqref{eq:conservedcharge:D} and \eqref{eq:conservedcharge:K}
	are conserved in time.
\end{exercise}

The conservation of all these charges relies on the vanishing divergence of the energy-momentum tensor, $\partial_\nu T^{\mu\nu}$, which itself relies on the equation of motion being satisfied.
This is certainly true in the absence of sources. But when source terms are added to the action, the equation of motion is modified.
In our example of the free scalar field theory, adding to the action \eqref{eq:freescalaraction} a source term of the form
\begin{equation}
	S_\text{source} = \int d^dx \, J(x) \phi(x),
\end{equation}
modifies the equation of motion to
\begin{equation}
	\partial^2 \phi + J = 0.
\end{equation}
Instead of the conservation equation \eqref{eq:Tcanonical:conservation} for the canonical energy-momentum tensor, we must replace it with
\begin{equation}
	\partial_\nu T^{\mu\nu}(x) = - J \partial^\mu \phi.
\end{equation}
In the presence of such a source, the charges \eqref{eq:conservedcharge:P}--\eqref{eq:conservedcharge:K} are not conserved anymore.
However, if the source is local, say $J(x) = \delta^d(x - x_\odot)$, so that
\begin{equation}
	\partial_\nu T^{\mu\nu}(x) = -\delta^d(x - x_\odot)
	\partial^\mu \phi(x_\odot),
\end{equation}
then we can determine the change in one of the charge --- say $P^\mu$ ---
between a time $x^0 < x_\odot^0$ anterior to the local source, and a time $x^0 > x_\odot^0$ posterior to it, and call this difference the momentum of the source, $P^\mu_\odot$. 
By our definition, this is equal to
\begin{equation}
	P^\mu_\odot
	= -i \int d^{d-1}\vec{x} \, T^{0\mu} \Big|_{x^0 > x^0_\odot}
	+ i \int d^{d-1}\vec{x} \, T^{0\mu} \Big|_{x^0 < x^0_\odot}.
\end{equation}
Since the two surfaces of integration meet at spatial infinity, they can be viewed as the two sides of a closed surface $\partial\Sigma$ surrounding the point $x_\odot$, hence
\begin{equation}
	P^\mu_\odot = -i \int\limits_{\partial\Sigma} d^{d-1} n_\nu
	T^{\mu\nu}.
\end{equation}
By the divergence theorem, this is equal to
\begin{equation}
	P^\mu_\odot = -i \int\limits_{\Sigma} d^dx \, \partial_\nu
	T^{\mu\nu}(x)
	= i \int\limits_{\Sigma} d^dx \, \delta^d(x - x_\odot)
	\partial^\mu \phi(x)
	= i \partial^\mu \phi(x_\odot).
	\label{eq:P:localsource}
\end{equation}
Note that this result does not depend on the choice of surface $\partial\Sigma$, as long as it encloses the point $x_\odot$.
In technical terms, $P^\mu$ is a \emph{topological charge}.
We chose in eq.~\eqref{eq:conservedcharge:P} to use a standard definition of $P^\mu$ in which the energy-momentum tensor is integrated along a surface of constant time. But if we work in Euclidean space, this is conformally equivalent to integrating over the surface of a sphere, or in fact over any other closed surface.

Eq.~\eqref{eq:P:localsource} is very important: it says that the charge $P^\mu$ associated with a local source for the field $\phi$ is equal to $i \partial^\mu \phi$. This is strikingly similar to the action of the generator \eqref{eq:P:fcts} on functions of the coordinates. In fact, it is easy to verify that the other charges \eqref{eq:conservedcharge:M}, \eqref{eq:conservedcharge:D}, and \eqref{eq:conservedcharge:K} also act on the classical field $\phi(x)$ exactly like the generators \eqref{eq:M:fcts}, \eqref{eq:D:fcts}, and \eqref{eq:K:fcts} respectively.
We worked here with the free scalar field theory as an example, but our discussion can be generalized to arbitrary classical field theories.
The important lesson is that a traceless energy-momentum tensor can be used to give a field-theoretical realization of the conformal generators discussed before.

%%%%%%%%%%%%%%%%%%%%%%%%%%%%%%%%%%%%%%%%%%%%%%%%%%%%%%%%%%%%%%%%%%%%%%

\section{Conformal quantum field theory}
\label{sec:quantum}

Let us now turn to quantum field theory and study the implications of conformal symmetry in that context. The standard approach to quantum field theory is to think of a classical field theory, in which we now have a basic understanding of what conformal symmetry does, and then quantize it by promoting the fields to operators acting on some Hilbert space.
But this is not the approach that we will take here. There will still be states and operators, but the latter will not necessarily be associated with fields appearing in a Lagrangian.

\subsection{Non-perturbative quantum field theory}

To define a quantum field theory non-perturbatively, we need the following ingredients:
\begin{enumerate}

\item
\textbf{Hilbert space}:
The Minkowski space-time is foliated into surfaces of equal-time,
and to each time slice we associate a Hilbert space of quantum states.

\item
\textbf{Local operators}:
There are a number of (in fact infinitely many) \emph{local} operators that act on this Hilbert space. For instance, let us take $\phi(x)$ to be an operator acting on the Hilbert space at time $t = x^0$.
We call this operator \emph{local} because we require that it commutes with any other local operator inserted at a distinct point $\vec{x}' \neq \vec{x}$ on the same time slice:%
\footnote{For operators of half-integer spin, the commutator must be replaced by an anti-commutator.} 
\begin{equation}
	\left[ \phi(x), \phi(x') \right]
	= 0,
	\qquad\qquad
	x^0 = x'^0, \qquad
	\vec{x} \neq \vec{x}'.
	\label{eq:locality:equaltime}
\end{equation}

\item
\textbf{Symmetries}:
One of the local operators of the theory is the energy-momentum tensor, and from it we can define conserved charges, including $P^\mu$ and $M^{\mu\nu}$ as in eqs.~\eqref{eq:conservedcharge:P} and \eqref{eq:conservedcharge:M}. In a generic QFT the energy-momentum tensor needs not be traceless, so the charges $D$ and $K^\mu$ cannot be considered. 
$P^\mu$ and $M^{\mu\nu}$ are conserved in time, so they are valid operators on all Hilbert spaces at every time $t$. 
Their value changes however every time an operator is inserted at some point $x$. In analogy with eq.~\eqref{eq:P:localsource}, we require that this change is encoded in the commutator
\begin{equation}
	\left[ P^\mu, \phi(x) \right] = i \partial^\mu \phi(x).
	\label{eq:commutator:P}
\end{equation}
Since this equation is solved by
\begin{equation}
	\phi(x) = e^{-i x \cdot P} \phi(0) e^{i x \cdot P},
	\label{eq:P:exponentiated}
\end{equation}
we say that $P^\mu$ is the generator of translations, which act as \emph{unitary} transformations on the operators (note that $P^\mu$ is Hermitian).

Lorentz transformations are similarly realized as unitary transformations, generated by the charge $M^{\mu\nu}$. 
We can choose to decompose local operators inserted at the origin of space-time into irreducible representations of the Lorentz group and denote these with $\phi^a(0)$, with $a$ standing for a collection of Lorentz indices, so that
\begin{equation}
	\left[ M^{\mu\nu}, \phi^a(0) \right]
	= i \left( \mathcal{S}^{\mu\nu} \right)^a_{~b} \phi^b(0),
\end{equation}
where $\left( \mathcal{S}^{\mu\nu} \right)^a_{~b}$ is a matrix that satisfies the Lorentz algebra. For a scalar operator, $\mathcal{S}^{\mu\nu}$ vanishes; for a vector operator with one Lorentz index, it is given by
\begin{equation}
	\left( \mathcal{S}^{\mu\nu} \right)_{ab}
	= \delta^\mu_a \delta^\nu_b
	- \delta^\mu_b \delta^\nu_a,
	\label{eq:spinop:vector}
\end{equation}
and so on.
When combined with eq.~\eqref{eq:P:exponentiated}, and requiring that $P^\mu$ and $M^{\mu\nu}$ satisfy the Poincaré algebra \eqref{eq:conformalalgebra}, this implies
\begin{equation}
	\left[ M^{\mu\nu}, \phi^a(x) \right] = 
	-i \left( x^\mu \partial^\nu - x^\nu \partial^\mu \right) \phi^a(x)
	+ i \left( \mathcal{S}^{\mu\nu} \right)^a_{~b} \phi^b(0).
	\label{eq:commutator:M}
\end{equation}
Note that $P^\mu$ and $M^{\mu\nu}$ are not local operators, but their commutator with any local operator is again local (the same will later be true of the generators $D$ and $K^\mu$).

\item
\textbf{Vacuum state:}
The Hilbert space includes a vacuum state $\ket{0}$, which we assume to be invariant under Poincaré transformations (and later conformal transformations),
\begin{equation}
	P^\mu \ket{0} = M^{\mu\nu} \ket{0} = 0.
	\label{eq:vacuuminvariance:P}
\end{equation}
Other states of the theory are obtained acting with products of local operators on the vacuum (see below for a more precise statement). There is in principle one Hilbert space for each time slice, but since time translation is a symmetry generated by $P^0$, the evolution operator $U(t) = e^{i t P^0}$ is unitary and all Hilbert spaces are equivalent. 
We also require that the vacuum is the lowest-energy state in the Hilbert space. This means that if we can construct an eigenstate of energy, 
\begin{equation}
	P^0 \ket{\Psi} = E \ket{\Psi},
\end{equation}
then its eigenvalue must satisfy $E \geq 0$.

\end{enumerate}
These four points essentially give the most general non-perturbative definition of quantum field theory. They are nearly equivalent to the so-called \emph{Wightman axioms} (but presented here without much mathematical rigor).
One notable difference is that the Wightman axioms do not rely on the existence of an energy-momentum tensor, but assume directly that the Poincaré transformations are realized as unitary transformations on the Hilbert space. 

The locality condition \eqref{eq:locality:equaltime} is often formulated in the Lorentz-invariant way
\begin{equation}
	\left[ \phi(x), \phi(y) \right] = 0
	\qquad
	\text{if}~(x-y)^2 > 0,
	\label{eq:causality}
\end{equation}
stating that local operators commute as long as they are space-like separated, which is known as the \emph{micro-causality} axiom.
Similarly, when it is possible to work with eigenstates of energy and momentum, 
\begin{equation}
	P^\mu \ket{\Psi} = p^\mu \ket{\Psi},
\end{equation}
then the Lorentz-invariant condition on the positivity of energy becomes
\begin{equation}
	p^0 \geq \left| \vec{p} \right|
	\qquad \Leftrightarrow \qquad
	p^2 \leq 0 ~\text{and}~ p^0 \geq 0,
	\label{eq:forwardcone}
\end{equation}
i.e.~the momentum $p^\mu$ in contained in the forward light cone.
Such eigenstates of $P^\mu$ can be constructed from the Fourier transform of local operators,
\begin{equation}
	\widetilde{\phi}(p)
	= \int d^dx \, e^{i p \cdot x} \phi(x)
	\label{eq:momentumspace}
\end{equation}
and acting on the vacuum. This implies that
\begin{equation}
	\widetilde{\phi}(p) \ket{0} = 0
	\qquad\qquad
	\text{if}~p^0 < \left| \vec{p} \right|,
\end{equation}
as well as its generalization to the product of multiple local operators,
\begin{equation}
	\widetilde{\phi}_1(p_1) \cdots \widetilde{\phi}_n(p_n) \ket{0} = 0
	\qquad\qquad
	\text{if}~p_1^0 + \ldots + p_n^0
	< \left| \vec{p}_1 + \ldots + \vec{p}_n \right|.
\end{equation}
This is sometimes called the \emph{spectral condition}.

\begin{exercise}
	Using the definition~\eqref{eq:P:exponentiated} and integration by parts, show that 
	$$ 
		P^\mu \widetilde{\phi}(p) \ket{0}
		= p^\mu \widetilde{\phi}(p) \ket{0}.
	$$
\end{exercise}

\subsection{Wightman functions}

Starting from these axioms, the next thing we can do is compute the vacuum expectation value of products of local operators,
\begin{equation}
	\bra{0} \phi_1(x_1) \cdots \phi_n(x_n) \ket{0}.
\end{equation}
This can be viewed as an overlap of the vacuum state, $\bra{0}$, with a  state created acting on the vacuum with a sequence of local operators. Note that these operators need not be ordered in time: the time-evolution operator is unitary, so it can go both ways. This object is therefore different from time-ordered correlation functions obtained from the path integral.

Correlators of this type are called \emph{Wightman functions}. They are the fundamental observables in non-perturbative quantum field theory. In fact, it is even possible to completely define a quantum field theory just by its Wightman functions: the Wightman reconstruction theorem states that the Hilbert space of a quantum field theory can be constructed from all its Wightman functions.
A convenient perspective is therefore to forget about the Hilbert space and focus on correlation functions.

The symmetry properties of these correlation functions are encoded in ``Ward identities'': given a conserved charge $G$ that annihilates the vacuum, $G \ket{0} = 0$ (this could be $P^\mu$ or $M^{\mu\nu}$), the following equation must be satisfied
\begin{align}
	& \bra{0} \left[ G, \phi_1(x_1) \right] \phi_2(x_2) \cdots 
	\phi_n(x_n) \ket{0}
	\nonumber \\
	& + \bra{0} \phi_1(x_1) \left[ G, \phi_2(x_2) \right] \cdots 
	\phi_n(x_n) \ket{0}
	\nonumber \\
	& + \ldots
	\nonumber \\
	& + \bra{0} \phi_1(x_1) \, \phi_2(x_2) \cdots 
	\left[ G, \phi_n(x_n) \right] \ket{0} = 0.
\end{align}
This is an equation that is obvious in the Hilbert space picture, but it is also valid as a differential equation for the Wightman function, since each commutator is again related to the local operator. Let us see some examples.

The simplest Wightman function involves a single scalar operator, 
\begin{equation}
	\bra{0} \phi(x) \ket{0}.
\end{equation}
In this case the Ward identity associated with translations implies
\begin{equation}
	\bra{0} \left[ P^\mu, \phi(x) \right] \ket{0} = 0
	\qquad\Rightarrow\qquad
	\frac{\partial}{\partial x^\mu}
	\bra{0} \phi(x) \ket{0} = 0,
\end{equation}
or in other words that the vacuum expectation value of the operator is a constant over all of space-time. Lorentz symmetry does not give more information about that constant, but it forbids vacuum expectation values for all operators transforming non-trivially under the Lorentz group.

Let us consider next a Wightman 2-point function of identical scalar operators,
\begin{equation}
	\bra{0} \phi(x) \phi(y) \ket{0}.
\end{equation}
In this case, translation symmetry tells us that
\begin{equation}
	\left( \frac{\partial}{\partial x^\mu}
	+ \frac{\partial}{\partial y^\mu} \right)
	\bra{0} \phi(x) \phi(y) \ket{0} = 0.
\end{equation}
If we think of the correlator as being a function of $x + y$ and $x - y$, then this Ward identities establishes that there is no dependence on the former, i.e.~
\begin{equation}
	\bra{0} \phi(x) \phi(y) \ket{0} = W(y - x),
	\label{eq:W}
\end{equation}
where $W$ denotes a function that is so far arbitrary.

In general, the consequences of translation symmetry are easier to see in momentum space, using the Fourier transform of the local operators. By eq.~\eqref{eq:momentumspace}, we can establish that
\begin{equation}
	\big[ P^\mu, \widetilde{\phi}(p) \big]
	= p^\mu \, \widetilde{\phi}(p),
	\label{eq:commutator:P:momentum}
\end{equation}
and therefore the Fourier transform of the Wightman 2-point function obeys
\begin{equation}
	(p^\mu + q^\mu)
	\bra{0} \widetilde{\phi}(p) \widetilde{\phi}(q) \ket{0} = 0.
\end{equation}
From this, we conclude that the 2-point function is proportional to a Dirac delta function:
\begin{equation}
	\bra{0} \widetilde{\phi}(p) \widetilde{\phi}(q) \ket{0}
	= (2\pi)^d \delta^d(p + q) \widetilde{W}(q).
	\label{eq:Wtilde}
\end{equation}
The numerical factor $(2\pi)^d$ is merely a convention. As the notation suggests, $\widetilde{W}$ is actually the Fourier transform of $W$,
\begin{equation}
	\widetilde{W}(q) = \int d^dx \, e^{i p \cdot x} W(x).
\end{equation}

Taking into account Lorentz symmetry, one can also establish that the Wightman function $W(x)$ can only depend on the Lorentz-invariant distance $x^2$, although there is a subtlety: this is a different function depending whether $x$ is space-like or time-like (future- or past-directed), as Lorentz transformations act separately on each of these regions.
In momentum space, the same arguments says that $\widetilde{W}(q)$ must be a function of $q^2$. In this case, the condition that only states of positive energy exist requires that $\widetilde{W}(q)$ vanishes unless $q^0 \geq \left| \vec{q} \right|$, and therefore we can unambiguously write
\begin{equation}
	\widetilde{W}(q)
	= 2\pi \, \theta\left( q^0 - \left| \vec{q} \right| \right)
	\rho(-q^2),
	\label{eq:spectraldensity}
\end{equation}
where $\rho$ is a function of the positive quantity $-q^2$, and $\theta$ is the Heaviside step function.%
\footnote{It satisfies $\theta(a) = 0$ for $a < 0$, and $\theta(a) = 1$ for $a > 1$.}

The fact that the Wightman 2-point function \eqref{eq:Wtilde} is proportional to a delta function raises an important concern: in spite of their names, Wightman functions are \emph{not} functions but rather distributions (this is also part of the Wightman axioms: they are in fact \emph{tempered distributions}). Note also that the same function computes the overlap between the two states 
\begin{equation}
	\widetilde{\phi}(q) \ket{0}
	\qquad\text{and}\qquad
	\widetilde{\phi}(-p) \ket{0}
\end{equation}
(Hermitian conjugation flips the sign of momenta). Therefore, the limit $p \to - q$ corresponds to the norm of either of these states. But this limit is clearly discontinuous, or the norm of the state infinite.
The resolution of this issue is that the objects $\phi(x)$ and its Fourier transform $\widetilde{\phi}(p)$ are not operators, but rather \emph{operator-valued distributions}. In other words, $\phi(x) \ket{0}$
and $\widetilde{\phi}(p) \ket{0}$ are \emph{not} states of the theory, as they have in fact infinite norm. Formally, these operator-valued distributions only make sense when they are integrated against test functions, defining
\begin{equation}
	\phi[f] = \int d^dx \, f(x) \phi(x),
\end{equation}
or 
\begin{equation}
	\widetilde{\phi}[\tilde{f}] = \int d^dp \, \tilde{f}(p)
	\widetilde{\phi}(p),
\end{equation}
where $f$ and $\widetilde{f}$ are Schwartz-class test functions (smooth functions decaying faster than any power at infinity).
When acting on the vacuum, these smeared operators give well-defined states, with finite norms. For instance, we have
\begin{align}
	\left\| \widetilde{\phi}[\tilde{f}] \ket{0} \right\|^2
	&= \int d^dp d^dq \tilde{f}^*(p) \tilde{f}(q)
	\bra{0} \widetilde{\phi}(-p) \widetilde{\phi}(q) \ket{0}
	\nonumber \\
	&= (2\pi)^{d+1}
	\int\limits_{q^0 > \left| \vec{q} \right|} \!\!
	d^dq \left| \tilde{f}(q) \right|^2
	\rho(-q^2).
	\label{eq:norm}
\end{align}
Test functions will not appear further in these lectures. For physicists, they are mostly an annoyance that we prefer to avoid. However, it is important to know that there exists a mathematically rigorous way of dealing with Wightman functions.
For one thing, this gives a proper justification of why it is always fine to take the Fourier transform between the position- and momentum-space representation, as tempered distributions always admit a Fourier transform.
But bear in mind that this is only true of Wightman functions, not of time-ordered correlators.

\subsection{Spectral representation}

The norm \eqref{eq:norm} is also giving away important information: it can only be positive for any test function $f$ if the function $\rho$ is positive,
\begin{equation}
	\rho(\mu^2) \geq 0,
	\qquad\quad
	\forall \, \mu^2 > 0.
\end{equation}
$\rho(\mu^2)$ is in fact the \emph{spectral density} encountered in standard quantum field theory textbooks, where we can often find it in the form
\begin{equation}
	\bra{0} \phi(x) \phi(y) \ket{0}
	= 2\pi \int \frac{d^dk}{(2\pi)^d}
	\int\limits_0^\infty d\mu^2 \,
	e^{i k \cdot (x - y)} \theta(k^0) 
	\delta(k^2 + \mu^2) \rho(\mu^2).
	\label{eq:spectralrepresentation}
\end{equation}
The spectral density is an essential tool in non-perturbative quantum field theory. It can for instance be used in the construction of the time-ordered correlation function
\begin{equation}
	\langle \phi(x) \phi(y) \rangle_T
	\equiv 
	\theta(x^0 - y^0) \bra{0} \phi(x) \phi(y) \ket{0}
	+ \theta(y^0 - x^0) \bra{0} \phi(y) \phi(x) \ket{0}.
\end{equation}
Unlike the Wightman function, this is not a tempered distribution, because the $\theta$ function is not differentiable at the origin.
Nevertheless, the time-ordered product admits the simple representation
\begin{equation}
	\langle \phi(x) \phi(y) \rangle_T
	= \int \frac{d^dk}{(2\pi)^d}
	\int\limits_0^\infty d\mu^2 \,
	e^{i k \cdot (x - y)}
	\frac{i}
	{ -k^2 - \mu^2 + i \varepsilon} \, \rho(\mu^2)
	\label{eq:KallenLehmann}
\end{equation}
where the limit $\varepsilon \to 0_+$ is understood.
This is the \emph{Källen-Lehmann representation} for the time-ordered 2-point function.

\begin{exercise}
	Derive the Källen-Lehmann representation.
	An elegant derivation is to first show that the 
	time-ordered 2-point function can be written as the difference
	between the Wightman function and the vacuum expectation value
	of a retarded commutator,
	$$
	\langle \phi(x) \phi(y) \rangle_T
	= 
	W(y - x)
	- \theta(y^0 - x^0)
	\bra{0} \left[ \phi(x), \phi(y) \right] \ket{0}.
	$$
	The next step is to Fourier transform both terms in $y$
	after setting $x = 0$.
	We know that the Wightman function \eqref{eq:Wtilde}
	has a nice Fourier transform
	$$
		\widetilde{W}(q)
		= 2\pi \int\limits_0^\infty d\mu^2 \,
		\theta(q^0) 
		\delta(q^2 + \mu^2) \rho(\mu^2),
	$$
	which can be equivalently written
	$$
		\widetilde{W}(q)
		= \int\limits_0^\infty d\mu^2 \,
		\theta(q^0) 
		\left[ \frac{i}{q^2 + \mu^2 + i \varepsilon}
		- \frac{i}{q^2 + \mu^2 - i \varepsilon} \right] \rho(\mu^2).
	$$
	The retarded commutator is only non-zero in the forward light cone
	in $y$, and therefore it also admits a Fourier transform
	that converges provided that we give an imaginary part to $q$.
	Compute this Fourier transform, and show that at real $q$
	it is equal to
	$$
		\int\limits_0^\infty d\mu^2 \,
		\left[ \theta(q^0) \frac{i}{q^2 + \mu^2 + i \varepsilon}
		+ \theta(-q^0) \frac{i}{q^2 + \mu^2 - i \varepsilon}
		\right] \rho(\mu^2).
	$$
	The difference between these last two integrals can then easily be
	turned into the Källen-Lehmann representation
	\eqref{eq:KallenLehmann}.
\end{exercise}

\begin{figure}
	\includegraphics[width=0.48\linewidth]
		{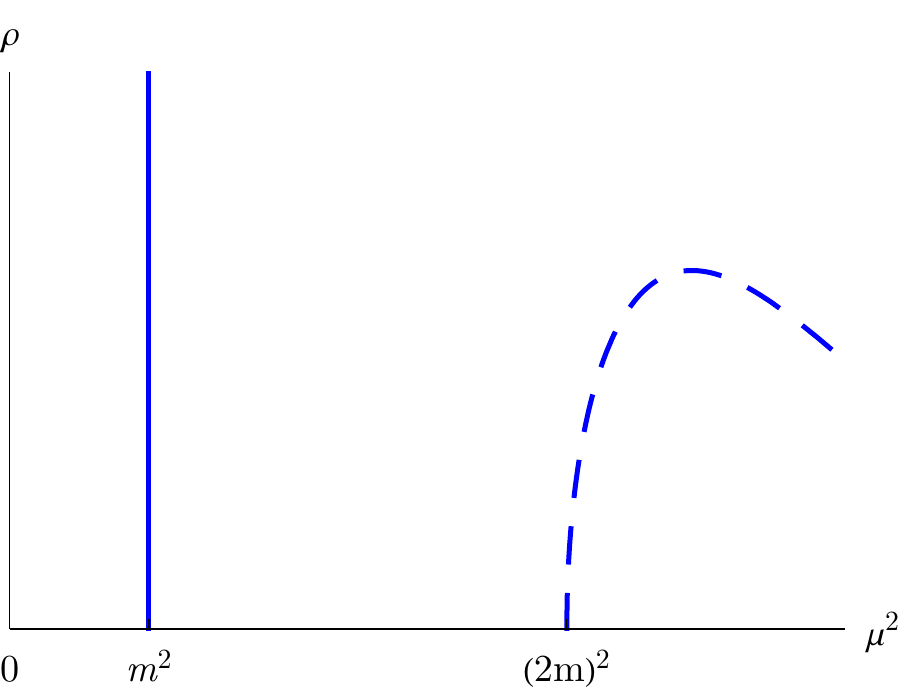}
	\hfill
	\includegraphics[width=0.48\linewidth]
		{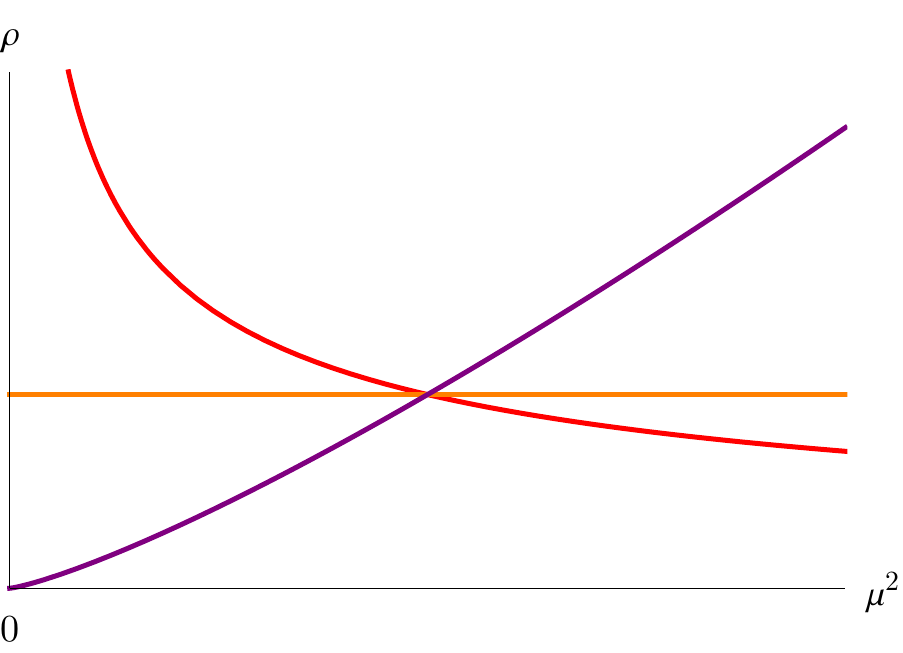}
	\caption{Left: spectral density in the theory of a free massive
	field, and typical contributions above the threshold
	for particle production (dashed line).
	Right: possible spectral densities in a scale-invariant theory.}
	\label{fig:spectraldensity}
\end{figure}

\noindent
The spectral representation for the 2-point function gives familiar results in non-interacting theories. A massive scalar field has for instance the spectral density
\begin{equation}
	\rho(\mu^2) = \delta(\mu^2 - m^2),
\end{equation}
and from this we recover the known massive propagator
\begin{equation}
	\langle \phi(x) \phi(y) \rangle_T
	= \int \frac{d^dk}{(2\pi)^d}
	e^{i k \cdot (x - y)}
	\frac{i}
	{-k^2 - m^2 + i \varepsilon}.
\end{equation}
In an interacting theory, the spectral density will get contributions corresponding to particle production above a certain threshold (see figure~\ref{fig:spectraldensity}).

The discussion applied so far to a generic quantum field theory without conformal symmetry. Let us now examine the role of scale and special conformal invariance, starting with the former.

\subsection{Scale symmetry}

The assumption of scale symmetry coincides with the existence of a third conserved charge besides $P^\mu$ and $M^{\mu\nu}$, namely the operator $D$.
We obtained before the commutator of a generic local operator with $M^{\mu\nu}$ assuming that it transforms in some irreducible representation of the Lorentz group at the origin $x = 0$
(an operator inserted at some other point is not in an irreducible representation because $P^\mu$ does not commute with $M^{\mu\nu}$).
Since $M^{\mu\nu}$ commutes with $D$, the same assumption can be made about scale: any local operator can be decomposed further into 
irreducible representations of the group of scale transformations, meaning that we can write
\begin{equation}
	\left[ D, \phi(0) \right] = -i \Delta \phi(0).
\end{equation}
$\Delta$ is called the \emph{scaling dimension} of the operator $\phi$ (each local operator of the theory has its own scaling dimension).
The factor of $i$ ensures that $\Delta$ is a real number when $\phi$ is a real operator.
As before, we can use eq.~\eqref{eq:P:exponentiated} to obtain the commutator at any other point $x$:
\begin{equation}
	\left[ D, \phi(x) \right]
	= -i \left( x^\mu \partial_\mu + \Delta \right) \phi(x).
	\label{eq:commutator:D}
\end{equation}
Note that this is consistent with the transformation rule \eqref{eq:freescalar:conformaltransformation:3} for a classical field:
the scaling dimension $\Delta$ coincides with the mass dimension of the operator $\phi$ in a free theory.

Using this new commutator and assuming that the vacuum state is invariant under scale transformations, a new Ward identity can be obtained for the Wightman 2-point function,
\begin{equation}
	\left( x^\mu \frac{\partial}{\partial x^\mu} 
	+ y^\mu \frac{\partial}{\partial y^\mu}  + 2\Delta \right)
	\bra{0} \phi(x) \phi(y) \ket{0} = 0,
\end{equation}
or equivalently, using eq.~\eqref{eq:W},
\begin{equation}
	\left( x^\mu \frac{\partial}{\partial x^\mu} + 2 \Delta \right) 
	W(x) = 0. 
\end{equation}
The corresponding condition on the momentum-space 2-point function
is obtained from the Fourier transform, using integration by parts:
\begin{equation}
	\left( -q^\mu \frac{\partial}{\partial q^\mu} + 2 \Delta - d \right)
	\widetilde{W}(q) = 0. 
	\label{eq:2ptWardIdentity:scale}
\end{equation}
The solution to this equation consistent with the form \eqref{eq:spectraldensity} is unique, up to a multiplicative constant $C$,
\begin{equation}
	\widetilde{W}(q)
	= 2\pi C \, \theta\left( q^0 - \left| \vec{q} \right| \right)
	(-q^2)^{\Delta - d/2},
	\label{eq:2pt:momentum}
\end{equation}
implying that the spectral density is a power of the energy,
\begin{equation}
	\rho(\mu^2) = C \, (\mu^2)^{\Delta - d/2}.
\end{equation}
This kind of spectral density, shown in figure~\ref{fig:spectraldensity}, is very different from that of a massive interacting theory: the operator $\phi$ creates states of all energies.
At the same time, its simplicity is striking: it is characterized by a single parameter $\Delta$, and a normalization constant that does not carry physical information (the operators can be re-defined to absorb this constant).

It is instructive to compute the time-ordered function using the Källen-Lehmann representation: performing the integral over $\mu^2$, we find
\begin{equation}
	\langle \phi(x) \phi(y) \rangle_T
	= \frac{i \pi C}
	{\sin\left[ \pi \left( \Delta - \frac{d}{2} \right) \right]}
	\int \frac{d^dk}{(2\pi)^d}
	e^{i k \cdot (x - y)}
	\left( k^2 - i \varepsilon \right)^{\Delta - d/2}.
\end{equation}
The term $\left( k^2 - i \varepsilon \right)^{\Delta - d/2}$ in the integral looks like the propagator for a massless scalar field raised to a non-integer power. This is in fact what we expect in perturbation theory when the $\beta$-function has a non-trivial fixed point: the renormalized 2-point function has logarithms that can be re-summed into a power controlled by the anomalous dimension $\gamma$ of the field $\phi$,
\begin{equation}
	\frac{i}{-q^2} \left[ 1 + \gamma \log(-q^2) + \ldots \right]
	\approx i (-q^2)^{-1 + \gamma}.
\end{equation}
In this case, the scaling dimension of the scalar operator corresponding to the renormalized field is
\begin{equation}
	\Delta = \frac{d-2}{2} + \gamma.
\end{equation}
In the limit $\gamma \to 0$, we recover the free propagator mentioned above.
Note however that the factor in front of the integral diverges in this limit, unless the coefficient $C$ satisfies
\begin{equation}
	C \propto \gamma = \Delta - \frac{d-2}{2}.
	\label{eq:C:freelimit}
\end{equation}
Assuming that it is the case (see below), the spectral density obeys%
\footnote{This limit can be verified by integrating both sides in $\mu^2$ and taking the limit $\gamma \to 0$ afterward.}
\begin{equation}
	\rho(\mu^2) \propto \gamma \, (\mu^2)^{-1 + \gamma}
	\xrightarrow{\gamma \to 0} \delta(\mu^2).
\end{equation}
This is precisely the spectral density that is expected in the free scalar field theory. In this case (and this case only!), the operator $\phi$ describes a massless scalar particle.

It turns out that $\Delta = \frac{d-2}{2}$ is the lowest possible value for $\Delta$: for any $\Delta$ below that value, the spectral density is not integrable in the limit $\mu^2 \to 0$.
One might also worry about the opposite limit $\mu^2 \to \infty$ in the integral: for any $\Delta > \frac{d}{2}$, the spectral density grows with $\mu^2$. However, remember that this spectral density is in fact a Wightman function, i.e.~a tempered distribution that should be understood as integrated against test functions that decay faster than any power at large $q^2$. Therefore, arbitrarily large values of $\Delta$ are possible, but there is a lower bound on $\Delta$ below which the states smeared with test functions have infinite norm.
The inequality
\begin{equation}
	\Delta \geq \frac{d-2}{2}.
	\label{eq:unitarybound:scalar}
\end{equation}
is known in the literature as the \emph{unitarity bound} for scalar operators.%
\footnote{It is usually derived using the action of special conformal transformations, but we have just seen here that it applies to scale-invariant theories as well.}
Note that any scalar operator that saturates this unitarity bound has
\begin{equation}
	\bra{0} \widetilde{\phi}(p) \widetilde{\phi}(q) \ket{0}
	\propto \delta(q^2),
\end{equation}
which implies
\begin{equation}
	q^2 \bra{0} \widetilde{\phi}(p) \widetilde{\phi}(q) \ket{0}
	= 0,
\end{equation}
or in position space
\begin{equation}
	\bra{0} \phi(x) \partial^2 \phi(y) \ket{0} = 0.
\end{equation}
Since this is true for any $x$ and $y$, it implies that
\begin{equation}
	\partial^2 \phi(x) = 0
\end{equation}
is true as an operator equation. Since this is the equation of motion for a free field, a theory in which $\Delta = \frac{d-2}{2}$ is a free field theory.

The simplicity of the momentum-space 2-point function in a scale-invariant theory also means that it can easily be Fourier transformed back to position space using
\begin{equation}
	W(x) = \int \frac{d^dq}{(2\pi)^d} e^{- i q \cdot x}
	\widetilde{W}(q).
\end{equation}

\begin{exercise}
	Perform the Fourier transform explicitly.
	You can use the fact that the integral is Lorentz
	invariant to determine that $W(x)$ is in fact a function of $x^2$.
	Moreover, since the integrand only has support for $q$
	in the forward light-cone, this defines a function of $x$
	that is analytic in $x$ as long as $\im x$ is contained in 
	the future light cone: the integrand is damped by the
	exponential $e^{q \cdot \im x}$ with $q \cdot \im x < 0$
	in that case
	(this domain of analyticity is known as the ``future tube'').
	This means that we are free to evaluate the integral at a point 
	$x = (i \tau, 0)$, and then use $\tau^2 = x^2$ to recover the 
	general solution.
	The integral at that point is convergent for all $\Delta$
	satisfying the unitarity bound, and you should find
	$$
	W(\tau) = C \frac{2^{2\Delta} \Gamma(\Delta)
	\Gamma\left( \Delta - \frac{d-2}{2} \right)}
	{(4\pi)^{d/2}} \,
	\tau^{-2\Delta}
	$$
\end{exercise}

\noindent
The result of this integral can be written as
\begin{equation}
	W(x) = \frac{C'}
	{\left[ -(x^0 + i \varepsilon)^2 + \vec{x}^2 \right]^\Delta}
\end{equation}
where the limit $\varepsilon \to 0_+$ is understood to make sense of the case in which $x^2 \leq 0$, namely
\begin{equation}
    W(x) = \frac{C'}{\left| x^2 \right|^\Delta} \times
    \left\{ \begin{array}{ll}
        e^{-i \pi \Delta} & \text{if}~x^0 < -\left| \vec{x} \right|,
        \\
        1 & \text{if}~-\left| \vec{x} \right| < x^0 < \left| \vec{x} \right|,
        \\
        e^{i \pi \Delta} & \text{if}~x^0 > \left| \vec{x} \right|.
    \end{array} \right.
\end{equation} 
The coefficient $C'$ and $C$ are related by
\begin{equation}
	C = \frac{(4\pi)^{d/2}}
	{2^{2\Delta} \Gamma(\Delta)
	\Gamma\left( \Delta - \frac{d-2}{2} \right)}
	C'.
	\label{eq:standardnormalization}
\end{equation}
Note that the proportionality factor is positive for all $\Delta$ satisfying the unitary bound \eqref{eq:unitarybound:scalar}. This implies that the 2-point correlation function is always decreasing with the distance and not the other way around.
It is in fact customary in conformal field theory to normalize the scalar operator $\phi$ so that $C' = 1$, in which case $C$ vanishes as in eq.~\eqref{eq:C:freelimit} in the limit $\Delta \to \frac{d-2}{2}$.
	
Finally, let us conclude the analysis of scale symmetry with a comment about one-point functions. We saw in the previous section that a constant vacuum expectation value for a scalar operator was compatible with Poincaré symmetry. However, the commutator \eqref{eq:commutator:D} requires then that $\Delta = 0$, which violates the unitarity bound.
We conclude that all one-point functions must vanish in a scale-invariant theory.

\subsection{Special conformal symmetry}

As with scale symmetry, the presence of special conformal symmetry is associated with the existence of the conserved charges $K^\mu$, organized in a $d$-dimensional vector.
Unlike $D$, however, $K^\mu$ does not commute with $M^{\mu\nu}$, so it cannot be diagonalized at the point $x = 0$.
Nevertheless, we can use the conformal algebra to establish that, if $\phi$ is a local operator with scaling dimension $\Delta$, then $\left[ K^\mu, \phi \right]$ has scaling dimension $\Delta - 1$:
\begin{align}
	\big[ D, [ K^\mu, \phi(0) ] \big]
	&= \big[ K^\mu, [ D, \phi(0) ] \big]
	+ \big[ [ D, K^\mu ], \phi(0) \big]
	\nonumber \\
	&= \big[ K^\mu, -i \Delta \phi(0) \big]
	+ \big[ i K^\mu, \phi(0) \big]
	= -i (\Delta - 1) [ K^\mu, \phi(0) ].
\end{align}
This is similar to the observation that the commutator $\left[P^\mu, \phi(x)\right]$ has scaling dimension $\Delta + 1$,
\begin{equation}
	\big[ D, [ P^\mu, \phi(0) ] \big]
	= -i (\Delta + 1) [ P^\mu, \phi(0) ],
\end{equation}
consistent with the fact that the derivative $\partial^\mu$ has mass dimension $+1$.
The fact that $K^\mu$ lowers the scaling dimensions appears to be in contradiction with our findings of the last section stating that $\Delta$ is bounded below: given any local operator, one can always construct other local operators with arbitrarily smaller scaling dimension.

The only way out of this apparent paradox is to assume that at some point the action of $K^\mu$ annihilates the operator. In other words, there must exist some local operator such that
\begin{equation}
	\left[ K^\mu, \phi(0) \right] = 0.
\end{equation}
We call this local operator a \emph{primary}. Any other local operator can be obtained acting on a primary with $P^\mu$, and we call it a \emph{descendant}. Since the action of $P^\mu$ coincides with taking derivatives, a primary operator is simply an operator that cannot be written as the derivative of some other operator.
Unless specified otherwise, we shall from now on only consider Wightman correlation functions of primary operators. Descendants will be explicitly denoted with a derivative.

The transformation of a primary operator away from the origin can once again be obtained from eq.~\eqref{eq:P:exponentiated}. Note that since the commutator of $P^\mu$ and $K^\mu$ involves both $D$ and $M^{\mu\nu}$, this transformation depends on the scaling dimension and on the Lorentz representation of the operators, i.e.~on the eigenvalues $\Delta$ and $\mathcal{S}^{\mu\nu}$. We find
\begin{equation}
	\left[ K^\mu, \phi(x) \right]
	= -i \left( 2 x^\mu x^\nu \partial_\nu - x^2 \partial^\mu 
	+ 2 \Delta x^\mu - 2 \mathcal{S}^{\mu\nu} x_\nu \right) \phi(x).
	\label{eq:commutator:K}
\end{equation}
This equation also defines the commutator of a momentum-space operator: using integration by parts in the definition \eqref{eq:momentumspace}, one can show that this amounts to replacing $\partial_\mu \to -i q_\mu$ and $x^\mu \to -i \partial/ \partial q_\mu$,
so that
\begin{equation}
	\big[ K^\mu, \widetilde{\phi}(q) \big]
	= \left[ 2 \frac{\partial^2}{\partial q_\mu \partial q_\nu} q_\nu
	- \frac{\partial^2}{\partial q_\nu \partial q^\nu} q^\mu
	- 2 \Delta \frac{\partial}{\partial q_\mu}
	+ 2 \mathcal{S}^{\mu\nu} \frac{\partial}{\partial q^\nu} \right]
	\widetilde{\phi}(q),
\end{equation}
or after permuting the derivatives with $q$,
\begin{equation}
	\big[ K^\mu, \widetilde{\phi}(q) \big]
	= \left[ 2 q^\nu \frac{\partial^2}{\partial q_\mu \partial q^\nu} 
	- q^\mu \frac{\partial^2}{\partial q_\nu \partial q^\nu}
	+ 2 (d - \Delta) \frac{\partial}{\partial q_\mu}
	+ 2 \mathcal{S}^{\mu\nu} \frac{\partial}{\partial q^\nu} \right]
	\widetilde{\phi}(q).
\end{equation}
This is now a second-order differential acting on the operator expressed in momentum space. 

The first thing we can do with this commutator is to examine the related Ward identity for the Wightman 2-point function. Remember that this function can be written as
\begin{equation}
	W(x) = \bra{0} \phi(0) \phi(x) \ket{0}.
\end{equation}
The commutator acts trivially on the operator inserted at the origin, so that we must have (note that $\mathcal{S}^{\mu\nu} = 0$ for scalar operators)
\begin{equation}
	\left( 2 x^\mu x^\nu \partial_\nu - x^2 \partial^\mu 
	+ 2 \Delta x^\mu \right) W(x) = 0.
\end{equation}
Let us check at space-like $x$:
using $W(x) = 1 /(x^2)^{\Delta}$, we have $\partial_\mu W(x) = - 2 \Delta W(x) x^\mu / x^2$, and therefore the differential equation is readily satisfied.
The same can be verified in momentum space: by definition, we have
\begin{equation}
	\widetilde{W}(q) = \bra{0} \phi(0) \widetilde{\phi}(q) \ket{0},
\end{equation}
where only the operator on the right is Fourier transformed while the one on the left is kept fixed at the origin in position space, and so the commutator above implies that
\begin{equation}
	\left[ 2 q^\nu \frac{\partial^2}{\partial q_\mu \partial q^\nu} 
	- q^\mu \frac{\partial^2}{\partial q_\nu \partial q^\nu}
	+ 2 (d - \Delta) \frac{\partial}{\partial q_\mu} \right]
	\widetilde{W}(q) = 0.
\end{equation}
With $\widetilde{W}(q) = (-q^2)^{\Delta - d/2}$, this equation is again satisfied.

\paragraph{Distinct operators:}

The fact that $W(x)$ and $\widetilde{W}(q)$ readily satisfy the constraint imposed by special conformal symmetry is very specific to identical scalar operators. In every other case, special conformal symmetry adds more constraints than Poincaré and scale symmetry alone.
The simplest example is that of a 2-point function of distinct scalar operators,
\begin{equation}
	\bra{0} \phi_1(x) \phi_2(y) \ket{0}.
\end{equation}
This is still a function of $(x-y)^2$ by Poincaré symmetry. But there are now two distinct scaling dimensions $\Delta_1$ and $\Delta_2$ corresponding to the operators $\phi_1$ and $\phi_2$, and the Ward identity for scale symmetry becomes (for simplicity setting $\phi_1$ at the origin)
\begin{equation}
	\left( x^\mu \frac{\partial}{\partial x^\mu}
	+ \Delta_1 + \Delta_2 \right)
	\bra{0} \phi_1(0) \phi_2(x) \ket{0} = 0.
\end{equation}
The solution is fixed up to a multiplicative constant to be
(assuming $x$ space-like for simplicity)
\begin{equation}
	\bra{0} \phi_1(0) \phi_2(x) \ket{0}
	= \frac{C_{12}}{(x^2)^{(\Delta_1 + \Delta_2)/2}}.
\end{equation}
The Ward identity for special conformal transformation is obtained from the commutator \eqref{eq:commutator:K}, giving
\begin{equation}
	\left( 2 x^\mu x^\nu \partial_\nu - x^2 \partial^\mu 
	+ 2 \Delta_2 x^\mu \right)
	\bra{0} \phi_1(0) \phi_2(x) \ket{0}
	= 0.
\end{equation}
Using $\partial_\mu \left[ (x^2)^{-(\Delta_1 + \Delta_2)/2} \right] = -(\Delta_1 + \Delta_2) x^\mu / x^2$, this implies
\begin{equation}
	\left( \Delta_2 - \Delta_1 \right) C_{12} 
	\frac{x^\mu}{(x^2)^{(\Delta_1 + \Delta_2)/2}} = 0.
\end{equation}
If the scaling dimensions are different ($\Delta_1 \neq \Delta_2$), then $C_{12}$ must vanish. This is an important lesson: in conformal field theory, only primary operators of identical scaling dimensions can have non-zero 2-point functions.

In fact, if there are several scalar operators with the same scaling dimension $\phi_i$ with $i = 1, \ldots, N$, then 
\begin{equation}
	\bra{0} \phi_i(0) \phi_j(x) \ket{0}
	= \frac{C_{ij}}{(x^2)^\Delta},
\end{equation}
where $C_{ij}$ is a symmetric $N \times N$ matrix. By unitarity, this matrix must be positive-definite: if this were not the case, then one could define a negative-norm state by taking an appropriate linear combination of the $\phi_i$ and smearing. Therefore, it is always possible to choose a basis of operators in which $C_{ij}$ is diagonal. Moreover, the operators can be normalized so that $C_{ij} = \delta_{ij}$. From now on, we will therefore always assume that the only non-zero 2-point functions are those involving identical operators.

\paragraph{Operators with spin:}

The other situation in which special conformal symmetry plays an essential role is when the operators carry spin.
Let us take the simplest example of a vector operator $A^\mu(x)$, and denote its 2-point function by
\begin{equation}
	W^{\mu\nu}(x) = \bra{0} A^\mu(0) A^\nu(x) \ket{0}.
\end{equation}
As with scalars, one can also take the Fourier transform of this tempered distribution, defining
\begin{equation}
	\widetilde{W}^{\mu\nu}(p) = \int d^dx \, e^{i p \cdot x}
	\bra{0} A^\mu(0) A^\nu(x) \ket{0}
	= \bra{0} A^\mu(0) \widetilde{A}^\nu(p) \ket{0}.
	\label{eq:vector:mixedrep}
\end{equation}
Again, this function corresponds to the momentum-space correlation function without the delta-function imposing momentum conservation, namely
\begin{equation}
	\bra{0} \widetilde{A}^\mu(p) \widetilde{A}^\nu(q) \ket{0}
	= (2\pi)^d \delta^d(p + q) \widetilde{W}^{\mu\nu}(q).
\end{equation}
By Lorentz symmetry, this function $\widetilde{W}$ of a single momentum can be decomposed into two different tensor structures multiplying scalar functions,
\begin{equation}
	\widetilde{W}^{\mu\nu}(p)
	= (p^\mu p^\nu - p^2 \eta^{\mu\nu}) \widetilde{W}_1(p)
	+ p^\mu p^\nu \widetilde{W}_0(p).
	\label{eq:vector:polarizations}
\end{equation}
Moreover, using scale symmetry and energy positivity, we can infer that the functions $\widetilde{W}_{1,0}$ are just powers of $p^2$ over the forward light cone,
\begin{equation}
	\widetilde{W}_{1,0}(p)
	= \theta\left( p^0 - \left| \vec{p} \right| \right)
	(-p^2)^{\Delta - d/2 - 1} C_{1,0},
\end{equation}
where $\Delta$ is the scaling dimension of the operator $A^\mu$ and $C_1$, $C_0$ two constants that cannot be related by scale and Poincaré symmetry only.

There is a good reason for using precisely the two tensor structures in eq.~\eqref{eq:vector:polarizations} and not, say, $\eta^{\mu\nu}$ and $p^\mu p^\nu$. Thanks to energy positivity, it is always possible to choose a Lorentz frame in which $\vec{p} = 0$.%
\footnote{This is similar to going to a massive particle's rest frame.}
In this frame, the momentum is invariant under the group $\SO(d-1)$ of spatial rotations, and therefore the 2-point function can be decomposed into irreducible representations of that group. The part proportional to $C_0$ only appears in the component $\widetilde{W}^{00}$, and it transforms like a scalar under rotations. Conversely, the part proportional to $C_1$ only has non-zero entries for spatial Lorentz indices $\widetilde{W}^{ij}$; it is in fact proportional to the identity in the $d-1$ subspace, i.e.~it is the invariant tensor for the vector representation of $\SO(d-1)$.
In particle physics language, we would call these two parts respectively \emph{longitudinal} and \emph{transverse}.

Being able to use irreducible representations of $\SO(d-1)$ is an advantage of working in momentum space: there is no obvious Lorentz frame in which such a decomposition can be made in position space since the 2-point function has support over all of Minkowski space-time. The disadvantage of working in momentum space is that the Ward identity for special conformal transformation is a second-order differential equation in $p$, while it is a first-order differential in position space.
This Ward identity can nevertheless be straightforwardly applied to eq.~\eqref{eq:vector:polarizations}, and it yields a relation between the longitudinal and transverse parts, i.e.~between the coefficients $C_0$ and $C_1$, given by (see exercise)
\begin{equation}
	C_0 = \frac{\Delta - d + 1}{\Delta - 1} \, C_1.
	\label{eq:vector:polarizationrelations}
\end{equation}
This is a very important consequence of special conformal symmetry: 
while in a scale-invariant theory the longitudinal and transverse polarizations are independent, in a conformal theory they are related.

\begin{exercise}
	Using the definition \eqref{eq:vector:mixedrep} for the function
	$\widetilde{W}^{\mu\nu}(p)$, show that it satisfies the special
	conformal Ward identity
	$$
	\left( 2 p^\beta
	\frac{\partial^2}{\partial p_\alpha \partial p^\beta}
	- p^\alpha
	\frac{\partial^2}{\partial p_\beta \partial p^\beta}
	+ 2 (d - \Delta) \frac{\partial}{\partial p_\alpha}
	+ 2 \mathcal{S}^{\alpha\beta} \frac{\partial}{\partial p_\beta} \right)
	\widetilde{W}^{\mu\nu}(p) = 0,
	$$
	where $\mathcal{S}^{\alpha\beta}$ is given by eq.~\eqref{eq:spinop:vector},
	and use this to prove the relation
	\eqref{eq:vector:polarizationrelations}.
\end{exercise}

\noindent
This has consequences on the possible values that $\Delta$ can take. As before, this 2-point function computes the norm of a state, and its positivity requires:
\begin{itemize}

\item
$\Delta > \frac{d}{2}$ so that the 2-point function is integrable at $p^2 \to 0$;

\item
$C_0$ and $C_1$ both positive, so that the norm is positive for any choice of external polarization vector (i.e.~the tensor $\widetilde{W}^{\mu\nu}$ must be positive-definite). This requires $\Delta - 1$ and $\Delta - d + 1$ to have the same sign.

\end{itemize}
The combination of these two conditions in $d > 2$ dimensions (spin is treated differently in $d = 2$) implies
\begin{equation}
	\Delta \geq d - 1.
\end{equation}
This is known as the unitarity bound for a vector operator.

As in the scalar case, something special happens when the unitarity bound is saturated ($\Delta = d - 1$). In this case the 2-point function has no longitudinal component, $C_0 = 0$,
and $\widetilde{W}^{\mu\nu}$ vanishes when contracted with $p_\mu$ or $p_\nu$. This implies that the longitudinal part of the state is null,
\begin{equation}
	p_\mu \widetilde{A}^\mu(p) \ket{0} = 0,
\end{equation}
or equivalently that $\widetilde{A}^\mu(p)$ is an operator that only creates transverse-polarization states.
The equivalent statement in position-space is 
\begin{equation}
	\partial_\mu A^\mu(x) \ket{0} = 0.
\end{equation}
In other words, $A^\mu$ is a conserved current.
The equivalence goes both way: any vector operator with $\Delta = d - 1$ is a conserved current, and any conserved current must have scaling dimension $\Delta = d - 1$. This also shows that conserved currents are  primary operators: they cannot be written as $\partial^2$ acting on another vector operator (that operator would have $\Delta = d - 3$, below the unitarity bound), nor as $\partial^\mu$ acting on a scalar operator $\phi$, because the conservation requirement would then imply $\partial^2 \phi = 0$, which is only possible if $\phi$ has scaling dimensions $(d-2)/2,$ and thus the current $\Delta = d/2$ (again below the unitarity bound).

The fact that 2-point functions of primary operators are completely fixed by conformal symmetry up to a choice of normalization is not specific to scalar and vector operators. In fact, any local operator specified by a representation under the Lorentz group and a scaling dimension defines an irreducible representation of the conformal group $\SO(d,2)$, and as such its 2-point function is fixed by group theory. This also explains on more general grounds why 2-point functions of distinct operators vanish.
The construction of all unitary representations of the conformal group in $d = 4$ dimensions was performed by Mack in 1975~\cite{Mack:1975je}, and similar constructions can be done in other dimensions. 
Some Lorentz representations are specific to a given dimension $d$, and others exist in any $d$, like the scalar and the symmetric, traceless representations with $\ell$ Lorentz indices (the vector discussed above is a special case corresponding to $\ell = 1$). All such symmetric tensors satisfy the unitarity bound
\begin{equation}
	\Delta \geq d - 2 + \ell,
	\label{eq:unitaritybound:symmetricops}
\end{equation}
and they are in general described by $\ell + 1$ distinct polarizations (irreducible representations of the rotation group), except when the bound is saturated, in which case there is just a single, transverse polarization and the operator is a higher-spin conserved current.
In terms of representations of the conformal group, generic operators are said to belong to \emph{long multiplets}, whereas special cases such as scalars with $\Delta = (d-2)/2$ or symmetric tensors with $\Delta = d - 2 + \ell$ are said to be in \emph{short multiplets} (they contain fewer descendants). 

\begin{exercise}
	Construct explicitly the 2-point function of a 2-index
	symmetric traceless operator $B^{\mu\nu}(x)$.
	As a starting point, let us decompose the momentum-space
	correlation functions into tensors that transform covariantly
	under rotations in the rest frame. Using the transverse projector 
	$$ \eta_\perp^{\mu\nu} = \eta^{\mu\nu} - \frac{p^\mu p^\nu}{p^2}$$
	satisfying $p_\mu \eta_\perp^{\mu\nu} = 0$, this can be done as
	\begin{align*}
		\bra{0} B^{\mu\nu}(0) \widetilde{B}^{\rho\sigma}(p) \ket{0}
		= \bigg[ & \frac{1}{2}
		\left( \eta_\perp^{\mu\rho} \eta_\perp^{\nu\sigma}
		+ \eta_\perp^{\mu\sigma} \eta_\perp^{\nu\rho} 
		- \text{traces} \right) C_2
		\\
		& + \frac{1}{4}
		\left( \eta_\perp^{\mu\rho} \frac{p^\nu p^\sigma}{p^2}
		+ \text{permutations} \right) C_1
		\\
		& + \frac{p^\mu p^\nu p^\rho p^\sigma}{(p^2)^2}
		C_0 \bigg]
		\theta\left( p^0 - \left| \vec{p} \right| \right)
		(-p^2)^{\Delta - d/2 + 1}.
	\end{align*}
	Then write down the Ward identity for special conformations
	(including the spin operator for a 2-index tensor, which 
	you have to determine),
	and show that it leads to the conditions
	\begin{align*}
		C_1 &= 2 \frac{\Delta - d}{\Delta} \, C_0,
		\\
		C_2 &= \frac{d}{d-1} \frac{(\Delta - d) (\Delta - d + 1)}
		{\Delta (\Delta - 1)} \, C_0.
	\end{align*}
	Argue that this gives rise to the unitarity bound $ \Delta \geq d$,
	in agreement with eq.~\eqref{eq:unitaritybound:symmetricops}.
	Conclude that the energy-momentum tensor $T^{\mu\nu}$
	is a primary operator with $\Delta = d$.
\end{exercise}

\subsection{UV/IR divergences and anomalies}

The discussion has been focused so far on Wightman functions. Besides having a Hilbert-space interpretation and satisfying conformal Ward identities that give strong constraints on their possible form, Wightman functions are also free of divergences.
In momentum space, regularity at small momenta (IR) is enforced by the unitarity bound, whereas the power-law growth at large momenta (UV) is compatible with the damping provided by test functions.
In position space, the apparent singularity at short distance (UV) is resolved by the $i \varepsilon$ that provides an unequivocal prescription for deforming any contour of integration.

These properties are not found in time-ordered products. Let us consider once again the scalar 2-point function. Using the standard CFT normalization \eqref{eq:standardnormalization} and the Källen-Lehmann representation, we find
\begin{equation}
	\langle \phi(x) \phi(y) \rangle_T
	= -i
	\frac{(4\pi)^{d/2} \Gamma\left( \frac{d}{2} - \Delta \right)}
	{2^{2\Delta} \Gamma(\Delta)}
	\int \frac{d^dk}{(2\pi)^d}
	e^{i k \cdot (x - y)}
	\left( k^2 - i \varepsilon \right)^{\Delta - d/2}.
	\label{eq:UVdivergence}
\end{equation}
This expression diverges whenever $\Delta = \frac{d}{2} + n$ with integer $n$ due to the $\Gamma$-function multiplying the integral, indicating that the Fourier transform does not exist.
In this case, the 2-point function has scaling dimension $d + 2n$, and is therefore compatible with contact terms of the form
\begin{equation}
	(\partial^2)^n \delta^d(x - y).
	\label{eq:contactterm}
\end{equation}
In path integral language, this is the situation in which the source field $J$ for the operator $\phi$ has scaling dimension $d - \Delta = \frac{d}{2} - n$, and therefore contact terms of the form $J (\partial^2)^n J$ can (and must) be added to the action. 
These contact terms are obviously covariant under Poincaré and scale transformations, so they can be added to the correlation function without affecting the Ward identities at separated points.
\eqref{eq:contactterm} is the only possible contact term appearing in a scalar 2-point function, but more terms can appear in other correlation functions.

When Fourier transformed to momentum space, all such contact terms become polynomials in the momenta. In the scalar 2-point function case, these are $(p^2)^n$. Polynomial terms are incompatible with the positive-energy condition of Wightman functions, as they have support over all causal regions. But there are allowed (and in fact required) in time-ordered products. Time-ordered products are defined in position space as the product of Wightman functions, which are tempered distributions, with step functions, which are not: therefore they do not necessarily have a Fourier transform. Contact terms can be seen as a way of ``fixing'' the time-ordered product so that they can be Fourier transformed. 

This can be understood by analytic continuation in scaling dimension $\Delta$. Let us assume that our scalar operator has scaling dimension $\Delta = \frac{d}{2} + n - \epsilon$, and take the limit $\epsilon \to 0$ (note that this $\epsilon$ is different from the one in the $i \varepsilon$ limit). Then the time-ordered 2-point function in momentum space takes the form
\begin{equation}
	\Gamma(-n + \epsilon) (p^2)^{n - \epsilon}
	+ Z (p^2)^n,
\end{equation}
where the first term with a pole in $\epsilon$ comes from the expression \eqref{eq:UVdivergence} valid at $x \neq y$, and the second term is the counterterm added to the action.
Choosing $Z \propto \epsilon^{-1}$ allows to cancel the divergence as $\epsilon \to 0$, but it also gives rise to a logarithm, 
\begin{equation}
	(p^2)^n \log\left( \frac{p^2}{\mu^2} \right).
\end{equation}
The fact that we need to introduce a dimensionful quantity is the sign of a \emph{conformal anomaly}: a 2-point function of this form does not  satisfy the Ward identity for scale transformations \eqref{eq:2ptWardIdentity:scale},
\begin{equation}
	\left[ -p^\mu \frac{\partial}{p^\mu} + 2n \right]
	(p^2)^n \log\left( \frac{p^2}{\mu^2} \right)
	= -2 (p^2)^n \neq 0.
\end{equation}
However, the anomalous term on the right-hand side is a polynomial in the momenta, corresponding to a contact term, indicating that the anomaly is local. In QFT language, this is typical of a (renormalized) UV divergence.

Note that the presence of contact terms is associated with a very special type of operator of (half-)integer scaling dimensions. Such operators are generically absent in an interacting conformal field theory, where the scaling dimensions take irrational values. 
However, an exception to this rule concerns conserved currents in even space-time dimension $d$. For instance, a conserved current $J^\mu$ in $d = 4$ has scaling dimensions $\Delta = 3$; it is therefore associated in the path integral language with a source $a_\mu$ that carries dimension 1. This source is also subject to a gauge symmetry, $a_\mu \sim a_\mu + \partial_\mu \alpha$, and therefore a possible contact term is $f_{\mu\nu} f^{\mu\nu}$, where $f_{\mu\nu} = \partial_\mu a_\nu - \partial_\nu a_\mu$. As in the scalar case, this term must be used as a counterterm to cancel a divergence arising in the time-ordered correlation function, leading generically to logarithms in correlation functions involving the transverse polarization of the current,
\begin{equation}
	\langle J^\mu(p) J^\nu(q) \rangle_T
	\propto (2\pi)^4 \delta^4(p + q)
	( q^\mu q^\nu - q^2 \eta^{\mu\nu} )
	\log\left( \frac{q^2}{\mu^2} \right).
\end{equation}
As in the scalar case, this is the manifestation of a UV divergence. However, the logarithm also implies that the limit $q^2 \to 0$ diverges: this is what we would call an IR divergence in quantum field theory. 
In the context of CFT, there is no clear distinction between UV and IR divergences, as the two are closely related. They both arise from an ambiguity in taking the Fourier transform. UV divergences can be cured with local counterterms, at the price of introducing a reference scale, but IR divergences are physical.

Note finally that time-ordered correlation functions involving a conserved current do not only have transverse polarizations: while the conservation condition implies the vanishing of the state
\begin{equation}
	\partial_\mu J^\mu(x) \ket{0} = 0,
\end{equation}
and thus of Wightman correlation functions constructed from that state, this is not true of the divergence of $J^\mu$ appearing in a time-ordered product: the conservation equation $\partial_\mu J^\mu(x) = 0$ is only true as an \emph{operator equation}, namely away from coincident points. On general grounds, one expects
\begin{align}
	\langle \phi_1(x_1) \cdots \phi_n(x_n)
	\partial_\mu J^\mu(y) \rangle_T
	&= - \delta^d(x_1 - y)
	\langle \delta\phi_1(x_1) \cdots \phi_n(x_n)  \rangle_T
	\nonumber \\
	& \quad - \ldots
	\nonumber \\
	& \quad -\delta^d(x_n - y)
	\langle \phi_1(x_1) \cdots \delta\phi_n(x_n)  \rangle_T,
    \label{eq:conservedcurrentWI}
\end{align}
where $\delta\phi_i$ indicates the charge of the field $\phi_i$ under $J^\mu$.%
\footnote{It is conventional to normalize the conserved current $J$ so that eq.~\eqref{eq:conservedcurrentWI} is true, which means that the 2-point function of $J$ cannot be arbitrarily normalized like a scalar operator in eq.~\eqref{eq:standardnormalization}. The same is true of the energy-momentum tensor, and of higher-spin conserved currents. For all these operators the normalization of the 2-point function involves a coefficient of physical relevance.}

%%%%%%%%%%%%%%%%%%%%%%%%%%%%%%%%%%%%%%%%%%%%%%%%%%%%%%%%%%%%%%%%%%%%%%

\section{Conformal correlation functions}
\label{sec:correlators}

The method presented in the last section using conformal Ward identities and the Wightman axioms could in principle be used to determine 3- and higher-point correlation functions.%
\footnote{The conformal Ward identities give differential equations and the Wightman axioms boundary conditions. Together, these are sufficient to constrain the conformal correlation functions. Note however that energy positivity alone is not sufficient: the micro-causality axiom is required as well (see ref.~\cite{Gillioz:2021sce} for a discussion at the level of 3-point functions).
Alternatively, for time-ordered or Euclidean correlators, the boundary condition can be given by an OPE limit~\cite{Osborn:1993cr} (see below).}
However, it is quite inconvenient, and it hides the simplicity of the result. To put things in perspective, note that the Wightman 3-point function of scalar operators in momentum space was only constructed in 2019~\cite{Gillioz:2019lgs, Bautista:2019qxj},%
\footnote{Results for the Euclidean momentum-space 3-point function appeared in 2013 already~\cite{Coriano:2013jba, Bzowski:2013sza}.}
whereas the position-space correlator has been known since the work of Polyakov in 1970~\cite{Polyakov:1970xd}.

\subsection{From Minkowksi space-time to Euclidean space}

For a start, let us go back to the Wightman 2-point function of scalar operators, given in position space by
\begin{equation}
	W(x) = \frac{1}{\left[ -(x^0 + i \varepsilon)^2
	+ \vec{x}^2 \right]^\Delta}.
\end{equation}
The very definition of this 2-point function with its $i \varepsilon$ prescription suggests that $x^0$ should be thought of as a complex variable: as a function of a complex $x^0$, $W$ is analytic in the upper-half complex plane. Taking $x^0$ to be purely imaginary, i.e.
\begin{equation}
	x^0 = i \tau,
	\qquad\qquad
	\tau > 0,
\end{equation}
we obtain the \emph{Schwinger function}
\begin{equation}
	\langle \phi(0) \phi(x_E) \rangle
	= \frac{1}{( \tau^2 + \vec{x}^2 )^\Delta}
	\equiv \frac{1}{(x_E^2)^\Delta}.
	\label{eq:2pt:Schwinger}
\end{equation}
We denote with $x_E = (\tau, \vec{x})$ the Euclidean vector that is contracted with the $d$-dimensional Euclidean metric. Schwinger functions will always be written using the ``average'' notation $\langle \cdots \rangle$, as opposed to the Wightman functions $\bra{0} \cdots \ket{0}$ that can be interpreted as vacuum expectation values; since the two types of functions cannot be confused, we shall denote the Euclidean coordinates by $x$ instead of $x_E$, even though the latter is implied.

Note that the Schwinger function transforms covariantly under the Euclidean conformal group $\SO(d + 1, 1)$ that is obtained by replacing the Minkowski metric by the Euclidean one. This includes both translations and $\SO(d)$ rotations, and as a consequence we have
\begin{equation}
	\langle \phi(0) \phi(x) \rangle
	\stackrel{\text{rotation}}{=} \langle \phi(0) \phi(-x) \rangle
	\stackrel{\text{translation}}{=} \langle \phi(x) \phi(0) \rangle,
\end{equation}
exhibiting a new symmetry under the exchange of the order of the two operators.
Also note that the Schwinger 2-point function is positive, and that it is not defined at the point $x_E = 0$, unlike the Wightman function whose $i \varepsilon$ prescription indicates how to approach any null point.

\begin{exercise}
	There is another way to get to the same result, starting from
	the momentum-space representation of the 2-point function. 
	The Wightman function is not well-suited to do so
	(at least not without exploring its analyticity properties 
	following from the micro-causality axiom),
	but the time-ordered function is: starting from the
	Källen-Lehmann representation~\eqref{eq:UVdivergence},
	which in momentum space becomes (using a notation in which the 
	momentum-conservation delta function is implicit)
	$$
	\langle \phi(-p) \phi(p) \rangle_T
	= -i
	\frac{(4\pi)^{d/2} \Gamma\left( \frac{d}{2} - \Delta \right)}
	{2^{2\Delta} \Gamma(\Delta)}
	\left( p^2 - i \varepsilon \right)^{\Delta - d/2},
	$$
	one can perform a Wick rotation in which $x^0$ and $p^0$
	are simultaneously rotated  in opposite directions
	in the complex plane, to arrive at the Euclidean result
	$$
	\langle \phi(-p_E) \phi(p_E) \rangle
	= 
	\frac{(4\pi)^{d/2} \Gamma\left( \frac{d}{2} - \Delta \right)}
	{2^{2\Delta} \Gamma(\Delta)}
	\left( p_E^2  \right)^{\Delta - d/2}.
	$$
	Perform the Fourier transform in $p_E$ to recover the 
	Schwinger function \eqref{eq:2pt:Schwinger}.
	Note that you will need to make assumptions about $\Delta$
	for the Fourier integral to converge. The result is 
	however analytic in $\Delta$, so you can argue a posteriori
	that it must be valid for all scaling dimensions.
\end{exercise}

This construction of a Euclidean function by analytic continuation from the Wightman function can in fact be generalized to any number of operators. Consider the $n$-point Wightman function parameterized as
\begin{equation}
	\bra{0} \phi_n(x_n - x_{n-1}) \phi_{n-1}(x_{n-1} - x_{n-2}) \cdots
	\phi_2(x_2 - x_1) \phi_1(x_1) \ket{0},
\end{equation}
and complexify all time components $x_i^0$. This function of several complex variables (and many more real ones) is in fact analytic in every upper-half complex plane in $x_1^0$ to $x_n^0$, because it can be written as the Fourier transform of a function that only has support when the dual variable $p_1^0$ to $p_n^0$ are all positive.%
\footnote{If we let all components of $x_i^\mu$ be complex, then the primary domain of analyticity is the so-called \emph{future tube}, defined by $\im x_i^0 > \left| \im \vec{x}_i \right|$.}
Therefore, going to purely imaginary times $x_i^0 = i \tau_i$, one obtains the Schwinger $n$-point function
\begin{equation}
	\langle \phi_n(x_n - x_{n-1}) \cdots
	\phi_2(x_2 - x_1) \phi_1(x_1) \rangle,
\end{equation}
in which the Euclidean times all satisfy $\tau_i > 0$, i.e.~the operators are ordered along the Euclidean time direction. 
This latter observation is however irrelevant because the ordering of operators does not matter in a Schwinger function: the analytic continuation can be performed starting with a configuration in which all real Minkowski times are equal, in which case the operators commute by micro-causality. 
The observation made on the 2-point function is therefore valid more generally: 
\begin{itemize}

\item
Schwinger functions are symmetric under the exchange of operators.

\end{itemize} 
The other observations made before are also true in general: 
\begin{itemize}

\item
Schwinger functions transform covariantly under the Euclidean conformal group $\SO(d+1, 1)$. This property will be very useful because it means that we can use finite conformal transformations that act nicely in Euclidean space (including $\infty$ as a point) to simplify our computations.

\item
Schwinger functions are not defined at coincident points. This is not a bug but a feature: using functions that are only defined at separated points means that we do not need to worry about contact terms and UV divergences.%
\footnote{It is in fact possible to describe contact terms (local and semi-local ones) using a generalization of the embedding formalism described in the next section~\cite{Nakayama:2019mpz}. But these contact terms are ambiguous: they carry more information than the Schwinger or Wightman functions themselves, and as such are unphysical.}
It means however that we cannot simply take the Fourier transform of these functions to obtain momentum-space Schwinger functions. 

\item
Schwinger functions enjoy a property called \emph{reflection positivity}: if the operators are organized in a configuration that is invariant under reflection across a plane (e.g.~four points at the corners of a square; or trivially any two points), then the correlator is positive.%
\footnote{This property would not be needed if we had been studying Euclidean conformal field theory from the start. Indeed, there are interesting critical fixed points in condensed matter or statistical physics that are described by CFTs that are not reflection-positive (often called non-unitary).
However, the conformal bootstrap described in section~\ref{sec:bootstrap} relies on this property in an essential way.}

\end{itemize}

We have seen that the Wightman functions let us define Schwinger functions by analytic continuation. But it turns out that the opposite is also true: the Osterwalder-Schrader theorem states that the properties of Schwinger functions listed above are sufficient to reconstruct Wightman functions.%
\footnote{There is in fact another property that is needed in the proof, called \emph{linear growth condition}. This property is very difficult to establish in quantum field theory. But within conformal field theory the linear growth condition is not necessary if one works instead with a set of ``Euclidean CFT axioms'', which are otherwise equivalent to the Osterwalder-Schrader axioms, and from which the Wightman axioms can be recovered, at least for correlation functions of up to 4 operators~\cite{Kravchuk:2020scc, Kravchuk:2021kwe}.}
This means that we can in fact focus on the Euclidean Schwinger functions for all our purposes, as any other physical observables can be reconstructed from them (we do not claim that this reconstruction is easy, though).

The only piece of information that we shall take for granted from the analysis of section~\ref{sec:quantum} is the unitarity bound on the scaling dimension of operators. Working out these bounds purely from the Schwinger functions is also possible~\cite{Minwalla:1997ka}, but we will not discuss this procedure here.

\subsection{From Euclidean space to embedding space}

Once we are dealing with correlation functions that transform under the Euclidean conformal group $\SO(d + 1, 1)$, it makes a lot of sense to make use of an analogy with the Lorentz group in $d+2$ dimensions to gain mileage.%
\footnote{This idea dates back to Dirac in 1936~\cite{Dirac:1936fq}.}
We already introduced in section~\ref{sec:classical} a set of coordinates in this  $(d+2)$-dimensional, \emph{embedding} space, and a metric obeying
\begin{equation}
	\eta_{MN} dX^M dX^N
	= (dX^\mu)^2 + (dX^{d+1})^2 - (dX^{d+2})^2.
\end{equation}
Note that the directions labeled with indices $\mu$ are all space-like: the time-like direction is $X^{d+2}$.

The question is how do we go from $X^M \in \mathbb{R}^{d+1,1}$ to $x^\mu \in \mathbb{R}^d$ without explicitly breaking the $(d+2)$-dimensional Lorentz symmetry. This can be done in two steps:
\begin{enumerate}

\item
Restrict our attention to the future light-cone $X^2 = 0$ with $X^{d+2} > 0$, which is an invariant subspace.

\item
Identify any two points related by a scale transformation on this light cone, i.e.~$X^M \sim \lambda X^M$ with $\lambda > 0$.

\end{enumerate}
This means that we are essentially considering a map between a point $x^\mu$ in $d$-dimensional Euclidean space and a light ray in a $(d + 2)$-dimensional Minkowski space-time.
To make the map explicit, we choose a section of the cone, which we will take to be $X^{d+1} + X^{d+2} = 1$,
and identify
\begin{equation}
	X^\mu = x^\mu,
	\qquad
	X^{d+1} = \frac{1 - x^2}{2},
	\qquad
	X^{d+2} = \frac{1 + x^2}{2}.
	\label{eq:embedding:section}
\end{equation}
Conformal transformations now act linearly in embedding space,
\begin{equation}
	X^M \to X'^M = \Lambda^M_{~N} X^N
\end{equation}
To obtain the action on $x^\mu$, we first map it to our preferred section of the cone using eq.~\eqref{eq:embedding:section}, apply the Lorentz transformation on $X^M$, then perform a (space-time dependent) rescaling $X'^M \to \lambda(X') X'^M$ to get back on the preferred section, and read off $x'^\mu = \lambda(X') X'^\mu$.

Local operators in Euclidean space must also be lifted to the embedding space, or at least to the null cone. On our preferred section, we declare
\begin{equation}
	\phi(X) \equiv \phi(x),
	\qquad 
	\left( X^2 = 0, ~ X^{d+1} + X^{d+2} = 1 \right).
\end{equation}
Then primary operators are defined on the rest of the cone by the scaling rule
\begin{equation}
	\phi(\lambda X) = \lambda^{-\Delta} \phi(X).
	\label{eq:embeddingconescaling}
\end{equation}
Note that this rule can only apply to \emph{primary} operators: descendants obtained acting with derivatives do not satisfy the same scaling property.
This choice gives the right transformation rules for primary operators under all infinitesimal conformal transformations. This can be verified explicitly (see the exercise below), or argued as follows: a conformal transformation is a composition of a $(d+2)$-dimensional Lorentz transformation, which locally acts on the operator like a rotation or a boost, followed by a position-dependent scale transformation to get back to the preferred section, which by the rule \eqref{eq:embeddingconescaling} amounts to a local scale transformation with weight given by the scaling dimension $\Delta$ of the operator. The combination of these two local transformations is precisely what we expect from the conformal transformation of an operator.

\begin{exercise}
	Verify that a Lorentz boost in the direction of $X^{d+1}$
	corresponds to a scale transformation in Euclidean space,
	in agreement with eq.~\eqref{eq:embeddingspacealgebra},
	and that operators transform accordingly. 
\end{exercise}

A bit more care is required to define operators that carry spin on the projective null cone, as the additional Lorentz indices in the directions of $X^{d+1}$ and $X^{d+2}$ imply that there are additional degrees of freedom that need to be constrained~\cite{Costa:2011mg}. This can be achieved by imposing transversality in embedding space, together with a gauge symmetry condition. But we do want to dive into this level of technical detail here. We will therefore restrict our attention to correlation functions of scalar primary operators only.

Once the transformation properties of local operators are clear, the construction of correlation functions of $n$ points $X_1$ to $X_n$ follows two simple rules:
\begin{itemize}

\item
The correlators must depend on Lorentz-invariant quantities in embedding space, i.e.~scalar products of the form $X_i \cdot X_j$. Since $X_i^2 = 0$ on the null cone, only scalar products with $i \neq j$ can appear.

\item
Applying a local scale transformation $X^M \to \lambda(X) X^M$ under which all primary operators $\phi_i$ with scaling dimension $\Delta_i$ satisfy
\begin{equation}
	\phi_i(X_i) \to \lambda(X_i)^{-\Delta_i} \phi_i(X_i),
\end{equation} 
the correlator must transform homogeneously as
\begin{equation}
	\langle \phi_1(X_1) \cdots \phi_n(X_n) \rangle
	\to \lambda(X_1)^{-\Delta_1} \cdots \lambda(X_n)^{-\Delta_n}
	\langle \phi_1(X_1) \cdots \phi_n(X_n) \rangle.
\end{equation}

\end{itemize}
These rules imply immediately that there cannot be one-point functions, as there simply is no corresponding Lorentz-invariant quantity on the projective null cone. The simplest non-trivial case is that of a 2-point function, which must obey
\begin{equation}
	\langle \phi(X_1) \phi(X_2) \rangle
	\propto (X_1 \cdot X_2)^{-\Delta}.
\end{equation}
Note that this is only consistent with the homogeneous scaling rule if both operators have the same scaling dimension, another property that was derived in the hard way in section~\ref{sec:quantum}.
To recover the dependence on the $d$-dimensional Euclidean coordinates, we simply use the identification \eqref{eq:embedding:section}, yielding
\begin{equation}
	X_1 \cdot X_2
	= -\frac{1}{2} (x_1 - x_2)^2.
\end{equation}
Upon fixing the proportionality factor in the above equation, one recovers the expected result
\begin{equation}
	\langle \phi(x_1) \phi(x_2) \rangle = 
	\frac{1}{\left[ (x_1 - x_2)^2 \right]^\Delta}.
\end{equation}

\subsection{3-point functions}

The embedding space formalism becomes very interesting when examining the 3-point function
\begin{equation}
	\langle \phi_1(X_1) \phi_2(X_2) \phi_3(X_3) \rangle,
\end{equation}
where the 3 scalar operators have now possibly distinct scaling dimensions $\Delta_1$, $\Delta_2$ and $\Delta_3$.
Solving the conformal Ward identities for this 3-point function would be an annoying task. Instead, by the above rules, we immediately know that this is a function of the 3 invariant quantities 
\begin{equation}
	X_1 \cdot X_2,
	\qquad
	X_1 \cdot X_3,
	\qquad
	X_2 \cdot X_3.
\end{equation}
Moreover, by the homogeneous scaling rule, the only possible form of the 3-point function is
\begin{equation}
	\langle \phi_1(X_1) \phi_2(X_2) \phi_3(X_3) \rangle
	\propto
	(X_1 \cdot X_2)^{\alpha_{12}}
	(X_1 \cdot X_3)^{\alpha_{13}}
	(X_2 \cdot X_3)^{\alpha_{23}}
\end{equation}
with the exponents satisfying 
\begin{align}
	\alpha_{12} + \alpha_{13} &= - \Delta_1,
	\nonumber \\
	\alpha_{12} + \alpha_{23} &= - \Delta_2,
	\\
	\alpha_{13} + \alpha_{23} &= - \Delta_3.
	\nonumber
\end{align}
This system of equations admits as unique solution
\begin{align}
	\alpha_{12} &= - \frac{\Delta_1 + \Delta_2 - \Delta_3}{2},
	\nonumber \\
	\alpha_{13} &= - \frac{\Delta_1 + \Delta_3 - \Delta_2}{2},
	\\
	\alpha_{23} &= - \frac{\Delta_2 + \Delta_3 - \Delta_1}{2}.
\end{align}
In terms of Euclidean coordinates, this can be written
\begin{equation}
	\langle \phi_1(x_1) \phi_2(x_2) \phi_3(x_3) \rangle
	= \frac{\lambda_{123}}
	{(x_{12}^2)^{\Delta_{12,3}}
	(x_{13}^2)^{\Delta_{13,2}}
	(x_{23}^2)^{\Delta_{23,1}}},
	\label{eq:3ptfunction}
\end{equation}
where we have introduced the compact notation
\begin{equation}
	x_{ij}^2 = (x_i - x_j)^2,
\end{equation}
and
\begin{equation}
	\Delta_{ij,k} = \frac{\Delta_i + \Delta_j - \Delta_k}{2}.
\end{equation}
Eq.~\eqref{eq:3ptfunction} is truly special. It should be compared with the most general 3-point function invariant under Poincaré and scale symmetry only: in this case, any term of the form
\begin{equation}
	(x_{12}^2)^\alpha
	(x_{13}^2)^\beta
	(x_{23}^2)^\gamma
\end{equation}
with
\begin{equation}
	\alpha + \beta + \gamma = \frac{\Delta_1 + \Delta_2 + \Delta_3}{2}
\end{equation}
satisfies the symmetry requirement, so that the 3-point function  could take the form
\begin{equation}
	\langle \phi_1(x_1) \phi_2(x_2) \phi_3(x_3) \rangle
	= \sum_i \frac{c_i}
	{(x_{12}^2)^{\alpha_i}
	(x_{13}^2)^{\beta_i}
	(x_{23}^2)^{\gamma_i}},
\end{equation}
with infinitely many free coefficients $c_i$. Instead, the conformal 3-point function \eqref{eq:3ptfunction} is fixed up to a \emph{unique} multiplicative coefficient $\lambda_{123}$.

To give another point of comparison, let us examine a 3-point function that involves a descendant operator. Since the focus is on scalar operators, let us act on the first operator in \eqref{eq:3ptfunction} with $\partial_\mu \partial^\mu$:
\begin{align}
	\langle \partial^2\phi_1(x_1) \phi_2(x_2) \phi_3(x_3) \rangle
	= 4 \lambda_{123} \bigg[ &
	\frac{\left( \Delta_1 - \frac{d-2}{2} \right) \Delta_{12,3}}
	{(x_{12}^2)^{\Delta_{12,3} + 1}
	(x_{13}^2)^{\Delta_{13,2}}
	(x_{23}^2)^{\Delta_{23,1}}}
	\nonumber \\
	& + \frac{\left( \Delta_1 - \frac{d-2}{2} \right) \Delta_{13,2}}
	{(x_{12}^2)^{\Delta_{12,3}}
	(x_{13}^2)^{\Delta_{13,2} + 1}
	(x_{23}^2)^{\Delta_{23,1}}}
	\nonumber \\
	& + \frac{\Delta_{12,3} \Delta_{13,2}}
	{(x_{12}^2)^{\Delta_{12,3} + 1}
	(x_{13}^2)^{\Delta_{13,2} + 1}
	(x_{23}^2)^{\Delta_{23,1} - 1}} \bigg].
	\label{eq:3pt:descendant}
\end{align}
Unlike the correlation function of primary operators, this is now the sum of three terms with distinct powers of the distances $x_{ij}^2$, all of which are individually consistent with Poincaré and scale symmetry.

The coefficients multiplying all three terms are peculiar: there are actually special cases in which this 3-point function takes the general form of eq.~\eqref{eq:3ptfunction}, i.e.~that of correlator involving primary operators.
These special cases are not accidental, but correspond to physically interesting situations:
\begin{itemize}

\item
If $\Delta_1 = \left| \Delta_2 - \Delta_3 \right|$, then either $\Delta_{12,3} = 0$ or $\Delta_{13,2} = 0$, and in both cases two terms on the right-hand side of eq.~\eqref{eq:3pt:descendant} vanish. This is a very special situation in which the primary 3-point function factorizes into a product of 2-point function, e.g.~when $\Delta_3 = \Delta_1 + \Delta_2$,
\begin{equation}
	\langle \phi_1(x_1) \phi_2(x_2) \phi_3(x_3) \rangle
	= \frac{\lambda_{123}}{(x_{13}^2)^{\Delta_1} (x_{23}^2)^{\Delta_2}}
	\propto \langle \phi_1(x_1) \phi_1(x_3) \rangle
	\langle \phi_2(x_2) \phi_2(x_3) \rangle.
\end{equation}
This situation is realized in generalized free field theory (definition in section~\ref{sec:bootstrap} below), where $\phi_3$ is a composite operator $\phi_3 \approx \phi_1 \phi_2$.

\item
If $\Delta_1 = \frac{d-2}{2}$, then the first two terms in eq.~\eqref{eq:3pt:descendant} vanish. As we saw in section~\ref{sec:quantum}, only a free scalar field satisfying the equation of motion $\partial^2 \phi_1(x) = 0$ can have such a scaling dimension. Since then obviously $\left[ K^\mu, \partial^2 \phi_1(x) \right] = 0$, it is natural that the correlator involving the equation of motion takes the form of a primary 3-point function. However, we also know that it must vanish identically, which implies that either the 3-point coefficient $\lambda_{123}$ vanishes, or that the additional condition $\Delta_1 = \left| \Delta_2 - \Delta_3 \right|$ is satisfied: this is for instance the case of a 3-point function involving the primary composite operator $\phi^2$,
\begin{equation}
	\langle \phi(x_1) \phi(x_2) \phi^2(x_3) \rangle
	= \frac{\lambda}{(x_{13}^2)^{(d-2)/2} (x_{23}^2)^{(d-2)/2}},
\end{equation}
which is non-zero but vanishes under the action of $\partial^2/(\partial x_1)^2$.

\end{itemize}

Finally, let us go back to Minkowski space-time through an analytic continuation in the opposite direction of what we did before. It is not hard to see that the Wightman function of scalar primary operators satisfies
\begin{equation}
	\bra{0} \phi_1(x_1) \phi_2(x_2) \phi_3(x_3) \ket{0}
	= \frac{\lambda_{123}}
	{(x_{12}^2)^{\Delta_{12,3}}
	(x_{13}^2)^{\Delta_{13,2}}
	(x_{23}^2)^{\Delta_{23,1}}},
	\label{eq:3ptfunction:Minkowski}
\end{equation}
where now the Minkowski distances between two points are defined by
\begin{equation}
	x_{ij}^2 = -(x_i^0 - x_j^0 - i \varepsilon)
	+ (\vec{x}_i - \vec{x}_j)^2.
	\label{eq:Minkowskidistance}
\end{equation}
Note that $x_{ij}^2 \neq x_{ji}^2$, and as a consequence, the Wightman 3-point function is not symmetric under the exchange of operators.
Nevertheless, for real operators satisfying $\phi_i(x)^\dagger = \phi_i(x)$, we must have
\begin{equation}
	\bra{0} \phi_1(x_1) \phi_2(x_2) \phi_3(x_3) \ket{0}
	= \bra{0} \phi_3(x_3) \phi_2(x_2) \phi_1(x_1) \ket{0}^*,
\end{equation}
which is only compatible with eq.~\eqref{eq:3ptfunction:Minkowski} if the coefficient $\lambda_{123}$ is real.
This property will in fact be essential in the conformal bootstrap discussed in section~\ref{sec:bootstrap}.

\subsection{4-point functions}

The embedding space technique can be used for higher-point functions as well, but things are getting more complicated.
To avoid dealing with four distinct operators and as many scaling dimensions, let us focus our attention on the case of 4 identical scalar operators,
\begin{equation}
	\langle \phi(X_1) \phi(X_2) \phi(X_3) \phi(X_4) \rangle.
\end{equation}
In this case there are 6 Lorentz invariants $X_i \cdot X_j$ with $i \neq j$. It is easy to see that a term of the form
\begin{equation}
	\frac{1}{(X_1 \cdot X_2)^\Delta (X_3 \cdot X_4)^\Delta}
\end{equation}
satisfies all the constraints of conformal symmetry. It is however not unique: so does
\begin{equation}
	\frac{1}{(X_1 \cdot X_3)^\Delta (X_2 \cdot X_4)^\Delta}.
\end{equation}
This shows right away that there is no hope of constraining the 4-point function as much as we did with the 3-point function.
In fact, one can construct two \emph{invariant} quantities out of the $X_i$:
\begin{equation}
	u = \frac{(X_1 \cdot X_2) (X_3 \cdot X_4)}
	{(X_1 \cdot X_3) (X_2 \cdot X_4)},
	\qquad\qquad
	v = \frac{(X_1 \cdot X_4) (X_2 \cdot X_3)}
	{(X_1 \cdot X_3) (X_2 \cdot X_4)}.
\end{equation}
These are called \emph{conformal cross-ratios}. 
Any function of $u$ and $v$ is conformally invariant, and the most general 4-point function can be of the form
\begin{equation}
	\langle \phi(X_1) \phi(X_2) \phi(X_3) \phi(X_4) \rangle
	\propto
	\frac{g(u,v)}{(X_1 \cdot X_2)^\Delta (X_3 \cdot X_4)^\Delta}.
\end{equation}
In terms of Euclidean coordinates, this can be written
\begin{equation}
	\langle \phi(x_1) \phi(x_2) \phi(x_3) \phi(x_4) \rangle
	= \frac{g(u,v)}{\left( x_{12}^2 x_{34}^2 \right)^\Delta}
	\label{eq:4pt}
\end{equation}
with
\begin{equation}
	u = \frac{x_{12}^2 x_{34}^2}{x_{13}^2 x_{24}^2},
	\qquad\qquad
	v = \frac{x_{14}^2 x_{23}^2}{x_{14}^2 x_{23}^2}.
	\label{eq:crossratios}
\end{equation}
To see that this is the most general result, let us consider the following argument based on a sequence of finite conformal transformations:
\begin{enumerate}

\item
Using translations, it is always possible to choose a reference frame in which $x_1 = 0$.

\item
Using a special conformal transformation, one can then send $x_4 \to \infty$ without moving $x_1$ away from the origin.

\item
A rotation can then be used to place $x_3$ along some chosen direction, followed by a scale transformation to get $x_3 = (0, \ldots, 0, 1)$, without touching the origin nor the point at infinity.

\item
Finally, there is still a subset of rotations that do not affect $x_3$, that can be used to move the point $x_2$ to the position $x_2 = (b, 0, \ldots, 0, a)$, shown later in figure~\ref{fig:conformalframes}.

\end{enumerate}
Note that the first 3 steps can be used to fix entirely the kinematics of a 3-point function, which explains why its only freedom is in the form of a multiplicative coefficient. Instead, with a 4-point function we are left with two quantities, $a$ and $b$, which are in one-to-one correspondence with the two conformal cross ratios.
It is in fact convenient to replace these two real numbers with a complex $z = a + i b$ and its conjugate $\bar{z} = a - i b$.
The cross-ratios are then related to $z$ and $\bar{z}$ by
\begin{equation}
	u = z \bar{z},
	\qquad\qquad
	v = (1 - z) (1 - \bar{z}).
\end{equation}
Schwinger functions are analytic at all non-coincident configurations of points, therefore the function $g(z)$ is a single-valued function over the complex plane minus the points $\{ 0, 1, \infty \}$. Note that $g$ is also subject to the \emph{crossing symmetry} of the 4-point function, which requires%
\footnote{It is standard in the CFT literature to call ``crossing symmetry'' the property of Schwinger functions to be symmetric under the exchange of operators, in analogy with scattering amplitudes. The two types of crossing are related, but they are not quite the same.}
\begin{equation}
	g(u, v) =
	g\left( \frac{u}{v}, \frac{1}{v} \right)
\end{equation}
upon exchanging the operators $\phi(x_1)$ and $\phi(x_2)$, as well as
\begin{equation}
	g(u,v) = \left( \frac{u}{v} \right)^\Delta g(v, u),
\end{equation}
upon $\phi(x_1) \leftrightarrow \phi(x_3)$.

The situation is more complicated in Minkowski space-time. One can still define cross ratios by eq.~\eqref{eq:crossratios}, with Minkowski distances defined through $i \varepsilon$ prescriptions as in \eqref{eq:Minkowskidistance}. But the Wightman functions are not analytic (they lie at the boundary of the domain of analyticity in complexified coordinates), and therefore $g$ is a multi-valued function. Its values can be reached by analytic continuation from a configuration in which all 4 operators live on a constant-time slice, in which case it coincides with the Schwinger function. For instance, keeping $x_1$, $x_3$ and $x_4$ on that time slice but letting the time component $b$ of $x_2$ become imaginary, one ends up in a configuration in which $z$ and $\bar{z}$ are both real but distinct. This procedure is in general tedious, and it is fair to say that the current understanding of conformal Wightman 4-point functions is still incomplete.%
\footnote{Working in momentum space is not helping, even though energy positivity reduces the number of non-trivial configurations: since the Ward identity for special conformal transformation is a second-order differential equation, the solution cannot be formulated in terms of invariants. The fact that there exist conformal invariants in position space is intimately connected with Ward identities being first-order differential equations.}

%%%%%%%%%%%%%%%%%%%%%%%%%%%%%%%%%%%%%%%%%%%%%%%%%%%%%%%%%%%%%%%%%%%%%%%

\section{State-operator correspondence and OPE}
\label{sec:OPE}

The construction of correlation functions could be continued beyond 4 points, but there is a good reason to stop here (at least in this course). We shall now see that any $n$-point function can be reduced to a $(n-1)$-point function using the operator product expansion (OPE). The procedure can be iterated until everything is expressed in terms of 2- and 3-point functions. The 4-point function will be examined as a typical situation in which this OPE can be applied, and higher-point functions will not be considered.

\subsection{The OPE in quantum field theory}

The operator product expansion in quantum field theory is the statement that when two local operators are ``sufficiently close'' to each other, they can be replaced by another local operator, or rather a sum of them:
\begin{equation}
	\phi_1(x) \phi_2(y) \xrightarrow{x \to y} \sum_i f_i(x-y) \phi_i(y),
	\label{eq:OPE:operators}
\end{equation}
It does not matter whether the operator $\phi_i$ on the right-hand side is inserted at $x$ or at $y$, or at the middle point $(x+y)/2$ since this expansion is valid in the limit in which $x$ and $y$ coincide. In fact, the proportionality factor $f_i$ might diverge in the limit $x \to y$, so this is to be understood in the sense of an asymptotic limit, whose radius of convergence is strictly-speaking zero.

In non-perturbative quantum field theory, the OPE can be formulated in terms of the Hilbert space: when the product of operators in eq.~\eqref{eq:OPE:operators} is acting on the vacuum, then the completeness of the Hilbert space implies that we can write
\begin{equation}
	\phi_1(x) \phi_2(y) \ket{0}
	= \sum_{\ket{\Psi}} \ket{\Psi}
	\bra{\Psi} \phi_1(x) \phi_2(y) \ket{0},
\end{equation}
where the sum is over states $\ket{\Psi}$ forming an orthonormal basis. Among these are states obtained acting with a local operator $\phi_i(y)$ on the vacuum, so that we have 
\begin{equation}
	\phi_1(x) \phi_2(y) \ket{0}
	= \sum_{i} f_i(x-y) \phi_i(y) \ket{0} + \ldots,
\end{equation}
where the object $f_i$ is related to a ratio of Wightman functions
\begin{equation}
	f_i(x-y) \approx
	\lim_{z \to \infty}
	\frac{\bra{0} \phi_i(z) \phi_1(x) \phi_2(y) \ket{0}}
	{\bra{0} \phi_i(z) \phi_i(y) \ket{0}}.
\end{equation}
This formulation remains however imprecise, and it is not quite obvious how to make it more rigorous using only general principles of quantum field theory.

In conformal field theory, however, the operator product expansion goes to a completely new level of rigor, thanks to the following observations:
\begin{itemize}

\item
2-point functions of primary operators are \emph{diagonal} (i.e.~only identical primaries have a 2-point function), which implies that states created by different primaries are orthogonal. The norm of primary states is also known in terms of the operators' normalization. 

\item
3-point functions are known (see section~\ref{sec:correlators}), which means that the proportionality coefficients $f_i$ are in fact fixed up to an overall multiplicative factor.%
\footnote{There is more than one factor if the operators $\phi_1$ and $\phi_2$ carry spin.}

\item
There is no other contribution to the OPE beyond those of local operators acting on the vacuum. This property is due to the state/operator correspondence that is discussed next.

\end{itemize}

\subsection{The state/operator correspondence}

In conformal field theory there exists a one-to-one correspondence between the states on a given time slice and local operators defined by their scaling dimensions and representations under the Lorentz group.

The fact that local operators define states is obvious in Minkowski space-time, by the Wightman axioms.%
\footnote{After appropriate smearing, as discussed in section~\ref{sec:quantum}.}
But it is also true in the analytic continuation to Euclidean space. To understand this, remember that a local operator inserted at a generic Minkowski coordinate $x$ can be expressed using eq.~\eqref{eq:P:exponentiated} as an operator living on the surface $x^0 = 0$ and evolved unitarily,
\begin{equation}
	\phi(x) \ket{0}
	= e^{i x^0 P^0} e^{-i \vec{x} \cdot \vec{P}} \phi(0) \ket{0}.
\end{equation}
Since the spectrum of $P^0$ is non-negative, this can be analytically continued to any value of $x^0$ in the upper-half complex plane; on the contrary, if $x^0$ had a negative imaginary part, then there would be states of arbitrarily high energy, with divergent norm. 
By going to purely imaginary values, $x^0 = i \tau$, we can therefore define a state of the theory through an operator insertion in Euclidean space, provided that this operator is inserted at Euclidean time $\tau > 0$.

This has an immediate interpretation in terms of the Euclidean path integral: 
a state on the Euclidean surface $\tau = 0$ (which is \emph{identical} to the original Minkowski time slice $x^0 = 0$) is defined by a path integral over all field configurations restricted to the region $\tau < 0$. This is valid with any number of operators inserted at separated points of Euclidean time $\tau < 0$ (including no operators, corresponding to the vacuum state). But the identification also works the other way around: a given state on the surface $\tau = 0$ defines a boundary condition for the path integral at $\tau < 0$, which might correspond to some number of local operators inserted in the ``past'' (or a superposition of such configurations).
In this way, any Euclidean correlation function in which the operators are ordered in Euclidean time $\tau$ can be given a Hilbert space interpretation.

Once we adopt the Euclidean path integral perspective, then it does not matter whether we use $P^0$ as the Hamiltonian, or another conformal generator, as long as the surface at $\tau = 0$ is part of the foliation that it defines. For instance, one can use the so-called conformal Hamiltonian $\frac{1}{2} (P^0 - K^0)$, which foliates Euclidean space as shown in figure~\ref{fig:foliations}.
We mentioned already in section~\ref{sec:classical} that this combination of generators is in fact equivalent to the dilatation generator $D$, in the sense that they are related by a rotation in the compact subgroup $\SO(d + 1)$ of the conformal group $\SO(d+1, 1)$.
This means that there exists a conformal transformation that maps the surface $\tau = 0$ to the unit sphere, and the other surfaces related by Hamiltonian evolution to spheres of different radii (see again figure~\ref{fig:foliations}).
Foliating Euclidean space with spheres centered at the origin and using the dilatation generator as a Hamiltonian is called \emph{radial quantization}. In comparison, the physically-equivalent choice of Hamiltonian  $\frac{1}{2} (P^0 - K^0)$ is often called \emph{N-S quantization}.

The importance of these two quantizations in conformal field theory is due to the fact that they have fixed points, unlike equal-time quantization: evolving back in ``time'', the path integral shrinks to a ball of arbitrarily small radius surrounding one of these fixed points (the origin in radial quantization, or the south pole in N-S quantization). This means that any state on the unit sphere, or on the plane $\tau = 0$, can be related by Hamiltonian evolution to a state \emph{localized} at the fixed point. It is therefore equivalent to a state created by a local operator inserted there.
This is in essence the argument showing that the state/operator correspondence goes both ways: a local operator defines a state, but any state also defines a local operator (or rather a superposition of local operators).

This correspondence provides a posteriori a justification for our choice to organize the operators into irreducible representations of the Lorentz group with a definite scaling dimension: the diagonalization can be performed at the level of the Hilbert space, and then radial quantization used to argue that each state corresponds to a local operator.
Note that there are both primary and descendant states, in the sense that the local operator does not have to be a primary.

Many introductory courses in conformal field theory begin in fact with radial quantization, as it provides a compelling picture in Euclidean space. The connection with unitary quantum field theory is not easy to establish, though. Hermitian conjugation in the unitary theory requires taking $x^0 = i \tau$ to $(x^0)^* = -i \tau$ , i.e.~it corresponds to a reflection across the surface of $\tau = 0$ in N-S quantization. In radial quantization, this becomes inversion $r \to r^{-1}$: correlation functions compute the overlap between states corresponding to a path integral inside the unit sphere with states corresponding to a path integral outside the sphere. The smearing of states required in Minkowski space-time is replaced by the limit $r \to \infty$ needed to define conjugate states in terms of local operators.
Once this limit is taken into account, the unitarity bounds can be derived using the interplay of generators $P_\mu$ and $K_\mu$, which are in fact conjugate to each other under inversion. This procedure reproduces the results of our section~\ref{sec:quantum}, but it might be less intuitive.

\subsection{The conformal OPE}

The lesson we learn from radial quantization is that every state in the Hilbert space can be written as a linear combination of states created by the action of a single local operator on the vacuum, including primaries as well as descendants. Applying this lesson to Wightman functions in Minkowski space-time, we can be tempted to write the complete Hilbert space as
\begin{equation}
	\mathscr{H} \stackrel{?}{=} \text{span}\big\{
	\ket{0}, 
	\phi(0) \ket{0}, 
	\partial_\mu \phi(0) \ket{0}, 
	\ldots
	\big\}.
	\label{eq:localstates}
\end{equation}
The problem is that these are not normalizable states. Instead, we can consider states created by primary operators inserted at any point $x$ in Minkowski space-time,
\begin{equation}
	\mathscr{H} \sim \text{span}\big\{
	\ket{0}, 
	\phi(x) \ket{0}, 
	\ldots
	\big\},
\end{equation}
where we implicitly understand that the local operators should be smeared against test functions. This gives a good description of the Hilbert space, but it is not very practical because primary operators inserted at different points (or smeared against distinct test functions) are not orthogonal to each other.
This leads us to consider yet another representation of the Hilbert space, using local operators in momentum space,
\begin{equation}
	\mathscr{H} \sim \text{span}\big\{
	\ket{0}, 
	\widetilde{\phi}(p) \ket{0}, 
	\ldots
	\big\}.
\end{equation}
States carrying different momenta are orthogonal to each other, and it is moreover sufficient to consider momenta $p$ inside the forward light cone.
This means that we can express the completeness of the Hilbert space through the resolution of the identity
\begin{equation}
	\mathds{1} = \ket{0} \bra{0}
	+ \sum_i \int\limits_{p^0 > \left| \vec{p} \right|} \!\! 
	\frac{d^dp}{(2\pi)^d}
	\frac{\widetilde{\phi}_i(p) \ket{0} \bra{0} \widetilde{\phi}_i(-p) }
	{2\pi C \, (-p^2)^{\Delta - d/2}}
	+ \ldots,
\end{equation}
where the sum is over all primary operators of the theory.
The denominator is simply the scalar 2-point function, eq.~\eqref{eq:2pt:momentum}, and states that carry spin have been omitted for simplicity: they could be included in the same way after inverting the  Lorentz tensor appearing in the 2-point function, which is positive-definite by unitarity.

This identity can be applied whenever we encounter a product of two local operators acting on the vacuum to rewrite the product as a sum of local operators. It takes its simplest form when the operators are all expressed in momentum space:
\begin{equation}
	\widetilde{\phi}_1(p_1) \widetilde{\phi}_2(p_2) \ket{0}
	= \sum_i \widetilde{f}_i \, \widetilde{\phi}_i(p_1 + p_2) \ket{0}.
	\label{eq:OPE:Minkowski}
\end{equation}
This is the conformal OPE in Minkowski space-time, which reduces higher-point functions to lower-point ones. It is trivial when applied to 2-point functions, as only the vacuum state appears in the sum (the identity operator can be interpreted as a primary operator in its own right). When applied to a 3-point function, it can be used to determine the proportionality factor $\widetilde{f}_i$ in terms of the scaling dimension, Lorentz representation, and 3-point coefficients $\lambda_{12i}$. Using this OPE further, a 4-point function can be decomposed into a sum of terms whose kinematics is completely determined by conformal symmetry.

This form of the OPE is however rarely used, for two reasons: our understanding of Wightman 3-point functions in momentum space is incomplete, and the convergence of this OPE is not particularly nice. By this, we mean that the OPE converges in the distributional sense, i.e.~only after smearing with smooth test functions.
In contrast, we will see next that the Euclidean OPE is \emph{absolutely} convergent!

Since we saw before that even Schwinger functions can be given a Hilbert space interpretation, we might as well formulate the OPE directly in Euclidean space. We write
\begin{equation}
	\phi_1(x_1) \phi_2(x_2) = \sum_i
	f_i\left( x_1 - x_2, \partial_{x_2} \right) \phi_i(x_2).
	\label{eq:OPE:Euclidean}
\end{equation}
Unlike the Minkowskian OPE \eqref{eq:OPE:Minkowski}, the convergence of this OPE is conditional: it is only true inside a correlation function in which the points $x_1$ and $x_2$ can be separated from all other insertions of local operators by a quantization surface, say a sphere in radial quantization. Since the center and radius of the sphere can be chosen at will thanks to translation and scale symmetry, this includes many (in fact nearly all) configurations of points, as we shall see below.

In the notation of eq.~\eqref{eq:OPE:Euclidean}, the object $f_i$ does not only depend on the distance between the two operators, but also on derivatives acting on the primary operator $\phi_i$. This is a compact way of including all descendants in the sum. Keep in mind however that there are infinitely many such descendants, so $f_i$ is itself an infinite series.
The only exception is the identity operator, which enters the OPE if and only if the two primary operators on the left-hand side of \eqref{eq:OPE:Euclidean} are identical: in a 2-point function, the OPE takes the form
\begin{equation}
	\langle \phi(x) \phi(0) \rangle = 
	\frac{1}{(x^2)^{\Delta_\phi}} \langle \mathds{1} \rangle,
\end{equation}
with $\langle \mathds{1} \rangle = 1$,
from which we deduce 
\begin{equation}
	f_\mathds{1}(x, \partial) = \frac{1}{(x^2)^{\Delta_\phi}}.
\end{equation}
When viewed as a primary operator, the identity does not have any descendant.
In all other cases, the value of $f_i(x, \partial)$ can be determined from the 3-point function: using the OPE in the scalar 3-point function \eqref{eq:3ptfunction}, one can for instance deduce that
\begin{equation}
	f_i(x, \partial)
	= \frac{\lambda_{12i}}{(x^2)^{\Delta_{12,i}}}
	\left[ 1 + a x^\mu \partial_\mu
	+ b_1 x^\mu x^\nu \partial_\mu \partial_\nu
	+ b_2 x^2 \partial^2 + \ldots \right]
\end{equation}
where
\begin{align}
	a &= \frac{\Delta_{i1,2}}{\Delta_i},
	\nonumber \\
	b_1 &= \frac{\Delta_{i1,2} (\Delta_{i1,2} + 1)}
	{2\Delta_i (\Delta_i + 1)},
	\\
	b_2 &= - \frac{\Delta_{i1,2} \Delta_{i2,1}}
	{4 \Delta_i (\Delta_i + 1) \left( \Delta_i - \frac{d-2}{2} \right)},
	\nonumber 
\end{align}
and so on. This expression is valid when $\phi_i$ is a scalar operator, but there are also operators with spin that enter the OPE, even though $\phi_1$ and $\phi_2$ are scalars: the operators being inserted at separated points $x_1 \neq x_2$ in space, their product carries angular momentum; for a point-like operator on the other side of the OPE, this angular momentum is realized as internal spin.
Note however that not all Lorentz representations appear in the OPE of two scalars, but only symmetric tensors. Moreover, when the operators are identical, there are only tensors with even numbers of Lorentz indices. The list of representations that appear can be determined from the group theory associated with Lorentz symmetry. On the contrary, any scaling dimension $\Delta$ can appear.

There are therefore still many unknowns in the OPE. We shall see in the next section that there are in fact always infinitely many primary operators in the sum. But when compared with the OPE in a generic quantum field theory, the conformal OPE is extremely rigid: in a scale-invariant QFT the coefficients $a$, $b_1$, $b_2$, $\ldots$, entering the definition of the coefficient $f_i(x, \partial)$ are all theory-dependent factors. In CFT, on the contrary, they are completely fixed by the kinematics. The only dynamical information is contained in the OPE coefficients $\lambda_{12i}$ multiplying the contribution of a primary and all of its descendants.

%%%%%%%%%%%%%%%%%%%%%%%%%%%%%%%%%%%%%%%%%%%%%%%%%%%%%%%%%%%%%%%%%%%%%%%

\section{The conformal bootstrap}
\label{sec:bootstrap}

Our approach to conformal field theory so far has been an \emph{algebraic} one:
in section~\ref{sec:quantum} we have defined primary operators as irreducible representations of the conformal group, characterized by their transformations under Lorentz symmetry and by a scaling dimension.
In section~\ref{sec:OPE} we have seen that these operators can be combined using the operator product expansion. The OPE coefficients play the role of the structure constants of this operator algebra.
The set of all scaling dimensions and representations under the Lorentz group of a theory, together with the OPE coefficients,
\begin{equation}
	\big\{ (\Delta_i, R_i), \lambda_{ijk} \big\}
	\label{eq:CFTdata}
\end{equation}
is called the \emph{CFT data}. A conformal field theory is completely defined by its CFT data: any correlation function can be computed from it, by repeated use of the OPE.

But does any CFT data define a good theory?
The answer to this question is no: the algebra of local operators must close, giving very constraining consistency conditions.%
\footnote{This is very similar in spirit to the classification of simple Lie groups, leading to the well-known families $\text{SU}(N)$, $\text{SO}(N)$, $Sp(2N)$ and a few exceptional groups.}
The precise formulation of this closure condition depends on whether we are talking about the Euclidean or Lorentzian conformal groups.
For a conformal field theory in Minkowski space-time, the crucial point is that the OPE must be consistent with the micro-causality condition~\eqref{eq:causality}. 
But the consequences of micro-causality are quite difficult to track down in practice.

Instead, there is a much simpler consistency condition for Schwinger functions in Euclidean space: since these functions are symmetric under the exchange of operators (a property that is in fact related to micro-causality), then the Euclidean OPE must be \emph{associative}. To illustrate this, let us focus on 4-point functions. Imposing associativity of the OPE on \emph{all} 4-point functions is in fact sufficient to guarantee associativity of higher-point functions. For simplicity let us consider as before the case of 4 identical scalar primary operators with scaling dimension $\Delta_\phi$:
\begin{equation}
	\langle \phi(x_1) \phi(x_2) \phi(x_3) \phi(x_4) \rangle
	= \frac{1}{(x_{12}^2 x_{34}^2)^{\Delta_\phi}}
	g(u, v).
	\label{eq:4pt:repeated}
\end{equation}

\subsection{Conformal blocks}

The simplest way to understand how this 4-point function can be expressed in terms of the CFT data is to use the OPE \eqref{eq:OPE:Euclidean} on it twice:
\begin{equation}
	\langle \phi(x_1) \phi(x_2) \phi(x_3) \phi(x_4) \rangle
	= \sum_i f_i\left( x_1 - x_2, \partial_{x_2} \right)
	f_i\left( x_3 - x_4, \partial_{x_4} \right)
	\langle \phi_i(x_2) \phi_i(x_4) \rangle
\end{equation}
There is a single sum on the right-hand side because the 2-point function is diagonal: it vanishes unless the two primary operators are identical. To put this in a convenient form, let us factor out the OPE coefficient $\lambda_i$ corresponding to the 3-point function $\langle \phi\phi\phi_i\rangle$, as well as a particular power of the distance, defining
\begin{equation}
	f_i(x, \partial)
	= \frac{\lambda_i}{(x^2)^{\Delta_\phi}}
	f'_i(x, \partial),
\end{equation}
The new function $f'_i$ is then of the form
\begin{equation}
	f'_i(x, \partial) = (x^2)^{\Delta_i/2}
	\left[ 1 + \ldots \right],
\end{equation}
where $\Delta_i$ is the scaling dimension of the operator entering the OPE (the \emph{internal} operator), not to be confused with the scaling dimension $\Delta_\phi$ of the operators in the original correlation function (the \emph{external} operators).
We have now
\begin{equation}
	\langle \phi(x_1) \phi(x_2) \phi(x_3) \phi(x_4) \rangle
	= \frac{1}{(x_{12}^2 x_{34}^2)^{\Delta_\phi}}
	\sum_i \lambda_i^2 g_i(u,v),
\end{equation}
where we have defined
\begin{equation}
	g_i(u,v) =
	f'_i\left( x_1 - x_2, \partial_{x_2} \right)
	f'_i\left( x_3 - x_4, \partial_{x_4} \right)
	\frac{1}{(x_{24})^{\Delta_i}}.
\end{equation}
This should be compared with the representation \eqref{eq:4pt:repeated} of the 4-point function, in terms of which
\begin{equation}
	g(u,v) = \sum_i \lambda_i^2 g_i(u,v).
\end{equation}
The $g_i$ are called \emph{conformal blocks}~\cite{Dolan:2000ut}: they represent the contribution of a single primary and of all its descendants to the 4-point function. They are conformally invariant,%
\footnote{In comparison, the correlation function is not invariant but \emph{covariant}.}
and must therefore be functions of the cross-ratios $u$ and $v$, even though this is not at all obvious from their definition. 

This definition is in fact not very practical: working out all the terms in the series of $f'_i$ is difficult, and there are also Lorentz indices that need to be contracted when the intermediate operator carries spin. But it is convenient to examine the limit $x_1 \to x_2$, $x_3 \to x_4$, or equivalently $u \to 0$: in this case, the leading term in the OPE shows that 
\begin{equation}
	g_i(u, v) = u^{\Delta_i/2} \left[ 1 + \ldots \right].
	\label{eq:conformalblocks:leadingterm}
\end{equation}
Primary operators with the lowest scaling dimensions give the leading contribution to the 4-point function in this limit. 
The operator with the absolute lowest scaling dimension is the identity, for which there are no descendants, so that%
\footnote{In this special case it does not make much sense to speak of an OPE coefficient, but it is conventional to take $\lambda_\mathds{1} = 1$.}
\begin{equation}
	g_\mathds{1}(u,v) = 1.
\end{equation}
Note that this statement is related to the more general \emph{cluster decomposition principle} in quantum field theory: since there is no absolute scale in CFT, the limit $x_1 \to x_2$ is equivalent to a limit in which the other points $x_3$ and $x_4$ are sent very far away, all the way to infinity, and in this case one expects on general grounds that the correlation function factorizes as
\begin{equation}
	\langle \phi(x_1) \phi(x_2) \phi(x_3) \phi(x_4) \rangle
	\xrightarrow{x_3, x_4 \to \infty}
	\langle \phi(x_1) \phi(x_2) \rangle
	\langle \phi(x_3) \phi(x_4) \rangle.
\end{equation}
This factorization is thus reproduced by the OPE in the limit $u \to 0$.

A better way of computing the conformal block is based on the Casimir invariant of the conformal group~\cite{Dolan:2003hv}.
From the group algebra, one can verify that the combination of generators~\eqref{eq:embeddingspacealgebra}
\begin{equation}
	\mathcal{C}_2 = -\frac{1}{2} J_{MN} J^{MN}
\end{equation}
commutes with all individual generators, e.g.
\begin{equation}
	\left[ \mathcal{C}_2, P^\mu \right] = 0.
\end{equation}
In group theory language, $\mathcal{C}_2$ is called the quadratic Casimir invariant.
Since it commutes with translations, its action on a state created by a local operator is the same no matter where the operator is inserted, or whether it is a primary or a descendant: for a scalar primary,
\begin{equation}
	\mathcal{C}_2 \, \phi(x) \ket{0}
	= \Delta (\Delta - d) \, \phi(x) \ket{0}.
\end{equation}

\begin{exercise}
	Compute the eigenvalue of the quadratic Casimir operator
	from the commutators of the conformal generators with $\phi(x)$.
	For simplicity, work at $x = 0$. 
\end{exercise}

\noindent
For a symmetric tensor with $\ell$ Lorentz indices, the eigenvalue contains an additional term that is simply the quadratic Casimir of the Lorentz/rotation group in $d$ dimensions,
\begin{equation}
	\mathcal{C}_2 \, \phi^{\mu_1 \ldots \mu_\ell}(x) \ket{0}
	= \big[ \Delta (\Delta - d) + \ell (\ell + d - 2) \big]
	\phi^{\mu_1 \ldots \mu_\ell}(x) \ket{0}.
	\label{eq:Casimireigenvalue}
\end{equation}

\begin{figure}
	\includegraphics[width=0.47\linewidth]{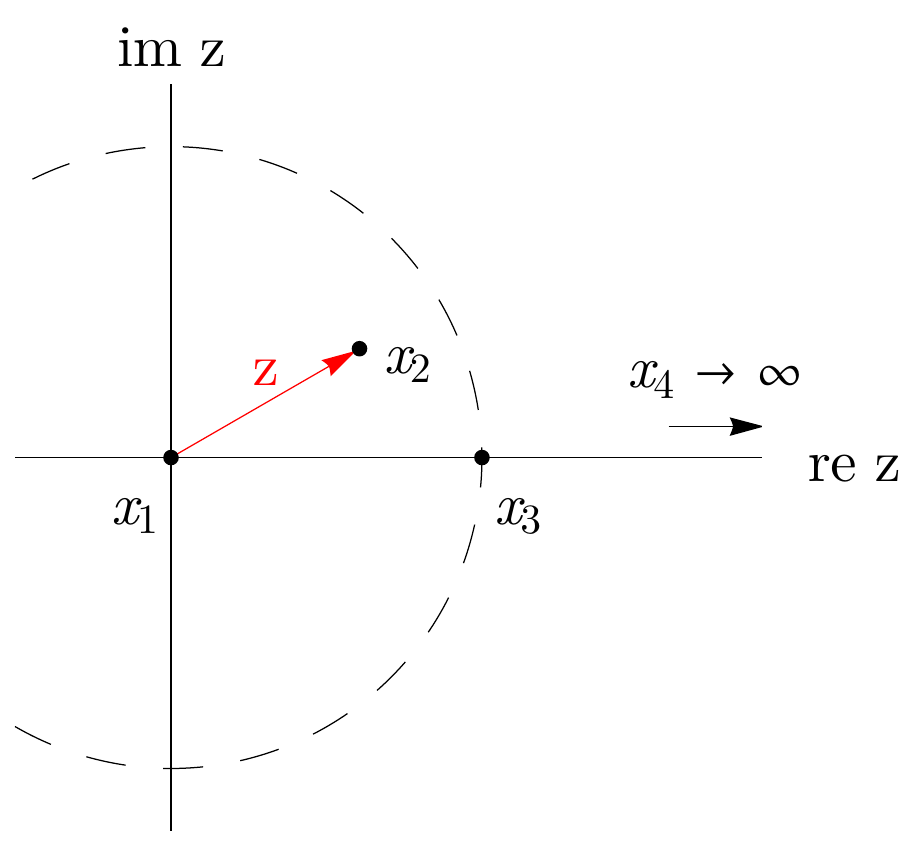}
	\hfill
	\includegraphics[width=0.47\linewidth]{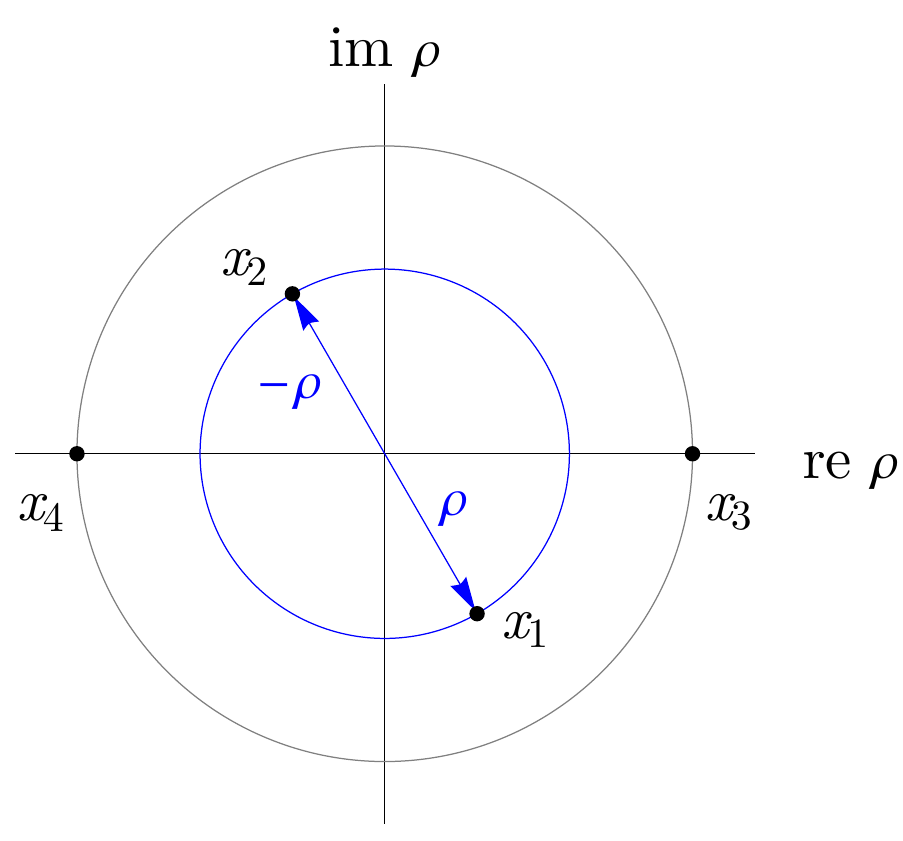}
	\caption{Two convenient conformal frames
	that describe the 4-point function.
	On the left, 3 points are mapped to $0$, $1$ and $\infty$
	using a conformal transformation, and the complex coordinate
	$z$ describes the position of the fourth point.
	Applying a conformal transformation in the plane,
	one can reach the configuration on the right, with operators
	placed at antipodal points on a pair of concentric circles,
	with the larger circle having unit radius; this configuration
	is parameterized by the complex coordinate $\rho$,
	satisfying $\left| \rho \right| \leq 1$.}
	\label{fig:conformalframes}
\end{figure}

The key idea is now to evaluate the value of this Casimir operator between states that are each created by two local operators, i.e.~compute the correlation function
\begin{equation}
	\langle \phi(x_1) \phi(x_2) \mathcal{C}_2 
	\phi(x_3) \phi(x_4) \rangle
\end{equation}
Applying the OPE once, this can be written
\begin{equation}
	\langle \phi(x_1) \phi(x_2) \mathcal{C}_2 \phi(x_3) \phi(x_4) \rangle
	= \sum_i f_i\left( x_3 - x_4, \partial_{x_4} \right)
	\langle \phi(x_1) \phi(x_2) \mathcal{C}_2 \phi_i(x_4) \rangle,
\end{equation}
where there are now two possible ways of computing each term on the right-hand side: either $\mathcal{C}_2$ acts to the right, and then the eigenvalue equation \eqref{eq:Casimireigenvalue} can be used, or it acts to the left. In the second case, the individual generators forming $\mathcal{C}_2$ must be commuted with $\phi(x_1)$ and $\phi(x_2)$ successively. This is most easily done in a convenient reference frame, for instance in the $z$-frame described in section~\ref{sec:correlators} and shown in figure~\ref{fig:conformalframes}. In this case, one finds that the action of $\mathcal{C}_2$ on each individual conformal block is a second-order differential operator in $z$ and $\bar{z}$, which we denote by $D_{z, \bar{z}}$. 
This means that every conformal block satisfies a differential equation of the form
\begin{equation}
	D_{z, \bar{z}} \, g_i(z, \bar{z})
	= \big[ \Delta_i (\Delta_i - d) + \ell_i (\ell_i + d - 2) \big]
	g_i(z, \bar{z}).
\end{equation}
Together with a boundary condition provided by  eq.~\eqref{eq:conformalblocks:leadingterm}, this is sufficient to determine the conformal blocks entirely. In even space-time dimensions, the solution is in the form of products of hypergeometric functions. In odd dimensions, there is no known closed-form solution, but the Casimir equation can be conveniently solved term-by-term in a series expansion.

Note that in our specific example where the external operators are all identical, the conformal blocks do not depend on the scaling dimension $\Delta_\phi$, but only on the scaling dimension $\Delta_i$ and spin $\ell_i$ of the internal operator.
This property is not true in general.

\subsection{OPE convergence}

So far we have been using the OPE in the 4-point function without worrying about its convergence. This is fine as long as the points $x_1$ and $x_2$ can be separated from $x_3$ and $x_4$ by a sphere. 
In terms of the coordinate $z$, this is obviously true for all $\left| z \right| < 1$ (the unit disk delimited by a dashed line in figure~\ref{fig:conformalframes}): then the product of operators $\phi(x_1) \phi(x_2)$ can be replaced by a sum of local operators at $x_1 = 0$ by the state-operator correspondence, and therefore the OPE converges since it is a Hilbert-space sum. But the domain of convergence is in fact much larger: radial quantization can be used around any point in space, not just $x_1$, and it is not hard to see that in most configurations one can draw a circle surrounding $x_1$ and $x_2$, but excluding $x_3$ and $x_4$. This is actually possible in all cases except when $z$ is real with $z > 1$.%
\footnote{In a generic frame, the only case in which the OPE $\phi(x_1) \times \phi(x_2)$ does not converge is when all four points are on a circle (this includes a line) and the points $x_1$ and $x_2$ are not consecutive, i.e.~they are separated by $x_3$ and $x_4$.}

This is even more easily seen in a different reference frame: using a conformal transformation in the plane of $z$, one can map the 4 points to a configuration depicted on the right-hand side of figure~\ref{fig:conformalframes}, with the pairs of points $(x_1, x_2) $ and  $(x_3, x_4)$ placed at antipodal points on two circles centered at the origin~\cite{Hogervorst:2013sma}. The map is given by
\begin{equation}
	\rho = \frac{z}{\left( 1 + \sqrt{1-z} \right)^2}
	\qquad\Leftrightarrow\qquad
	z = \frac{4\rho}{(1 + \rho)^2}.
\end{equation}
It takes all of the complex $z$ plane to the unit disk in $\rho$, with the half-line $z > 1$ being mapped to the unit circle. In this frame, it is now obvious that the OPE converges in radial quantization unless $\left| \rho \right| = 1$.

\begin{figure}
	\centering
	\includegraphics[width=0.4\linewidth]{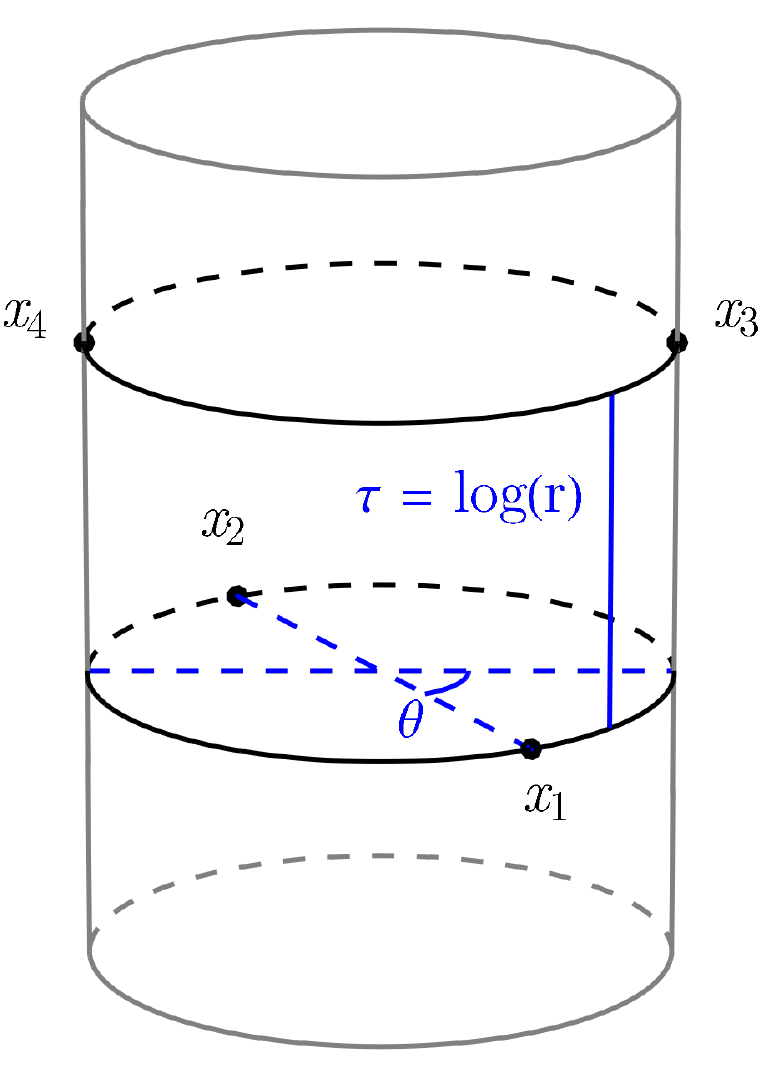}
	\caption{The $\rho$-coordinate configuration
	of figure~\ref{fig:conformalframes} as seen on
	the radial quantization cylinder. 
	The two segment $x_1 - x_2$ and $x_3 - x_4$ form an angle $\theta$,
	and they are separated in cylinder time by $\tau = \log(r)$.
	$\theta$ and $\tau$ are related to the flat-space
	configuration by $\rho = e^{\tau + i \theta} = r e^{i\theta}$.}
	\label{fig:rho}
\end{figure}

This conformal frame is also very useful for understanding the convergence properties of the OPE. It can be seen as a configuration on the radial quantization cylinder, as in figure~\ref{fig:rho}, with coordinates
\begin{equation}
	x_1 = (\tau, \vec{n}),
	\qquad
	x_2 = (\tau, -\vec{n}),
	\qquad
	x_3 = (0, \vec{n}'),
	\qquad
	x_4 = (0, -\vec{n}'),
\end{equation}
where $\tau = \log(r)$ is the ``time'' component, and the ``space'' components are unit vectors $\vec{n}$ and $\vec{n}'$ that parameterize a direction on the sphere $S^{d-1}$, and form an angle $\theta$ between them. The connection with the flat-space configuration is through
\begin{equation}
	\rho = r \, e^{i \theta}.
\end{equation}
A conformal block corresponds to the projection of this configuration onto intermediate primary and descendant states with scaling dimensions $\Delta_i + n$ ($n$ being an integer) and spin $j$ (whose range is determined by the spin $\ell$ of the primary). On general grounds, one can therefore expect the conformal block to take the form
\begin{equation}
	g_i(r, \theta)
	= \sum_{n,j} B_{n,j} r^{\Delta_i + n}
	\mathcal{C}_j^{(d-2)/2}(\cos\theta),
	\label{eq:radialexpansion}
\end{equation}
where $\mathcal{C}_j^{(d-2)/2}(\cos\theta)$ is a Gegenbauer polynomial obtained from the contraction of the symmetric and traceless tensor $\vec{n}^{\mu_1} \cdots \vec{n}^{\mu_j}$ with $\vec{n}'^{\mu_1} \cdots \vec{n}'^{\mu_j}$.
The coefficients $B_{n,j}$ are positive by unitarity because they are the norm of eigenstates of definite spin (labeled by $n$ and $j$). They are in fact rational functions of $\Delta_i$.
This representation is very useful: on the one hand, it gives an efficient way of evaluating the conformal block at any given level of precision by truncating the series in $n$, since by assumption $r < 1$; on the other hand, it shows that the $g_i$ are nice (i.e.~analytic) functions of scaling dimension $\Delta_i$ for all values of $\Delta_i$ above the unitarity bound.

Beyond individual conformal blocks, the $\rho$ coordinates also show how the OPE converges as a function of $r$: in a configuration with $r \ll 1$, the series is dominated by operators of low $\Delta$ (both primary and descendants), and a good approximation of the 4-point function can be obtained from a truncated OPE.
For instance, a configuration in which the operators are placed on the corners of a square corresponds to $z = \frac{1}{2}$, or $\left| \rho \right| = (1 + \sqrt{2})^{-2} \approx 0.17$. Even though the point $x_1$ and $x_2$ are not obviously close to each other in this configuration, $r = \left| \rho \right|$ is a small number and the convergence of the OPE is very fast.

This is a fantastic property of the \emph{Euclidean} OPE. The situation is not as nice in Minkowski space-time. 
The power-law dependence in $r$ seen in eq.~\eqref{eq:radialexpansion} is the consequence of exponential damping in Euclidean ``time'' $\tau$ ($r^{\Delta_i} = e^{\tau \Delta_i}$ with negative $\tau$), following from Hamiltonian evolution in radial quantization.
In Minkowski space-time, independently of the choice of quantization, the evolution is always \emph{unitary}, e.g.~$e^{i E t}$: there is in principle no reason why operators of higher scaling dimension should contribute less than those of lower dimension in a generic configuration.
This does not mean that the Minkowskian OPE is uninteresting, but it is more difficult to harness than the Euclidean one.

\subsection{The crossing equation and simple solutions}

We have so far only discussed one particular OPE, in which the points $x_1$ and $x_2$ were assumed to be closed to each other in a conformal sense. But since Schwinger functions are symmetric under the exchange of operators, there is no reason not to consider different OPEs, e.g.~between the operators $\phi(x_1)$ and $\phi(x_3)$, or $\phi(x_1)$ and $\phi(x_4)$.
In the vast majority of configurations of 4 points, the space can be cut in half by a sphere in three inequivalent ways. Even in the special case in which the 4 points are on a circle, there are still two inequivalent ways of surrounding pairs of points with a sphere. This means that we expect different OPEs to converge in all cases.%
\footnote{An equivalent statement is that Schwinger function can be obtained by analytic continuation from multiple Wightman functions with distinct orderings of operators.}

\begin{figure}
	\centering
	\Large
	$\sum\limits_i$ \quad
	\parbox[c]{45mm}
	{\includegraphics[width=\linewidth]{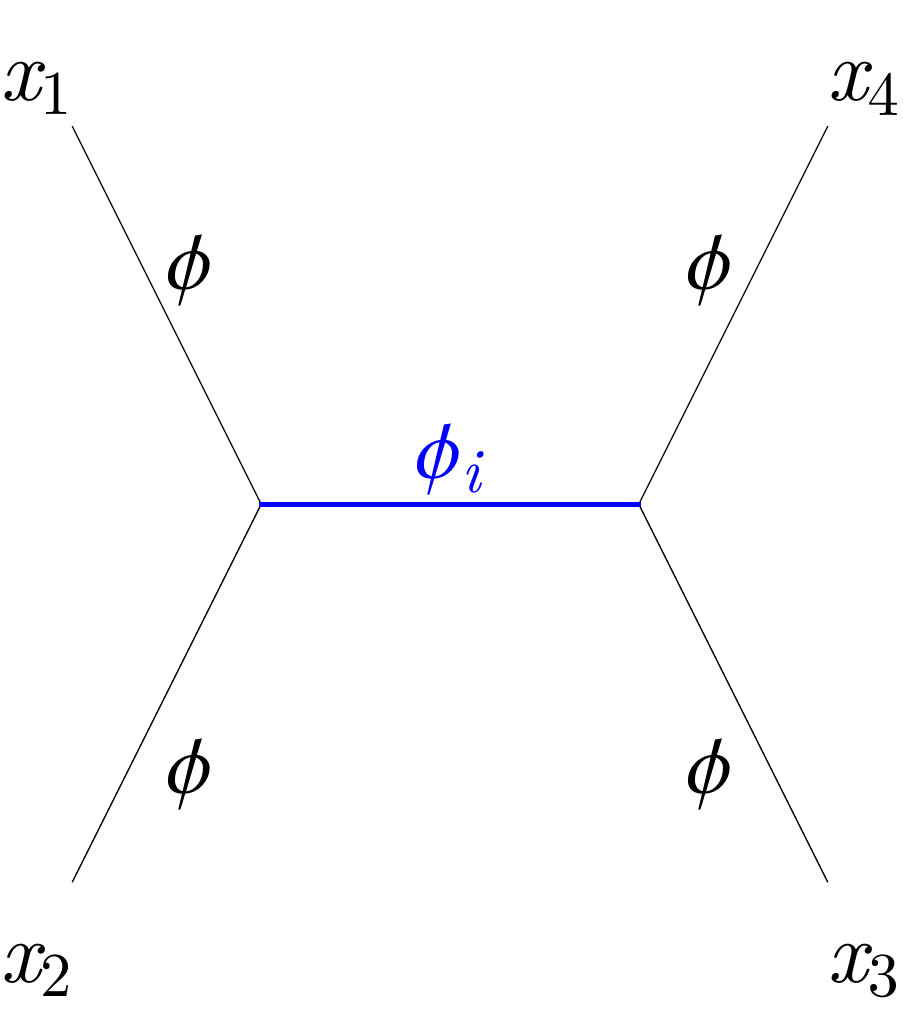}}
	\quad $= ~ \sum\limits_j$  \quad
	\parbox[c]{45mm}
	{\includegraphics[width=\linewidth]{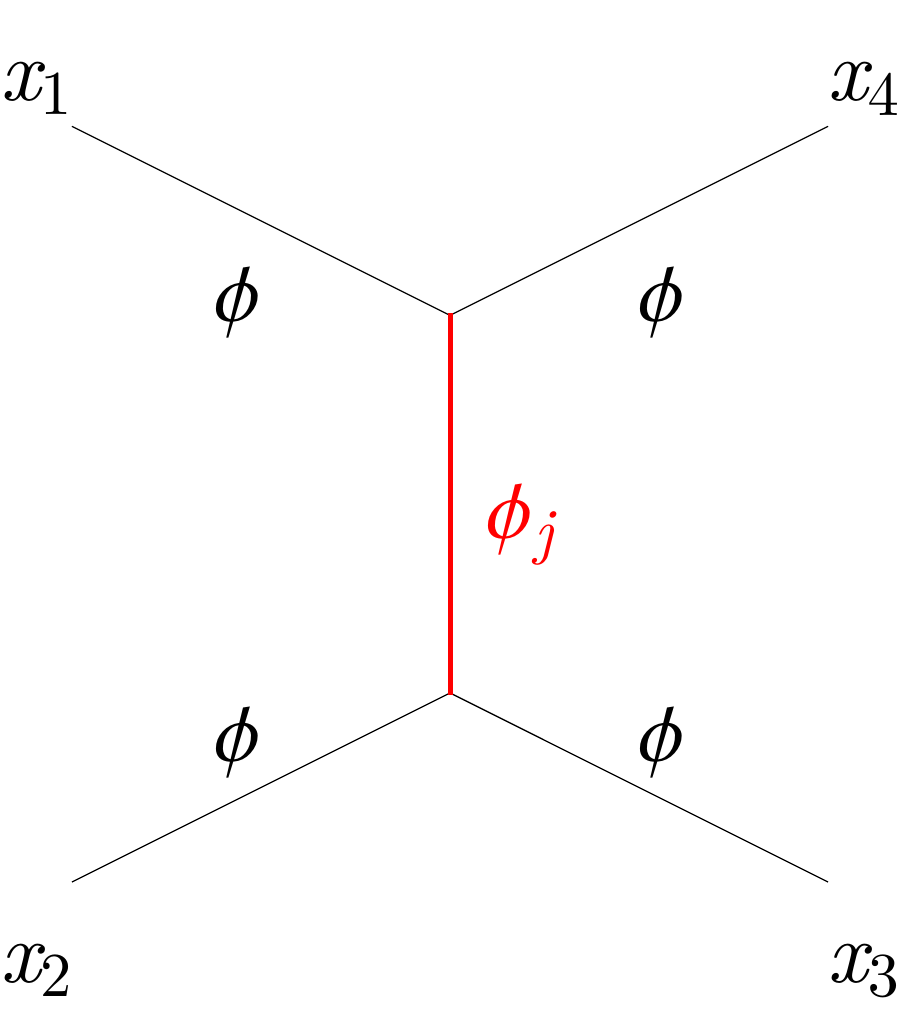}}
	\caption{Diagrammatic representation of the crossing
	equation at the core of the conformal bootstrap,
	relating two distinct OPEs.}
	\label{fig:crossing}
\end{figure}

This implies that besides the expansion
\begin{equation}
	g(u,v) = \sum_i \lambda_i^2 g_i(u,v),
\end{equation}
one can write a similar expansion for the 4-point function in a configuration related by crossing symmetry, e.g.
\begin{equation}
	\left( \frac{u}{v} \right)^{\Delta_\phi} g(v, u)
	= \left( \frac{u}{v} \right)^{\Delta_\phi} 
	\sum_i \lambda_i^2 g_i(v,u).
\end{equation}
Since the two are equal, we must have
\begin{equation}
	\sum_i \lambda_i^2 g_i(u,v)
	= \left( \frac{u}{v} \right)^{\Delta_\phi} 
	\sum_i \lambda_i^2 g_i(v, u).
	\label{eq:crossingequation}
\end{equation}
This equation is represented pictorially in figure~\ref{fig:crossing}. Note that in the case of identical external operators the two sums are over the same set of primary operators, but in general these could be different sums.

This is a relatively simple equation, but it is quite difficult to solve. We saw above that the OPE is dominated by intermediate operators of low $\Delta_i$ in the limit $u \to 0$, but this only applies to the left-hand side: taking the same limit on the right-hand side, one reaches the boundary of the domain of convergence of the OPE.
Using the known expression for the conformal blocks in terms of hypergeometric functions, it can be shown that the limit $z \to 1$ approaches a branch cut around which
\begin{equation}
	g_i(z, \bar{z}) \propto \log(1 - z).
\end{equation}
This means that the right-hand side of the crossing equation is dominated by terms of the form
\begin{equation}
	\left( \frac{u}{v} \right)^{\Delta_\phi} 
	g_i(v, u) \propto
	z^{2\Delta_\phi} \log(z),
\end{equation}
in the limit $z \to 0$.
A finite sum of terms of this form cannot reproduce the leading constant contribution from the identity operator on the left-hand side.
This is the first wisdom contained in the crossing equation \eqref{eq:crossingequation}: it can never be satisfied block-by-block, but only by an infinite sum of conformal blocks, and there must therefore be an infinite number of primary operators in the OPE.

Before looking into the clever way in which the conformal bootstrap deals with this problem, let us examine the simplest solution to crossing symmetry that we know of.
A possible function that transforms covariantly under the conformal group and is crossing symmetric is the following combination of 2-point functions:
\begin{align}
	\langle \phi(x_1) \phi(x_2) \rangle
	\langle \phi(x_3) \phi(x_4) \rangle
	+ \langle \phi(x_1) \phi(x_3) \rangle &
	\langle \phi(x_2) \phi(x_4) \rangle
	\nonumber \\
	+ & 
	\langle \phi(x_1) \phi(x_4) \rangle
	\langle \phi(x_2) \phi(x_3) \rangle,
\end{align}
corresponding to
\begin{equation}
	g(u,v) = 1 + u^{\Delta_\phi}
	+ \left( \frac{u}{v} \right)^{\Delta_\phi}.
\end{equation}
Performing an expansion at small $u$ and matching each term with a hypothetical primary operator, one finds an infinite spectrum of operators characterized by their spin $\ell$ and scaling dimension
\begin{equation}
	\Delta_i = 2\Delta_\phi + 2n + \ell
\end{equation}
where $n = 0, 1, 2, \ldots$
This defines a valid 4-point correlation function in a theory called \emph{generalized free field theory} (or sometimes mean free theory, or Gaussian theory). It is similar to a free theory in the sense that the OPE $\phi \times \phi$ contains operators that have the same scaling dimension and spin as \emph{composites} of the field $\phi$, of the schematic form
\begin{equation}
	\left[ \phi (\partial^2)^n \partial^{\mu_1} \ldots 
	\partial^{\mu_\ell} \phi \right].
\end{equation}
The free scalar theory is a special realization of this in which only operators with $n = 0$ are present in the OPE (the other vanish by the equation of motion $\partial^2 \phi = 0$, or they are descendants), and these operators are higher-spin conserved current.
Generalized free field theory is usually considered non-local: its definition through the above 4-point function does not include an energy-momentum tensor.%
\footnote{Also it can be defined through the non-local Lagrangian 
$\mathscr{L} = \frac{1}{2} \phi (\partial^2)^{d/2 - \Delta_\phi} \phi$.}
It is nevertheless a physically interesting case because the large-$N$ limit of gauge theory takes precisely this form: if one considers the 4-point function of a gauge-invariant composite operator such as the fermion bilinear $\bar{\psi}_a \psi_a$, then its decomposition into conformal blocks is given at leading order in $1/N$ by a sum of ``double-trace'' operators with scaling dimensions $2 \Delta_\phi + 2n + \ell$, whereas ``single-trace'' operators (among which the energy-momentum tensor) only enter at sub-leading order in $1/N$.

\subsection{The numerical bootstrap}

The revolutionary idea that gave birth to the modern conformal bootstrap appeared in 2008~\cite{Rattazzi:2008pe}.%
\footnote{The ``original'' conformal bootstrap was developed in two dimensions~\cite{Belavin:1984vu}, but it relies on very different techniques than the ``modern'' conformal bootstrap valid in any $d$. It leverages in particular the infinite Virasoro algebra that is specific to $d = 2$.}
The starting point is to rewrite the crossing equation \eqref{eq:crossingequation} as
\begin{equation}
	\sum_i \lambda_i^2 \left[ v^{\Delta_\phi} g_i(u,v)
	- u^{\Delta_\phi} g_i(v,u) \right] = 0,
\end{equation}
and to realize that it can be viewed as an equation in an infinite vector space
\begin{equation}
	\sum_i \lambda_i^2 \, \vec{F}_i(\Delta_\phi, \Delta_i, \ell_i) = 0,
	\label{eq:crossingequation:vector}
\end{equation}
where the $\lambda_i^2$ are positive coefficients (remember that $\lambda_i$ must be real in a unitary quantum field theory).
The components of the vector $\vec{F}_i$ correspond to $v^{\Delta_\phi} g_i(u,v) - u^{\Delta_\phi} g_i(v,u)$ evaluated at an infinite set of points $(u,v)$, or equivalently as the coefficients of a Taylor expansion around some preferred point.%
\footnote{Traditionally the numerical bootstrap uses a Taylor expansion around the point $u = v = \frac{1}{4}$, which corresponds to $z = \frac{1}{2}$, i.e.~the symmetric point that has the fastest converging OPEs.}
$\vec{F}_i$ has infinitely many components, but the equation is also true if we truncate it to any finite subset. Some interesting results can already be obtained from considerations about a 2-dimensional subset, but typically the more stringent bounds that are shown below require scanning over spaces with a large number of dimensions.

\begin{figure}[t]
	\centering
	\includegraphics[width=0.6\linewidth]{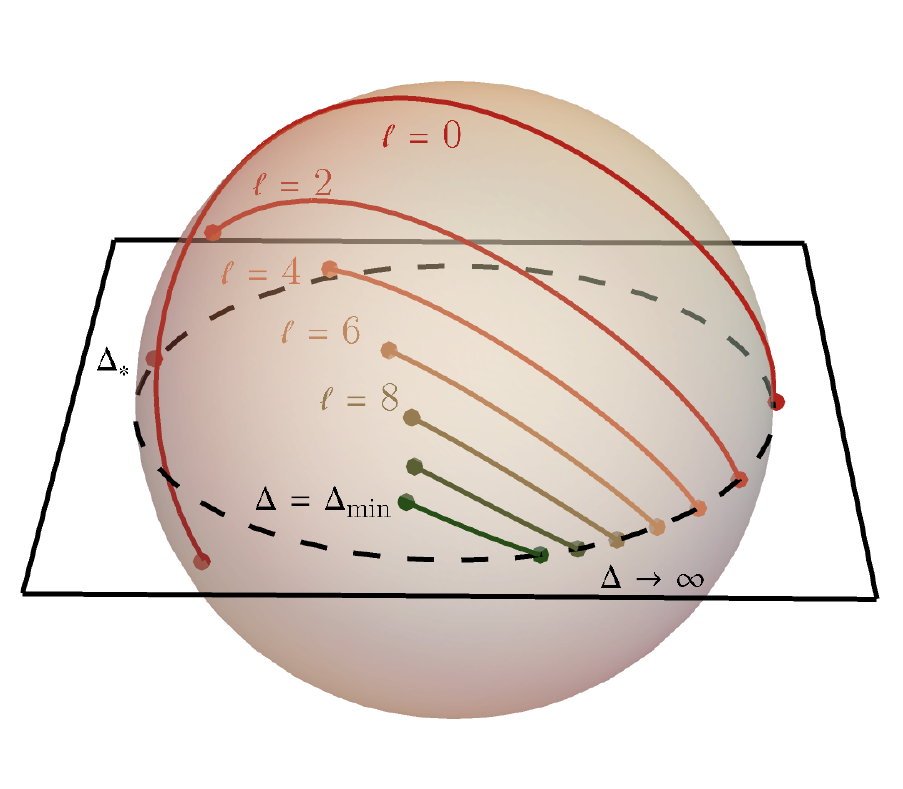}
	\caption{A toy example of the behavior of vectors $\vec{F}_i$
	in a 3-dimensional space.
	For each spin $\ell_i = 0, 2, 4, 6, \ldots$, $\vec{F}_i$
	draws a smooth curve on the unit sphere as $\Delta_i$ varies
	between the unitarity bound $\Delta_\text{min}$ and infinity.
	All curves are contained in the upper hemisphere, except
	the scalar curve $\ell = 0$ that enters the lower hemisphere
	when $\Delta < \Delta_*$.
	This means that a putative theory with no scalar primary operator
	satisfying $\Delta < \Delta_*$ is excluded, as the crossing equation 
	\eqref{eq:crossingequation:vector} cannot be satisfied:
	all vectors $\vec{F}_i$ point on the same side of
	the horizontal \emph{separating plane},
	so they cannot add up to zero.
	Note that the curves shown here are made up
	and do not correspond to actual functionals acting on
	conformal blocks.}
	\label{fig:separatingplane}
\end{figure}
\begin{figure}[t]
	\centering
	\includegraphics[width=0.8\linewidth]{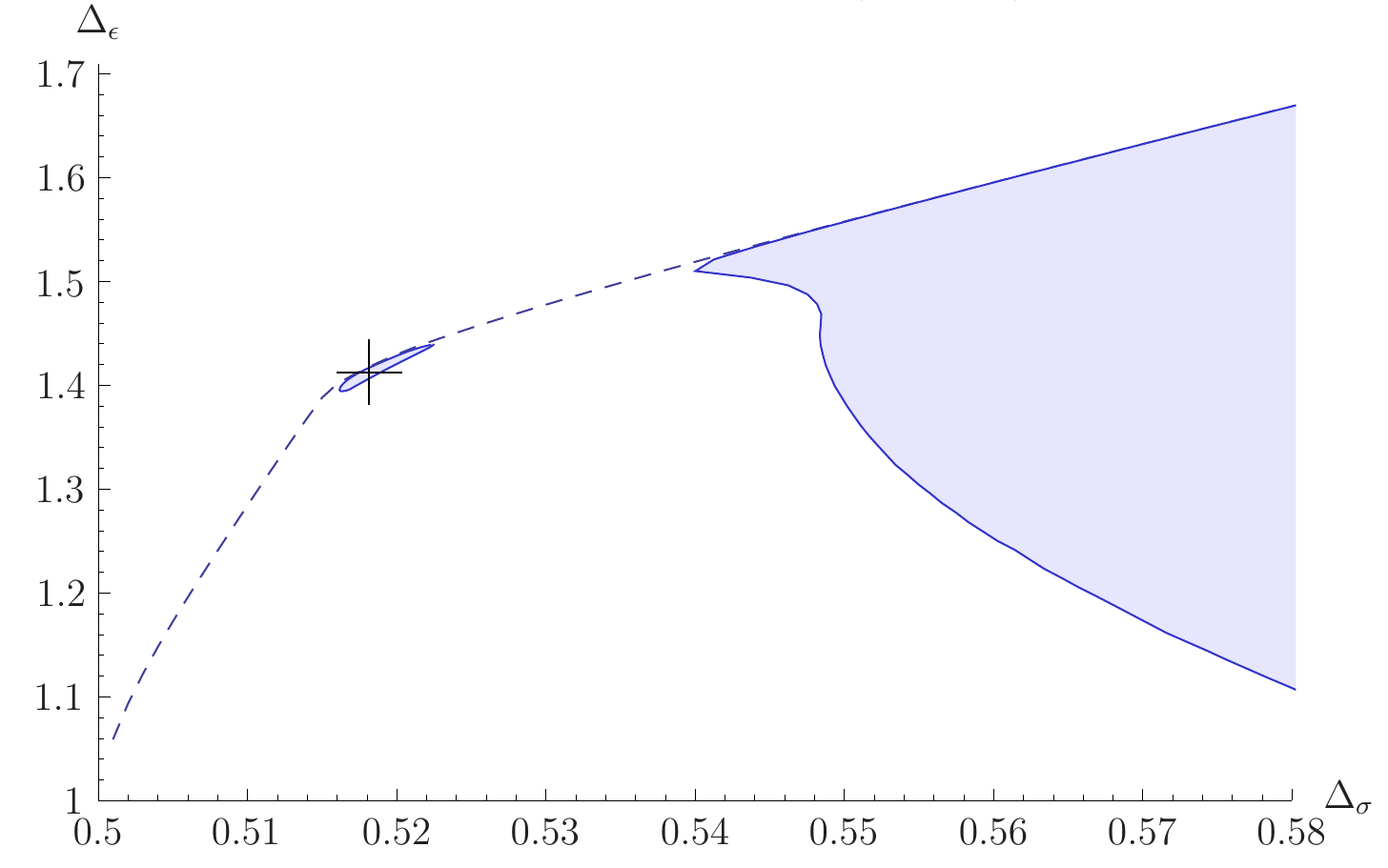}
	\caption{Figure taken from ref.~\cite{Kos:2014bka}, showing
	with a dashed line an upper bound on the
	scaling dimension $\Delta_\epsilon$ of the first operator in the OPE of
	two identical scalar operators with scaling dimension $\Delta_\sigma$,
	obtained from the conformal bootstrap procedure.
	Without making any theory-specific assumptions,
	this bound displays a ``kink'' close to the Ising CFT
	discussed in section~\ref{sec:Ising}, and marked with a cross.
	The blue region is the remaining allowed region after an 
	additional assumption is made, namely that the OPE takes
	the form of eq.~\eqref{eq:IsingOPE}.
	Note that this figure is outdated:
	by now the size of the island
	has shrunk to something smaller than the uncertainty
	of the best Ising model Monte-Carlo simulations.}
	\label{fig:Ising:island}
\end{figure}

The norm of the vector $\vec{F}_i$ is irrelevant in this equation, as it gets multiplied by a positive factor $\lambda_i^2$ that we do not know. But the direction in which the vector $\vec{F}_i$ points is crucial: if all vector $\vec{F}_i$ point in the same general direction, for all spins $\ell_i$ and all $\Delta_i$ above the corresponding unitarity bound, then the equation does not have a solution, because no non-trivial linear combinations of $\vec{F}_i$ can ever sum to zero.
A toy example of this mechanism is shown in figure~\ref{fig:separatingplane}.
This observation is at the core of all numerical bootstrap algorithms, which go along the following lines:
we begin with making a hypothesis about the scaling dimension of $\phi$, and of the operator with the lowest scaling dimension appearing in the OPE $\phi \times \phi$;%
\footnote{We know that there exists a crossing symmetric solution for all values of $\Delta_\phi$, namely the generalized free field theory one. Therefore it is not sufficient to make a hypothesis on $\Delta_\phi$ alone.} 
then we study the behavior of the vector $\vec{F}_i$ under this assumption and try to find a separating plane so that all vectors point on the same side of that plane; if such a plane is found, the equation cannot be satisfied, which implies that the hypothesis is wrong.
Iterating over this strategy allows to exclude entire regions in the space of possible parameters. Figure~\ref{fig:Ising:island} shows the result of this procedure in $d = 3$ dimensions. 

\subsection{Example: the Ising model in 3 dimensions}
\label{sec:Ising}

Very often, a lot of mileage is gained when one can input additional assumptions about some target theory.
Let us for instance consider the theory given by the action
\begin{equation}
	S = \int d^3x \left[
	-\frac{1}{2} \partial_\mu \phi \partial^\mu \phi
	- \frac{1}{2} m^2 \phi^2 - \frac{g}{4!} \phi^4 \right].
	\label{eq:IsingLagrangian}
\end{equation}
in which $g > 0$ so that the potential is bounded below.
Note that the free scalar field has mass dimension $\left[ \phi \right] = \frac{1}{2}$ in $d = 3$, and therefore
\begin{equation}
	\left[ m^2 \right] = 2,
	\qquad\qquad
	\left[ g \right] = 1.
\end{equation}
Since both $m^2$ and $g$ have positive mass dimensions, they are \emph{relevant} operators: they determine the dynamics in the low-energy limit (IR), but their importance decreases at high energy (UV): at energies $E \gg g, |m|$, this theory approaches the free, massless scalar field.
On the other hand, at low energies the physics depends obviously on $g$ and $m$: when $m^2 \gg g^2$, this is a theory of a massive scalar of mass $m$; when $m^2 \ll -g^2$, then the potential has two minima at
\begin{equation}
	\langle \phi \rangle = \pm \sqrt{- \frac{6m^2}{g}},
\end{equation}
with excitations of mass $\sqrt{2} |m|$ around it. Clearly, this theory has two phases, and there is therefore an intermediate value of $m^2$ where a phase transition must happen (or working in units set by $g$, a critical value of the dimensionless ratio $m^2/g^2$). Note that this theory has a $\mathbb{Z}_2$ symmetry corresponding to $\phi \to -\phi$, which is spontaneously broken in one phase and not in the other.

It turns out that the theory describing the IR physics exactly at the phase transition is a conformal field theory. Unlike the two phases surrounding it, it admits excitations of arbitrarily small energy (but they are not particles). How do we know that? Feynman diagram computations cannot be trusted in the IR: the approximation given by the (asymptotic) perturbative series is valid in the UV, but it breaks down in the IR, as in QCD. One way of understanding the phase transition is to examine the theory in $d \neq 3$ dimensions: the same theory in $d = 4 - \varepsilon$ has a perturbative fixed point, called the Wilson-Fisher fixed point. The position of the fixed point depends on the renormalization scheme, but there are other quantities that are scheme-independent. This is for instance the case of the anomalous dimension of the operator $\phi$.
In the limit $\varepsilon \to 1$,
the state-of-the-art, many-loop computation gives%
\footnote{See ref.~\cite{Henriksson:2022rnm} for a recent, comprehensive review of perturbative computations in the $\varepsilon$ expansion.}
\begin{equation}
	\gamma_\phi \approx 0.0182
	\qquad\Leftrightarrow\qquad
	\Delta_\phi = \frac{d-2}{2} + \gamma_\phi \approx 0.5182.
\end{equation}
This estimate of the scaling dimension coincides with that of a statistical physics model: the \emph{Ising model} is a theory of classical spins $\sigma_i = \pm 1$ on a lattice with nearest-neighbor interactions, characterized by the Hamiltonian
\begin{equation}
	H = - J \sum_{\langle ij \rangle} \sigma_i \sigma_j.
\end{equation}
This model has a critical value of $J$ at which the continuum limit is described by a CFT. At this value, Monte-Carlo simulations indicate that the correlation between two spins decreases with a power of the distance given by
\begin{equation}
	\Delta_\sigma \approx 0.5181.
\end{equation}
The two theories have completely different microscopic descriptions: one is a quantum field theory describing particles in Minkowski space-time, and the other is a simple theory on a Euclidean lattice. 
Even more surprisingly, the same critical exponent is found in experiments, such as the critical point of water and other liquids. This is an example of \emph{universality}.

The explanation for this coincidence is that there are not many candidate conformal field theories that can describe the phase transition of the $\phi^4$ theory and of the Ising model. These two theories have in common:
\begin{itemize}

\item
A global $\mathbb{Z}_2$ symmetry (respectively $\phi \to -\phi$ and $\sigma_i \to - \sigma_i$) that is broken in one phase and unbroken in the other;

\item
Exactly two relevant operators, both of them scalars, which we will denote $\sigma$ (odd) and $\epsilon$ (even under $\mathbb{Z}_2$).

\end{itemize}
\begin{table}
	\centering
	\begin{tabular}{|c|c||c|c||c|}
		\hline
		UV primary & $\Delta$ & IR primary
		& $\Delta$ & $\mathbb{Z}_2$
		\\ \hline
		$\phi$ & 0.5 & $\sigma$ & 0.51815 & odd
		\\
		$\phi^2$ & 1 & $\epsilon$ & 1.14126 & even
		\\
		$\phi^3$ & 1.5 & -- & -- & --
		\\
		$\phi^4$ & 2 & $\epsilon'$ & $> 3$ & even
		\\
		\hline
		$T^{\mu\nu} \sim \partial^\mu \phi \partial^\nu \phi$ & 3 &
		$T^{\mu\nu}$ & 3 & even
		\\
		\hline
	\end{tabular}
	\caption{List of primary operators in the free UV limit of the
	theory described by the Lagrangian~\eqref{eq:IsingLagrangian},
	and corresponding primaries in the IR CFT.}
	\label{tab:Isingops}
\end{table}
This second point requires some clarifications. In the lattice model, in addition to $J$, the phase diagram is characterized by an interaction with an external magnetic field, corresponding to the additional term $\delta H = - \mu \sum \sigma_i$ in the Hamiltonian. In the quantum field theory, this claim is supported by the list of relevant primary operators in the UV, given in the first column of table~\ref{tab:Isingops}. Two of these primaries can reasonably be expected to still be primary operators at the interacting fixed point: $\phi$ and $\phi^2$ have sufficiently low scaling dimensions to start with. On the contrary, $\phi^4$ is the relevant operator that triggers the renormalization group flow in the UV, so we expect it to become irrelevant in the IR (otherwise the flow would continue). $\phi^3$ is special: as soon as the coupling $g$ is turned on, the equation of motion $(-\partial^2 + m^2) \phi = \frac{1}{3!} \phi^3$ implies that it is not an independent primary operator anymore, but rather a descendant of $\phi$.
Among operators that carry spin, the only relevant one is the energy-momentum tensor $T^{\mu\nu}$: this operator exists in any QFT, and therefore it can be expected to be there in the IR as well, with an unchanged scaling dimension.

The conformal bootstrap philosophy is quite orthogonal to this discussion, in the sense that it does not care about the microscopic details (the lattice model or the Lagrangian theory).
It relies instead purely on symmetry arguments. Besides conformal symmetry that is built in the method, the $\mathbb{Z}_2$ transformation properties of $\sigma$ and $\epsilon$ imply that their OPEs have the following schematic form,
\begin{align}
	\sigma \times \sigma &= \mathds{1} + \epsilon + T^{\mu\nu} + \ldots
	\nonumber \\
	\sigma \times \epsilon &= \sigma + \ldots
	\label{eq:IsingOPE}
	\\
	\epsilon \times \epsilon &=
	\mathds{1} + \epsilon + T^{\mu\nu} + \ldots
	\nonumber
\end{align}
where the dots indicate contributions from all (infinitely many) irrelevant primary operators.
Using these properties as an input, and studying all 4-point functions involving $\sigma$ and $\epsilon$, one is able to make the allowed region of parameters ($\Delta_\sigma, \Delta_\epsilon$) shrink to an island surrounding the value known experimentally, see figure~\ref{fig:Ising:island}. With the more recent numerical analysis, the size of this allowed island has shrunk to become much smaller than the uncertainty of Monte-Carlo simulations, so that the current best theoretical prediction for the critical exponents of the Ising model stems from the conformal bootstrap ($\Delta_\sigma \approx 0.5181489$).

\begin{figure}
	\centering
	\includegraphics[width=0.8\linewidth]{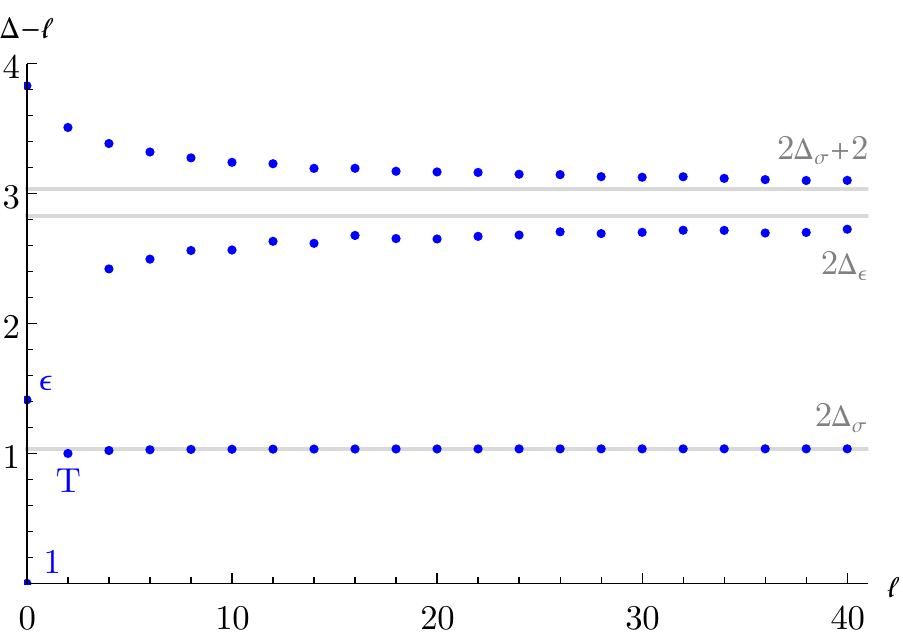}
	\caption{Spectrum of operators entering the 
	$\sigma \times \sigma$ OPE in the 3-dimensional Ising model,
	for spins up to 40
	and scaling dimensions of the order of the spin, 
	based on data taken from ref.~\cite{Simmons-Duffin:2016wlq}.}
	\label{fig:Ising:spectrum}
\end{figure}

This is very impressive, and it illustrates admirably how the concept of \emph{universality} in statistical physics works: given some very general assumptions 
(conformal symmetry, only two relevant operators, and a $\mathbb{Z}_2$ symmetry), the conformal bootstrap essentially establishes that there is a unique theory and provides excellent numerical results with rigorous error bars. Moreover, at the intersection between the allowed and disallowed regions (i.e.~on the boundary of the island), the crossing equation \eqref{eq:crossingequation:vector} must be satisfied in a peculiar way: many vectors must lie precisely on the separating plane, and only they can have non-zero OPE coefficients for the equation to be satisfied. Reading out the scaling dimensions $\Delta_i$ and spins $\ell_i$ associated with these vectors, one gets a numerical estimate of the spectrum of primary operators entering the OPE. Figure~\ref{fig:Ising:spectrum} shows such an estimate for the Ising model. The spectrum of operators is shown there as a function of the spin $\ell_i$ and of the difference $\Delta_i - \ell_i$ (called the ``twist'') so that an underlying structure clearly appears: the operators are organized into \emph{conformal Regge trajectories}. 

This observation matches the state-of-the-art analytical understanding of the crossing equation~\cite{Hartman:2022zik, Bissi:2022mrs}: every operator with a low scaling dimension $\Delta_i$ (hence low spin by the unitarity bound) is related by crossing symmetry to a family of operators with scaling dimensions $2\Delta_i + 2n + \ell$. This is what we saw in generalized free field theory, but it works reasonably well in the Ising model, as seen in the figure~\ref{fig:Ising:spectrum}: for $\Delta - \ell < 4$, there are 3 families of operators that are aligned with the values $\Delta_i - \ell_i = 2 \Delta_\sigma$, $2 \Delta_\epsilon$, and $2 \Delta_\sigma + 2$.
The equivalence between these families and operators of low scaling dimension has been established rigorously in the limit of infinite spin $\ell \to \infty$, and the leading correction to their twist, proportional to a power of the spin $\ell$, is well understood~\cite{Fitzpatrick:2012yx, Komargodski:2012ek}. In theories that are sufficiently close to generalized free fields, the equivalence goes even further~\cite{Alday:2016njk, Simmons-Duffin:2016wlq, Caron-Huot:2017vep}.

\subsection{Other conformal bootstrap results}

The reach of the conformal bootstrap is not limited to the Ising model.
Our understanding of many conformal field theories has been significantly improved by the conformal bootstrap. This is particularly true of theories in 2 and 3 dimensions. In 3 dimensions, there are generalizations of the Lagrangian \eqref{eq:IsingLagrangian} to $N$ scalar fields $\phi_a$, with interactions of the form $\left( \sum \phi_a^2 \right)^2$ and an $\text{O}(N)$ symmetry, for which the bootstrap has made compelling physical predictions:
\begin{itemize}

\item
The $\text{O}(2)$ model describes the superfluid transition of liquid helium. There is an unresolved discrepancy between the best measurement of this transition (an experiment performed in the space shuttle) and the best Monte-Carlo simulations. In this case, the theoretical bounds set by the conformal bootstrap are consistent with the Monte-Carlo result~\cite{Chester:2019ifh}, calling for a new experiment.

\item
The $\text{O}(3)$ model describes the critical behavior of Heisenberg magnets. These are statistical physics models in which the magnetization is isotropic: it can point in any direction in space, without a preferred direction. In practice, however, it seems very difficult to achieve the required level of isotropy on a lattice or in a solid: the $\text{O}(3)$ symmetry tends to be spontaneously broken to its cubic subgroup (the finite group of symmetries of the cube).
The crucial question here is whether the 4-index operator $\mathcal{O}_{ijkl}$ that can trigger a renormalization group flow from the $\text{O}(3)$ model to a CFT with cubic symmetry is relevant or not. Since its scaling dimension is accidentally very close to 3, this is hard to determine from lattice simulations. 
The conformal bootstrap has now established on rigorous grounds that the scaling dimension of this operator is below 3, and hence that the $\text{O}(3)$ symmetry naturally tends to be broken~\cite{Chester:2020iyt}.

\end{itemize}
There are many more interesting results and exciting open questions related to the conformal bootstrap, and this is not the place to list them exhaustively (the list would not stay up-to-date for long). For a recent summary, see for instance the \emph{Snowmass 2022 White Paper} on the numerical conformal bootstrap~\cite{Poland:2022qrs}.

%%%%%%%%%%%%%%%%%%%%%%%%%%%%%%%%%%%%%%%%%%%%%%%%%%%%%%%%%%%%%%%%%%%%%%%

\section{Conclusions}
\label{sec:conclusion}

There are many more known CFTs that we have not mentioned so far:
\begin{itemize}

\item
Free massless theories are conformal. This is the case of the free boson and free fermion in any number of dimensions, but also of the $n$-form gauge theory in $d = 2n + 2$ dimensions (e.g.~the free vector theory in $d = 4$). There are also theories with any number of free fields.
We are used to describe the Hilbert space of free theories as a Fock space (with states of a given ``particle number''), but there is also a conformal basis for them in terms of primary and descendants that turns out to be useful in the effective field theory approach, or in Hamiltonian truncation.%
\footnote{This is a numerical technique that studies renormalization group flows using the CFT data as input: the Hilbert space of a UV CFT is truncated by some cutoff before a relevant deformation is introduced; the new Hamiltonian can then be diagonalized numerically in the truncated basis, and IR observables measured; iterating the process while varying the cutoff gives a sense of how the observables depend on the scale~\cite{Fitzpatrick:2022dwq}.}

\item
There are theories in which the $\beta$-function has a perturbative fixed point. Typical examples are deformations of the free scalar theory with a potential of the type $\phi^n$, such as $\phi^4$ in $d = 4 - \varepsilon$ dimensions (it is possible to make sense of non-integer dimensions in perturbation theory): given the action
\begin{equation}
	S = \int d^dx \left[
	- \frac{1}{2} \partial_\mu \phi \partial^\mu \phi
	- \frac{g}{4!} \phi^4 \right],
\end{equation}
the $\beta$-function for $g$ can be computed to be
\begin{equation}
	\beta_g = \mu \frac{dg}{d\mu}
	= - \varepsilon + \frac{3g}{(4\pi)^2}
	+ \O(g^2),
\end{equation}
and it vanishes when
\begin{equation}
	\frac{g_*}{(4\pi)^2} = \frac{\varepsilon}{3}.
\end{equation}
Higher-order corrections are negligible in the limit $\varepsilon \ll 1$.
There are similar fixed points with a $\phi^3$ interaction in $d = 6 + \varepsilon$ dimensions, or with a $\phi^6$ interaction around $d = 3$.

\item
A similar type of fixed point can be found in the $\beta$-function of $SU(N_c)$ gauge theories with $N_f$ fermions (in the fundamental representation), which is given at leading order in the gauge coupling $\alpha = g^2/(4\pi)^2$ by
\begin{equation}
	\beta_\alpha = \mu \frac{d\alpha}{d\mu}
	= -\frac{2}{3} \alpha^2 \left(11 N_c - 2 N_f \right)
	+ \O(\alpha^3).
\end{equation}
When $N_f = \frac{11}{2} N_c$, the leading order term in this $\beta$-function vanishes, so the next-to-leading term becomes important. Around that value, the first two terms are of similar importance, and one finds a perturbative fixed point when $N_f \lesssim \frac{11}{2} N_c$. This is called the (Caswell-)Banks-Zaks fixed point. A theory in this situation is asymptotically free like QCD, but it approaches an interacting conformal field theory at low energy.
For an $SU(3)$ gauge theory like QCD, this critical value is at $N_f = 16.5$. There is strong evidence from lattice simulation that a theory with $N_f = 16$ is conformal. On the other hand, a theory with low $N_f$ (QCD has 3 light quarks) is clearly confining, meaning that its low-energy limit is a theory of massless Goldstone bosons (if the quarks are massless), or of massive pions. There is a critical value $N_f^* \approx 12$ above which we expect an interacting conformal field theory in the low-energy limit. The domain $N_f^* \leq N_f \leq \frac{11}{2} N_c$ is called the \emph{conformal window}. Note that gauge theories with different gauge groups and/or fermions coupling differently to the gauge fields (i.e.~transforming in different representations) can also have a conformal window. It is even possible to engineer gauge theories with perturbative UV fixed points using not only fermions but also scalars.

\item
There are also theories with extended supersymmetry in which the $\beta$-function is exactly zero at all orders in perturbation theory, for instance $\mathcal{N} = 4$ supersymmetric Yang-Mills. More generally, in theories with sufficiently many supersymmetries, it is often sufficient to engineer the matter content to make the $\beta$-function zero at leading order, and then non-renormalization theorems ensure their vanishing at all orders. 
When combined with supersymmetry, the conformal algebra gets extended to a bigger and more rigid structure. All superconformal algebras have been classified and are restricted to dimensions $d \leq 6$ (see refs.~\cite{Eberhardt:2020cxo, Argyres:2022mnu} for recent reviews).

\item
In two dimensions, we mentioned in section~\ref{sec:classical} that there are many more Killing vectors and hence conformal generators. In theories that have an energy-momentum tensor, this gives rise to the infinite-dimensional Virasoro algebra, which has been used to solve completely a class of \emph{minimal models}. 
One can also place a 2-dimensional CFT on a torus and study the modular properties of its partition function in a bootstrap setup. This is a vast topic that has its own literature~\cite{DiFrancesco:1997nk, Schellekens:1996tg, Polchinski:1998rq, Ginsparg:1988ui, Gaberdiel:1999mc, Ribault:2014hia}.

\item
Finally, another example of CFT is the family of theories obtained by putting a quantum field theory in $(d+1)$-dimensional anti-de-Sitter (AdS) space-time. AdS admits a compactification with a spherical boundary, and correlation functions on that boundary are isomorphic to that of a $d$-dimensional conformal field theory.%
\footnote{Another type of correspondence applies to late-time correlators in de Sitter space-time~\cite{Baumann:2022jpr}.}
There are two quite distinct situations:
On the one hand, it is possible to consider theories of quantum gravity that approach AdS space-time asymptotically; these are dual to a true CFT with an energy-momentum tensor (and some peculiar properties).
On the other hand, one can also place a local quantum field theory in a fixed AdS background, and study the CFT correlators on its boundary; this CFT does not necessarily have an energy-momentum tensor, and it is in fact very similar to the generalized free field theory (or a large-N theory).
Studying the limit of infinite AdS radius in this latter approach reveals interesting connections with flat-space scattering amplitudes. This observation is at the heart of the recently-revived S-matrix bootstrap techniques~\cite{Kruczenski:2022lot}.

\end{itemize}
This non-exhaustive list shows how important conformal field theory has become in theoretical physics.
But as particle physicists know, nature is certainly not scale-invariant, so why should we care so much about CFT?

Sometimes the best motivation to study a subject is its beauty, and we hope that this modest introduction to the topic is reflecting at least part of its appeal.
But there are also very practical reasons for particle physicists to get interested in CFT. For one, some of the conformal bootstrap results presented here provide a unique glimpse into the dynamics of strongly-coupled quantum field theory, one that is impossible to reach with perturbation theory.
Moreover, the constraints on possible CFTs in Minkowski space in $3+1$ space-time dimensions that will be derived in the future will certainly teach us important lessons about gauge theories such as QCD, grand unified theories, or various beyond-the-standard-model scenarios.
So far, generic bootstrap bounds have remained relatively weak and featureless in $3 + 1$ dimensions. The CFTs that are known to exist are gauge theories, and this provides an additional difficulty since the information on the gauge group is encoded in the CFT data in a complicated way.
This should not however be taken as a failure of the conformal bootstrap, but rather as a challenge for the next generation of theoretical physicists!

\vspace{5mm}

\setcounter{secnumdepth}{0}
\section{Acknowledgments}

I am indebted to the Institute for Theoretical Physics at the University of Bern (Mikko Laine and Thomas Becher in particular) for giving me the opportunity to teach a graduate class on my favorite topic. I would like to thank the students and postdocs who took part in the class for their involvement and their critical questions, 
as well as Johan Henriksson, Slava Rychkov, Flip Tanedo, and Luca Vecchi for their feedback on the notes.
I would like finally to say a big thanks to all members of the conformal bootstrap community for several enjoyable years of research together, which I wished could have continued.

\vspace{10mm}

\begin{center}
	\parbox{0.85\linewidth}{\itshape
	[Courses] are fantastically good for learning physics. The lecturer learns a lot of physics. After my first few studies, just about everything I learned about physics came from teaching it. I don’t know if the students learned a lot, but I certainly did. So I consider teaching physics very important.}
	
	\vspace{2mm}
	
	\parbox{0.85\linewidth}{\raggedleft
	Leonard Susskind \cite{Susskind}}
\end{center}

\bibliography{Bibliography}
\bibliographystyle{utphys}

\end{document}